\documentclass[twocolumn]{aastex63}

\usepackage{graphicx}
\usepackage{float}

\usepackage{aas_macros}
\usepackage{color}
\usepackage{url}

\usepackage{enumitem}
\usepackage{booktabs}
\usepackage{multirow}

\newcommand{\bt}[1]{  {\normalfont #1}   }

\newcommand{\rb}[1]{   {\normalfont #1}   }

\newcommand{\Fig}[1]{Fig.~\ref{#1}}

\shorttitle{Planes of satellites I}  
\shortauthors{Santos-Santos et al.}

\begin{document}


\title{
Planes of satellites around simulated disc galaxies: I.- Finding  high-quality planar configurations from positional information   and  their comparison to  MW/M31 data}

\author{Isabel Santos-Santos}
\affiliation{Department of Physics and Astronomy, University of Victoria, Victoria, BC, Canada V8P 5C2}
\affiliation{Departamento de F\'isica Te\'orica, Universidad Aut\'onoma de Madrid, E-28049 Cantoblanco, Madrid, Spain}


\author{Rosa Dom\'inguez-Tenreiro}
\affiliation{Departamento de F\'isica Te\'orica, Universidad Aut\'onoma de Madrid, E-28049 Cantoblanco, Madrid, Spain}
\affiliation{Centro de Investigaci\'on Avanzada en F\'isica Fundamental, Universidad Aut\'onoma de Madrid, E-28049 Cantoblanco, Madrid, Spain}

\author{H\'ector Artal}
\affiliation{Intelligent Light, Rutherford, NJ 07070, USA}

\author{Susana E. Pedrosa}
\affiliation{Instituto de Astronom\'ia y F\'isica del Espacio, CONICET-UBA, 1428, Buenos Aires, Argentina}
\affiliation{Departamento de F\'isica Te\'orica, Universidad Aut\'onoma de Madrid, E-28049 Cantoblanco, Madrid, Spain}

\author{Lucas Bignone}
\affiliation{Departamento de Ciencias F\'isicas, Universidad Andr\'es Bello, Santiago, Chile}

\author{Francisco Mart\'inez-Serrano}
\affiliation{Dassault Systemes S.E.,  28020 Madrid, Spain}

\author{M.\'Angeles G\'omez-Flechoso}
\affiliation{Departamento de F\'isica de la Tierra y Astrof\'isica, Univ. Complutense de Madrid, Madrid, Spain}

\author{Patricia B. Tissera}
\affiliation{Departamento de Ciencias F\'isicas, Universidad Andr\'es Bello, Santiago, Chile}
\affiliation{Millenium Institute of Astronomy, Universidad Andr\'es Bello, Santiago, Chile}

\author{Arturo Serna}
\affiliation{Departamento de F\'isica Aplicada, Univ. Miguel Hern\'andez, Elche, Spain}

\begin{abstract}

We address the 'plane of satellites problem' by studying planar configurations around two disc galaxies with no late major mergers, formed in zoom-in hydro-simulations.
Due to the current lack of good quality kinematic data for M31 satellites, we use only positional information. So far, positional analyses of simulations are unable to find planes as thin and populated as the observed ones. Moreover, they miss systematicity and detail in the plane-searching techniques, as well as in the study of the properties and quality of planes, both in simulations or real data.
To fill this gap, i) we extend the 4-galaxy-normal density plot method  \citep{Pawlowski13} in a way designed to efficiently identify the best quality planes (i.e., thin and populated) without imposing extra constraints on their properties,
and ii), we apply it for the first time to simulations. 
Using zoom-in simulations  allows us to mimic MW/M31-like systems regarding the number of satellites involved as well as the galactic disc mass and morphology, in view of possible disc effects.

At all timesteps analyzed in both simulations we find satellite planar configurations that are compatible, along given time intervals, with all the spatial characteristics of observed planes identified using the same methodology.
However, the fraction of co-orbiting satellites within them is in general low,
suggesting time-varying satellite membership.
We conclude that high-quality positional planes of satellites are  not infrequent in $\Lambda$CDM-formed disc galaxies with a quiet assembly history.
Detecting kinematically-coherent, time-persistent  planes demands considering 
the full six-dimensional phase-space information of satellites.

\end{abstract}




\section{Introduction}\label{intro}


The so-called ``small scale problems in $\Lambda$CDM refer to the discrepancies between the predictions 
for dwarf galaxies in the standard cosmological model
 as first revealed by dark matter (DM)-only cosmological simulations,
 and the actual observed properties dwarfs present 
 \citep[see][for a review]{Bullock17_review}.
 Among them, the planar configurations satellite galaxies show around their hosts (``Planes of satellites problem", see \citealt{Pawlowski18} for a review), 
observed in the local Universe, have long been considered 
 as one of the most challenging.

The high degree of anisotropy of Milky Way (MW) satellite positions, which appear forming a common plane approximately perpendicular to the Galactic disc, was noted several decades ago 
  \citep{Lynden76,Kunkel76} and first quantified by \citet{Kroupa05}  
with the then-known 11 ``classical"\footnote{Fornax, LMC, SMC, Draco, Leo II, Carina, Sculptor, Sextans I, Leo I, Sagittarius dSph, Ursa Minor.} satellites.
With the addition of globular clusters, streams and newly discovered fainter satellites (specially with SDSS, \citealt{York00}),  
 the anisotropy yet but increased,
as  these objects contributed further to a
``vast polar structure (VPOS)" around the MW
 \citep{Pawlowski12}.
Anisotropy among  Andromeda's (M31) satellites 
 was first noted by \citet{Koch06} and \citet{Metz07}, and later confirmed with a larger sample of satellites including those
 recently discovered  with the PAndAS survey  \citep[Pan-Andromeda Archaeological Survey,][]{Ibata13}.
It was found that
 a majority of satellites are lopsided towards the MW's side \citep{McConnachie06}.
In addition,
 M31 satellites do not define only one main planar structure, like in the MW, but 
 a thin plane of satellites including approximately half of the satellite sample was singled out \citep[][hereafter 'Ibata-Conn-14' plane]{Conn13,Ibata13,Pawlowski13}. 
Finally,  new star clusters and dwarf galaxy candidates have been recently found in other nearby galactic systems in the Local Universe like CenA or the M101 group of galaxies.
Studies suggest as well an anisotropical 3D-spatial distribution \citep{Tully15,Muller17,Muller18}.

Additionally,
proper motion data has revealed that  a high fraction
of MW satellites present orbital angular momentum vectors mostly perpendicular to the Galactic disc  axis
\citep{Metz08,Pawlowski13b,GaiaHelmi18,Fritz18}.
In particular,   \cite{Fritz18} used recent proper motion data from GAIA DR2
to calculate the orbital poles of  objects orbiting within 420 kpc around the MW. 
According to their results, approximately   $\sim40\%$ (lower limit) of
the confirmed
 MW satellites
present orbital poles within an area of $10\%$  of the sphere around the normal direction to the ``VPOS", which they define as a co-orbitation criteria.
Also,   claims for co-rotation of satellites in the M31 'Ibata-Conn-14' plane  have   been made based on the  direction  of radial (i.e. line-of-sight) velocities \citep{Metz07,Ibata13} as no proper motion data is yet available.
It has been shown, however, that these are generally not a representative measure of the true 3D-velocities \citep{Buck16}. 


Flattened  spatial configurations of satellites in the MW and M31 have been well studied and quantified using information from only the three-dimensional positions of the satellites.
In particular,
 these positional analyses have used 
 sampling techniques like 'bootstrapping' \citep{Metz07} or the '4-galaxy-normal density plot' method  \citep[hereafter 4GND plot,][]{Pawlowski13}
 to statistically show the existence of predominant planar configurations of satellite positions in the MW and M31. 
These planes are then accurately
characterized by their normal vectors, axis ratios and root-mean-square heights ($\Delta$RMS),
computed from the eigenvalues of the Tensor of Inertia (ToI)
plane-fitting technique \citep{Metz07,Pawlowski13,Pawlowski15}.
In particular, the 4GND plot method consists in fitting planes to subsamples of 4 different satellites, and projecting the normal vectors on the sphere, creating a density map. 
The over-density regions that appear as a consequence of the accumulation of normal vectors broadly point in the normal direction to a predominant planar configuration of satellites 
Following this method, \citet{Pawlowski13}   detected
 the specific subsamples of satellites that mostly contribute to planar configurations in the MW and M31,
defining
  the ``VPOS-3" plane of satellites in the MW and the  ``GPoA" plane in M31.  

In a recent paper,   \citet*[][hereafter Paper I]{SantosSantos19},
have extended the 4GND plot method 
 to allow an identification,  systematic cataloging, and more detailed quality study, of  
the planar configurations of satellites in the MW and M31 systems. 
Rather than
 deriving 
 a unique plane  of satellites  per over-density 
 in the 4GND plot found with the previous method, the extension yields a \textit{collection} of planes of satellites,
each with an increasing number of members.
 In this way, it is possible to identify the highest-quality planes   in terms of the ToI parameters and the number of satellites considered. 
New to previous findings, in Paper I it was shown that
  \textit{two} distinct planes of satellites are present in M31: the  ``GPoA" and another plane (labelled ``M31-2-18" in Paper I). The two planar structures present very similar characteristics
and are, interestingly, oriented perpendicularly to each other.
\\

Since the very advent of these discoveries, theoretical studies  have  tried to assess the
 frequency of planar satellite configurations like those observed in cosmological simulations within the $\Lambda$CDM, paradigm.
These studies 
have mostly 
made use of
large volume
 DM-only simulations and pay attention to determining the significance\footnote{Significance is understood as the inverse of the probability of occurrence of a particular plane of satellites versus an isotropical distribution. } 
of the observed planes. In particular, \citet{Libeskind09,Wang13,Bahl14,Cautun15} have analyzed different versions of the large volume Millenium DM-only simulation \citep{Springel05Mill,BoylanKolchin09}, populating subhalos with galaxies following semi-analytic models. Using different methods for plane-identification, they find that  planes of subhalos as thin and even thinner than the ones observed in the MW and in M31 are expected in $\Lambda$CDM.
However, they acknowledge that 
these are  not the mean case found, but only consistent with the tail of the predicted  flattening distribution 
\citep[see ][for a different view]{Pawlowski14a}.
\bt{On the other hand,
\citet{Cautun15} show that not accounting for the `look-elsewhere effect' results in  an important overestimation of the significance
of planes in the MW and M31 systems of a factor of 30 and 100 respectively. Indeed, 
according to that work,
a $\sim$ 10\% of MW-like mass halos in $\Lambda$CDM simulations have planes of satellites that are more prominent than those observed in the MW or M31 systems,
presenting a large diversity when characterized by their thickness and 
 number of satellites.}  
\bt{On the other hand, a different approach has been followed by \citet{Buck15}. Instead of a large volume they use several DMO zoom-in simulations to show that a thin plane as the one around M31 with 15 satellites \citep{Ibata13} is not a challenge for the $\Lambda$CDM paradigm.  
However, neither the  VPOS-3 (with 24 satellites) or  VPOSall (27) planes in the MW are recovered in their analysis.
}

\bt{While some insight has been gained from collisionless N-body simulations, these experiments do not allow the formation of galactic disks.
Having well-behaved massive  MW-like  simulated discs, however, may be critical to
the planes of satellites issue.
Indeed,   the dynamical effect of a disc potential on satellite planes could change the frequencies alluded to  above, due,  for example, to the torques that satellites suffer from galaxy discs --except when they are on planar or polar orbits, or far away from the disc plane \citep{Danovich15,Welker18a}--. Another effect is that galaxy discs preferentially destroy satellites on radial orbits that pass near them \citep{Sawala2017,GarrisonKimmel17}.
\citet{Riley2019}
 show that massive disc potential effects cause the velocity anisotropy parameter, $\beta$,  to decrease as compared to less massive discs. 
Including these relevant effects in planes of satellites studies is thus necessary for a fair comparison with results from the MW/M31 disc galaxy systems. 
Another relevant point is that  there are $\gtrsim 30$ confirmed satellites in the MW and M31 \citep[][]{McConnachie12}:
a proper comparison demands as well the analysis of 
\rb{simulated}
MW-like discs surrounded by  as many resolved satellites.}

\bt{
Meeting all the previous requirements is currently  a situation not found in large volume hydrodynamical simulations. For example, in their analysis of the EAGLE simulation,  \citet{Shao19} analyze planes of 11 satellites around central galaxies of any morphology and compare them to the MW 'classical' plane. 
Thus the motivation to analyze zoom-in high-resolution hydrodynamical simulations in order to study planes of satellites.}

\bt{A few studies exist using zoom-in hydro-simulations  \citep{Libeskind07,Gillet15,Ahmed17,Maji17a}.}
  In general, the method used    for optimal-plane-searching in these 
  studies  has consisted in fitting a vast number of pre-defined planes with a  given constant thickness to the satellite sample. These planes are forced to pass through the center of the main galaxy \citep[e.g.,][]{Buck15,Gillet15}.
\bt{In this way, planes of satellites have been found. The best ones, however,  are not thin and populated enough so as to reproduce the VPOS-3 and VPOSall planes in the MW \citep[see e.g.,][]{Maji17a}; or even the Ibata-Conn-14 plane in M31 \citep[see e.g.,][]{Gillet15}. }
\bt{
On top of that, these studies using zoom-in hydro-simulations lack systematicity and detail, 
as well as uniformity in their methodology in order to apply the same analysis to both simulations and observations.
}
\\




\bt{
To fill this gap,}
the focus of this  study  is to introduce and apply a methodology suitable to analyze both observational data and simulations.
\bt{We aim to gain further insight on  the properties of planes of satellites one can find in  well-behaved massive disc galaxies (like the MW and M31),
 therefore formed in zoom-in high-resolution cosmological hydrodynamical simulations,  where a high enough number of resolved satellites can  be identified.}
This is an important step, whose outcomes are needed  previously to attempting any  physical interpretation of the origin and/or evolution  of planes.

Specifically, in this paper we develop  
\bt{a detailed analysis of planar configurations of satellites from positional information}
by applying 
 for the first time 
 the 4GND plot method  \citep{Pawlowski13} and its extension (introduced in Paper I) to simulations. 
\bt{For the reasons explained above, we focus on 
 a  set of zoom-in hydrodynamical simulations  where well-behaved MW-like disc galaxies and their satellite systems form. Moreover, as the effects of galaxy companions either in binary systems or groups add complexity to this analysis,
  \bt{we consider only  isolated galaxies}.
 As  mentioned previously, 
this method allows us to identify the constituent satellites forming  planar spatial configurations and analyze their quality  
 in terms of their population $N_{\rm sat}$ (or, equivalently,  the fraction of satellites  $f_{\rm sat}$) and the  ellipsoid of concentration axes ($a, b,c$, with $a > b > c$).
 Planes of high quality are those with a high $f_{\rm sat}$ and a low $c/a$,  meaning they  are populated and thin.  
These analyses are done over the entire galaxy's evolution after halo virialization.
This  allows to arrive to important insights into the different kind of planar structures one can find in 
simulations of MW-like disc galaxies, and how they compare to the observed planes of satellites.
}

 We note that  in the "planes of satellites problem", the persistence issue  
(i.e., is there a same group of satellites  that is spatially distributed in a planar-like configuration across time?) 
   is closely related to the kinematical character of planes.
 Therefore,   in a forthcoming paper  (Santos-Santos et al. in prep, Paper III) ,
the full six-dimensional phase-space information on satellite motions will be used to carry out
 kinematically-based analyses as an optimal methodology to address satellite plane persistence.

\bt{
The paper is organized as follows. 
 In Section \ref{sims} we introduce the simulations analyzed, while in Section \ref{satsamples}  their corresponding satellite samples and some of their properties are presented. 
  Section \ref{PlaSearch}  describes the method used for positional-plane-searching and plane quality analysis.  
Results obtained 
are reported in Section \ref{sec_simsn4p} and \ref{PQA}.
 In Section \ref{Coorbit} we assess the possible co-orbitation of satellites within the planes they form.
In  Section \ref{Discu} we discuss  implications of our results.
 Finally, Section \ref{conclu} summarizes the results and exposes the conclusions reached.  
 }

\section{Simulations}\label{sims}

We have chosen to analyze planes of satellites orbiting around  isolated,
 simulated galaxies 
that resemble the MW system. 
We therefore demand the simulation to  meet the following requirements:
\begin{itemize}[noitemsep,nolistsep]
\item[(a)] to contain a central galaxy with a thin gaseous and stellar disc at redshift $z\sim0$, with a large radial extent ($R=15-25\,$kpc).
\end{itemize}
We note that very thin disks are not that common in hydro-simulations yet.

Moreover, as
 merger events could destabilize the 
  galaxy+satellites system,
complicating
clean plane detections as well as  the possibility of reaching conclusions concerning the origin of these planes,
we as well require:
\begin{itemize}[noitemsep,nolistsep]
\item[(b)] an overall quiet assembly history, i.e., free of  major merger events, especially at late times.
\end{itemize}
 This is in line with the current understanding of the MW's disc formation and accretion history \citep{Belokurov18,Helmi18}. 

 To allow for a proper statistical comparison of the results obtained with those coming from observations of the MW and M31 at $z=0$, which harbour (at least) around 30 satellites each,\footnote{
Given the resolution we can afford currently in hydro-simulations, this implies that satellite mass functions are biased towards more massive satellites when compared to   the MW or M31 mass functions.
}
 the system must also:
\begin{itemize}[noitemsep,nolistsep]
\item[(c)] host a numerous ($\sim30$) satellite population around the central galaxy.
\end{itemize}
Finally, in order to accurately compute the  center of mass and the orbital  angular momentum of the baryonic component of a simulated satellite, and rely on it as physical, 
\begin{itemize} [noitemsep,nolistsep]
\item[(d)]
we demand satellite objects must include more than 50  baryonic particles. 
\end{itemize}

We have pre-analyzed a  set of different zoom-in cosmological  hydro-simulations,
  finding among them two that reach the previous prerequisites:
Aquarius-C$^\alpha$ and  PDEVA-5004. 
Both simulations follow the ``zoom-in" technique, 
but make  use of very different 
initial conditions,
codes and physics prescriptions.
This fact will allow us to reach conclusions that are independent of simulation modelling.

\subsection{Codes and Host galaxies}

\subsubsection*{Aquarius-C$^\alpha$  (Aq-C$^\alpha$)}
\bt{The initial conditions of this simulation come from}
the Aquarius Project
\citep{Springel08},
a suite of high-resolution
dark matter simulations of Milky Way-sized halos, formed in a $100 h^{-1}\, \rm Mpc$  $\Lambda$CDM, cosmological box
with parameters:
$\Omega_m$= 0.25; $\Omega_b$= 0.04; $\Omega_\Lambda$= 0.75; $\sigma_8$=0.9; $n_s$=1;
  $H_0$ =  73 $ \rm km s^{-1}\,Mpc^{-1}$.
%
In this project we analyze a new re-simulation
of the so-called ``Aquarius-C" halo (hereafter Aq-C$^\alpha$), 
 including the hydrodynamic     and subgrid models  described in \citet{Pedrosa15}.  
 These include a multiphase model for the interstellar medium and a supernovae feedback scheme, where  energy from  both Type Ia and Type II SNe is considered  \citep[see][for more details]{Scannapieco05,Scannapieco06}.
The initial mass resolution of baryonic and dark matter particles is $m_{\rm bar}=4.1\times10^5 \,\rm M_{\odot}$,
and    $m_{\rm dm}=2.2\times10^6 \,\rm M_{\odot}$, respectively.

  This galaxy presents a long period during which there is no merger, namely from $z\approx1.5$ to $z\approx0.15$. Soonly after, a massive satellite galaxy collides with the disc, and it further suffers   a  very close encounter with a another massive object  at $z=0$.
Therefore, the analysis we will describe in the following sections has been carried out up to $z=0.18$.
 Properties of this galaxy measured at \bt{this $z$ are}
 $M_\star=7.6\times10^{10}\,\rm M_\odot$, $M_{\rm gas}=5.6\times10^{10}\,\rm M_\odot$,
$M_{\rm vir}=1.5\times10^{12}\,\rm M_\odot$ and $R_{\rm vir}\simeq 219 \,\rm kpc$.
It is roughly more massive and larger than PDEVA-5004, \bt{as we show in the next section}.


\bt{ The halo mass growth history follows a standard two-phase process:
first a fast one with high mass growth rates and then a slower one where this rate is lower.
An important time scale for halo evolution is its collapse or virialization time when it gets decoupled from global expansion. This moment can be identified
as the time when 
the radius enclosing the $z = 0$ halo mass stabilizes. Or, almost equivalently, as the time when 
the time derivative of the virial mass growth reaches a low value, indicating the end of the fast phase of mass assembly.
For this halo this happens
 between a Universe age of 
T$_{\rm vir, AqC} \simeq 6-7 $  Gyr. In this case a 25\% of the mass is accreted after collapse,  with around a 10\% in the last merger event near  $z=0$ (not analyzed in this paper).
}

\subsubsection*{PDEVA-5004}
The PDEVA code is the OpenMP parallel version of the DEVA code,
an AP3M+SPH code specially designed so that conservation laws (e.g. momentum, energy, angular momentum and entropy) hold accurately  \citep{Serna03}.
It includes the detailed chemical feedback and cooling methods implemented by
\citet{MartinezSerrano08}
  as well as
  innefficient star formation parameters   in order to mimick the effects of energy feedback on regulating star formation \citep{Agertz:2011}, which are assumed to work on sub-grid scales
 \citep{Serna03,domenech12}.
 In particular, star formation is implemented following a Kennicutt-Schmidt law, with a
$\rho_\star=1\times10^{-25} \rm g\,cm^{-3}$
  density threshold and
  $c_\star=0.008$
   efficiency.
The following $\Lambda$CDM, parameters are assumed:
$\Omega_{\Lambda}$ = 0.723, $\Omega_{m}$ = 0.277, $\Omega_{b}$ = 0.04, and $h$ = 0.7; in a 10 Mpc per side periodic box.

Several simulations have been run with this code, yielding a suite of different galaxies. The one used in this project is PDEVA-5004, previously studied in \citet{MartinezSerrano09,domenech12,Obreja13,DT14,dominguez15,DT16proc,DT17},
where satisfactory consistency with observational data has been found in all the comparisons addressed.
This galaxy has a remarkably thin gaseous and stellar disc  and a relatively quiet history after virialization.
At redshift $z=0$ it has the following properties: $M_\star=3.05\times10^{10}\,\rm M_\odot$, $M_{\rm gas}=8.6\times10^{9}\, \rm M_\odot$,
$M_{\rm vir}=3.44\times10^{11}\, \rm M_\odot$, 
\bt{$R_{\rm vir}\simeq 183  \rm\,kpc$.}
The mass resolution of baryonic and dark matter particles  is $m_{\rm bar}=3.94\times10^5 \, \rm M_{\odot}$,
and    $m_{\rm dm}=1.98\times10^6\, \rm M_{\odot}$, respectively. Particle masses do not change
during the simulation.

 \bt{The halo growth history can be found in Figure 1 of \citet{DominguezTenreiro2017}
 where again a two-phase process can be  clearly distinguished
with T$_{\rm vir, 5004} \simeq 7 $ Gyr,   
even if mergers around this event refrain us from a clean identification.
Only a 20 \% of the virial mass is assembled after this time, and no major mergers show up.
}

\bt{
It is worth noting that the mass growth histories for both Aq-C$^\alpha$ and PDEVA-5004  are standard for galaxy halos  with a quiet history after virialization, as demanded by  the selection criteria above. 
}



\section{Satellite samples}\label{satsamples}

\subsection{Identification}
The samples of satellites in each simulation have been selected following the same criteria,
this is, 
 to take all objects with stars  ($M_\star>0$)
\bt{within a distance of 350 kpc from their host} 
  that are bound to the  host galaxy, with a resolution limit of presenting at least 50 baryonic particles
 ($M_{\rm bar}\approx1\times 10^7 M_\odot$).
This selection has been made at redshift 
$z\sim0.5$,  to include satellites that may end up accreted by the disc at $z=0$.
To prove if a given object is indeed a satellite (i.e., is bound to its host) we have computed its orbit.
\bt{This fixed sample of satellites has then been followed back and forth in time.}

 \bt{ For a proper comparison with MW results, we take into account Galactic obscuration --which prevents us from observing satellites orbiting in the plane of the disc of the MW--, by applying an observational bias  to the simulated satellite sample  at each  timestep.  }
Following  \citet{Pawlowski16},
we have chosen it to hide objects with projected positions on the sphere at  latitudes  $|b|<12^{\circ}$ (angular distance as measured from the plane of the galaxy disc).
 This is a first approximation to try to mimick the obscuration effects of the MW's disc; however, we acknowledge that a more precise model to account for Galactic obscuration should depend on satellite distances and luminosities.

The selection of satellites in  Aq-C$^\alpha$ has been done using a Friends-of-Friends algorithm  to identify  structures, and then the SubFind halo finder  \citep{Springel01} to  construct  subhalo catalogs at each timestep. 
Particle  IDs have been used to trace back in time the selected  satellites.
The tool used for the selection of satellites in PDEVA-5004 has been \texttt{IRHYS} (by H. Artal, under development). This visualization and analysis tool permits the selection of objects \bt{(i.e., satellites)} as sets of particles,
\bt{and enables}  to trace them back and forth in time. 

\begin{figure}
\includegraphics[width=\linewidth]{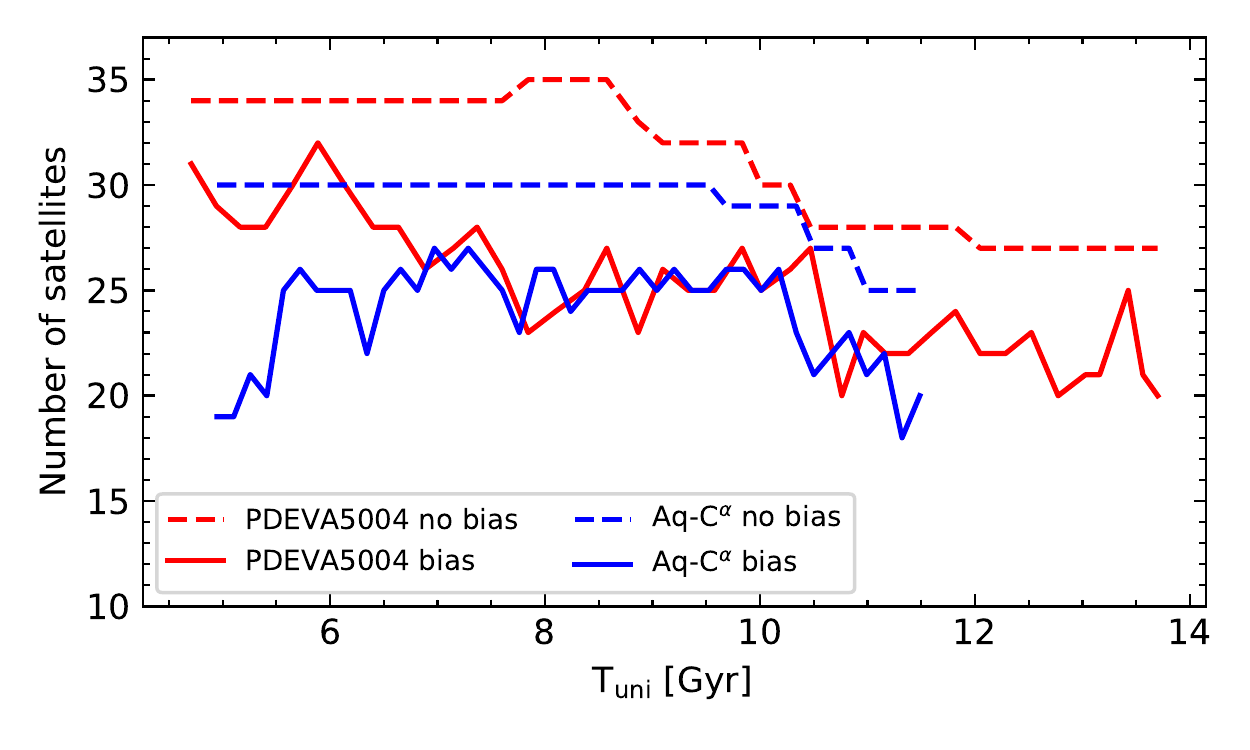}
\caption{
The total number of satellites  $N_{\rm tot}$ in the two simulated samples as a function of the Universe age T$_{\rm uni}$.
Red: PDEVA-5004; blue: Aq-C$^\alpha$.
\bt{A dashed line shows results when all satellites are considered ('no bias'); a solid line shows results when the observational Galactic obscuration bias is applied ('bias'), hiding satellites orbiting in the plane of the disc at latitudes $|b|<12^{\circ}$. }
}
\label{ntotvsTuni}
\end{figure}

A total number of $N_{\rm tot}$ = 30 (35) satellites have been detected in Aq-C$^\alpha$ (PDEVA-5004) at selection time ($z\sim0.5$),
 of which  25 (27)  survive until   the last analyzed timestep, respectively. 
\bt{In Figure \ref{ntotvsTuni} such numbers are plotted as a function of the Universe age T$_{\rm uni}$.
 $N_{\rm tot}$ changes because satellites disappear as they are accreted by the central disc galaxy. Also satellites are not considered during periods where they orbit at distances $>450$ kpc (this happens with a couple of backsplash galaxy cases). In the case when obscuration in the plane of the disc is considered ('bias'), $N_{\rm tot}$  varies additionally  because satellites go into and out of the avoidance volume.
}

\subsection{Satellite radial distances  and their distributions}

The suite of Aq-C$^\alpha$ and PDEVA-5004  satellites presents different properties and evolutionary histories that reflect in a variety of orbits. We find some satellites that progressively lose angular momentum and are eventually accreted by the disc, some that follow  orbits
where successive apocenter and pericenter distances do not show important variations, and some 
\bt{backsplash galaxies,}
that have just recently been captured by the halo and orbit at long distances occasionally even outside the virial radius.
Interestingly, when all radial distance histories are plotted together,  a coincidence of pericenters 
is clear at certain moments.
In this way,
as the simulations evolve,  there are moments of maximum spatial expansion and moments of a maximum compression of the satellite systems.

\begin{figure*}
\centering
\includegraphics[width=0.8\linewidth]{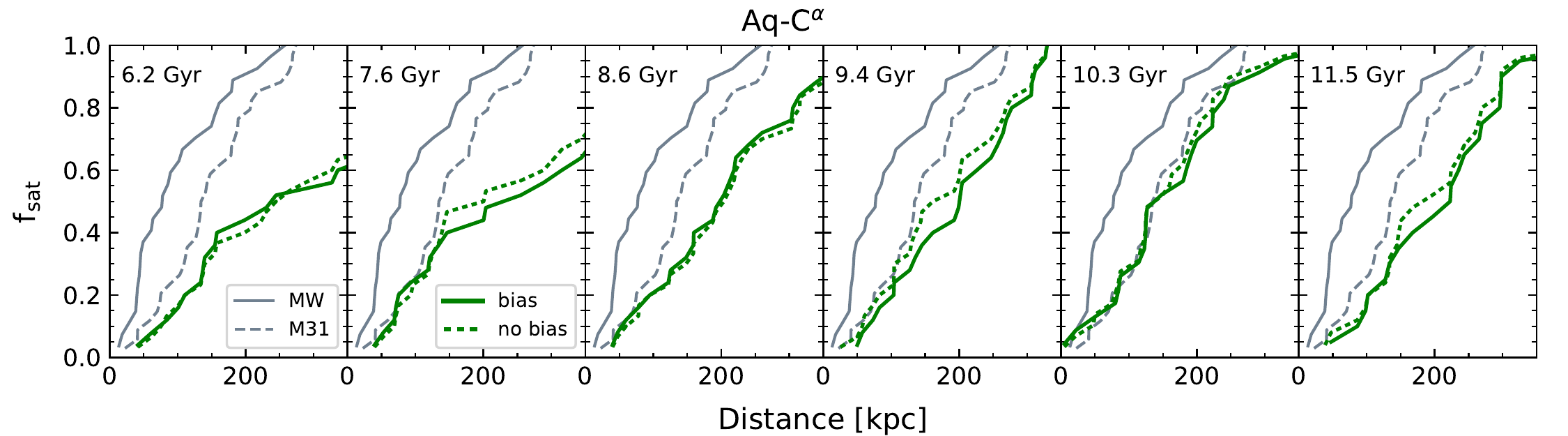}
\includegraphics[width=0.8\linewidth]{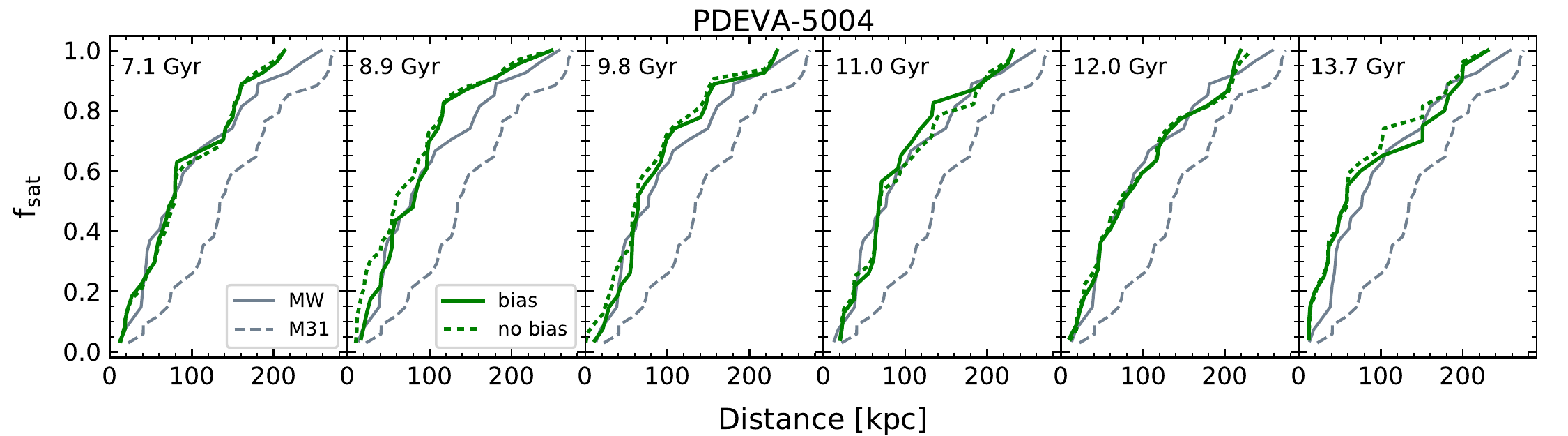}
\caption{
The radial distribution of satellites at different Universe ages.
 Top: Aq-C$^\alpha$. Bottom:  PDEVA-5004.  
  The green solid line shows the distribution for all the satellites present at the given timestep, while the dotted line takes into account the observational bias for Galactic obscuration. Gray lines show the distributions of satellites in the MW (solid) and in M31 (dashed) at $z=0$. 
}
\label{distpdevaaqc}
\end{figure*}

These effects can be observed when analyzing the evolution of the radial distribution of satellites with cosmic time. 
Figure \ref{distpdevaaqc} (top panel: Aq-C$^\alpha$, bottom panel: PDEVA-5004)
 shows
the fraction of the total number of satellites within a certain distance from the center of the main galaxy,
 compared to the MW and M31 distributions at $z=0$\footnote{The sample of MW and M31 satellites used is that described at the beginning of Section 5.1.}. Two colored lines are shown per panel, which respresent the results obtained using all the satellites, or taking into account the observational obscuration bias, at each timestep.
The distributions change with time, showing periods where there is a higher concentration of satellites at shorter distances (corresponding precisely to the moments of maximum collective approach) and others where there is a higher expansion of the system.
As an example, the PDEVA-5004 system (in the case where all satellites are considered, i.e., `no bias')
is more compact (higher $f_{\rm sat}$ within 100 kpc) at $T_{\rm uni}=8.9$ Gyr, and more expanded at   $T_{\rm uni}=11$ Gyr. 
On the other hand, in the Aq-C$^\alpha$  system
it is not until  $T_{\rm uni}\sim9$ Gyr that the complete sample of satellites is within  a distance of   $\sim350$ kpc. A moment of maximum compactness is  $T_{\rm uni}\sim10.3$ Gyr.
Curiously, PDEVA-5004's satellite radial distribution resembles very well that of the MW
for a long period of time, while Aq-C$^\alpha$'s is similar to that of M31 at $T_{\rm uni}\sim 10$ Gyr.
\bt{These resemblances are kept when using $N_{\rm sat}$ instead of $f_{\rm sat}$ in the case that  total satellite numbers of simulations and observations are matched (see Sections \ref{NsatA} and \ref{PeakasQ}).
We explore the effect of radial compactness on the quality of planes of satellites in Section \ref{RadialCompact}. }

\subsection{Mass distribution of satellites}\label{sec_massdis}

Satellites in Aq-C$^\alpha$ (PDEVA-5004) show baryonic masses ranging between
$M_{\rm bar} = 8.6 \times 10^6 - 8.9 \times 10^8\,\rm M_\odot$
($M_{\rm bar} = 3.9 \times 10^7 - 1.8 \times 10^8 \,\rm M_\odot$).
\bt{
This differs from the mass range of confirmed MW/M31 satellites. Indeed, the objects with  lowest stellar masses considered in this work have
 $ M_{*} \sim8\times10^6\,\rm M_\odot$  (Aq-C$^\alpha$) 
and   $ M_{*} \sim1\times10^7\,\rm M_\odot$  (PDEVA-5004)
}
 (see requirement  (d)  above)
 while observed MW/M31 satellites reach as low as \bt{$M_{*}\sim5\times 10^2\rm M_\odot$} (e.g., SegueI),
 \bt{with 13 out of 27 galaxies in the MW presenting masses lower than $M_{*}<5\times10^4$M$_\odot$}\footnote{Stellar masses for observed galaxies are calculated from the luminosity values in \citet{McConnachie12}, using the mass-to-light ratios from \citet{Woo08} according to galaxy morphological type.}.
In the last years, cosmological hydrodynamical simulations have reached a very high mass and spatial resolution, however, it is still not high enough 
as to produce 
\bt{as many}
resolved bound objects with masses as low
as those at the low-mass end of the MW and M31 satellite 
mass functions\footnote{
\bt{
For example, \citet{Buck19} (i.e., a higher resolution re-simulation of NIHAO MW-like galaxies with $m_{\rm gas}\approx5\times10^4$ M$_{\odot}$), produce one case with 20 satellites at $z=0$, reaching a low mass end of $M_{*}=1.5\times10^5$ M$_{\odot}$.
\citet{Wetzel16}'s Latte MW simulation presents a mass resolution of $m_{\rm gas}\approx7\times10^3$ M$_{\odot}$,  producing a number of \rb{13 satellites at $z=0$ that reach} a lowest mass of $M_{*}=8\times10^4$M$_{\odot}$.
\citet{Ahmed17}'s sample of MW-like galaxies with $m_{\rm gas}\approx3\times10^4$ M$_\odot$
present a large number of satellites at $z=0$ \rb{with minimum mass} of $M_{*}=2\times10^4$M$_{\odot}$.
}
}.

These differences between the simulated and observed mass distributions are not expected to  introduce a determinant bias on the formation or not of planes of satellites as analyzed in this work,
as it was shown in Paper I with the observed MW and M31 satellites
(which span a wider mass range than simulated ones)
 that stellar mass is not a satellite property determining its membership or not to the highest quality positionally-detected planes found at $z=0$ \bt{with the 4GND plot method}\footnote{This was quantitatively confirmed by finding low correlation coefficients  between the total stellar mass of a satellite and its   number-contribution, $C_{\rm p, s}$ 
(see definition in Paper I and in Sec. \ref{N4p}) 
 to the main planar configurations of satellites (i.e., the highest quality planes). }. 
Additionally, from the empirical side, the fact that
\bt{objects of different mass scales, like}
 globular clusters and stellar streams, \bt{seem to be} as well within the observed VPOS plane of satellites in the MW \citep{Pawlowski12}, 
 \bt{supports our findings.}
This is an  important result in view of the rather narrow satellite baryonic mass range we can currently afford in hydrodynamical simulations.
\bt{Therefore, we are allowed to meaningfully compare  planes from  the observed   MW/M31 satellite samples  and those of our  simulations,  even if the masses of the involved satellites span different mass ranges.}



\section[Searching for planes of satellites from a positional analysis]{Searching for planes of satellites from a positional analysis}
\label{PlaSearch}
\sectionmark{Positional analysis}


To search for planar positional configurations of satellites in our simulations we have  followed the 
\textit{4-galaxy-normal density plot} (4GND plot) method
 presented in
\citealt{Pawlowski13},
extended as explained below (see also Paper I).
This method allows to find  if there is a subsample out of a given sample of $N_{\rm tot}$ satellites that defines 
a planar arrangement  in terms of the outputs of a fitting technique based  on the Tensor of Inertia
\citep[ToI,][]{Metz07,Pawlowski13}.
Satellite planes are searched for through a regression method 
that minimizes orthogonal distances from the points to the optimal plane solution.
 Apart from the plane (or equivalently, its normal vector $\vec{n}$),
  the outputs  of the regression can be characterized by the following parameters
  in terms of the corresponding  ellipsoids of concentration \citep{Cramer}:
\begin{itemize} [noitemsep]    
\item $N_{\rm sat}$: the number of satellites in the subset (or, the fraction of satellites it involves 
f$_{\rm sat} \equiv N_{\rm sat}/N_{\rm tot}$); 
\item $c/a$: the ellipsoid short-to-long axis ratio;
\item $b/a$:  the ellipsoid intermediate-to-long axis ratio;
\item $\Delta$ RMS: the root-mean-square thickness perpendicular to the best-fitting plane;
\item D$_{\rm CG}$: the distance from the center-of-mass of the central galaxy to the plane.
\end{itemize}
 These outputs are used to quantify the quality of the best fitting structures to a subsystem of $N_{\rm sat}$ satellites.
First of all,
 assuming $c/a<1$, 
 $b/a$ indicates whether the distribution is planar ($b/a \sim 1$), or filament-like ($b/a<<1$). 
High quality planes are those that involve many satellites and are thin, therefore demanding high $f_{\rm sat}$ and low $c/a$ (or equivalently low $\Delta$RMS, a quantity that most often is correlated with $c/a$ once the system acquires its stable size).  
 Low D$_{\rm CG}$ planes pass near the disc center, a requirement to be asked to
 a physically  consistent    satellite system
  when its gravitational potential  minimum  lies approximately at this center.  
 Finally, $\vec{n}$ allows to visualize the plane orientation \bt{ with respect to a given reference frame, for example 
the host galaxy disc plane. }
 At the end of this section the quantification of plane quality, as well as how to compare the qualities of two or more planes, is described with more detail (see also  Paper I).

\subsection{Method: 4-galaxy-normal density plots}\label{N4p}

We have applied the  4GND plot  method to each timestep of the two simulations. A thorough description follows
 \citep[see also Section 2.4 in ][and Paper I]{Pawlowski13}.

\begin{enumerate} [leftmargin=*]
\item 
\bt{A plane is fitted}
(using the ToI method)
to the positions of every combination of 4 different satellites taken from the total  sample 
 of  $N_{\rm tot}$ satellites. 
As 3 points always define a plane, 4 is the lowest possible amount to take into consideration under the condition of making the number
of combinations\footnote{The number of such combinations is given by
$\frac{\rm N_{tot }!}{N_{\rm pl }! (N_{\rm tot} - N_{\rm pl})!}$,  where N$_{\rm pl}$ is the number of satellites  included in  the planes. }
\rb{high enough  to get a good outcome signal.}

\item
The  axes sizes $a,b,c$
 and normal vector directions (i.e., 4-galaxy-normals)  
 of the planes fitted to each combination of 4 satellites,  
 are stored. 
Then, all the 4-galaxy-normals 
are plotted in a galactocentric coordinate system such that the central disc's spin points towards the south pole,
 and a density map 
(i.e. 2D-histogram) 
 is drawn from their projections on a regularly-binned sphere   with N$_{\rm bin}$ bins. 
 Spherical projections are shown with Aitoff diagrams in Galactic (longitude $l$, latitude $b$) coordinates  
in a $l=[-90^\circ,+90^\circ]$ projection because opposite  normal  vectors are  equal.
\bt{As in \citealt{Pawlowski13},}
each normal is weighted by $w=\log\left( \frac{a+b}{c} \right)$, to emphasize
 planar arrangements of satellites over filament-like or spherical-like ones.
In these plots, over-densities  (\bt{or density peaks,} i.e., areas of 4-galaxy-normal vector accumulation) are signaling  groups of 4 satellites contributing to a same dominant planar space-configuration.
 As an illustration, in Figs. \ref{aqcdp} and \ref{pdevadp} we show examples of  4GND plots for the Aq-C$^\alpha$ and PDEVA-5004 simulations, respectively.
In some cases these over-dense areas are more extended and in others more concentrated.
\bt{(Note that the expectation from a random distribution of satellites is a density map with equal density in each bin). }

\item  
 Density peaks are differentiated and isolated. 
A set of N$_{\rm peaks}$ \bt{density} peaks are selected around   the highest density bins of  the 4GND plots, with the requirement that they are separated more than 15$^\circ$ from the center of all the (N$_{\rm peaks}$ - 1)  over-densities. 
The specific peak location in ($l,b$) is given by the center of the corresponding high-density bin\footnote{ As we show in Figure \ref{fixfrac}, changing the size of bins does not modify our results and conclusions.}.

\item   
\bt{
We determine how much a given satellite $s$ contributes to the $p$-th specific peak (i.e, its respective contribution-number, $C_{\rm p, s}$, with $p$=1, ..., N$_{\rm peak}$ and $s$=1, ..., $N_{\rm tot}$). 
To this end we select  all the 4-galaxy-normals  that fall within an aperture angle of 15$^{\circ}$\footnote{We take the same  angle as that used in \citealt{Pawlowski13}'s analysis.}
 around the $p$-th   peak location.
 Each \bt{of the 4 satellites} contributing to 
 these 4-galaxy-normals
 is counted as contributing the 4-galaxy-normal's \textit{weight} to the peak. 
This has been normalized using  $C_{N, all}$, the total weighted number of 4-galaxy-normals, included those that are not within 15$^{\circ}$ of some peak center, such that the sum $\sum_{p, s}$ $C_{\rm p, s}=1$. 
Such normalization is necessary  for a meaningful comparison of results at different timesteps
(where $N_{\rm tot}$ varies),
 and also, with observational data.
At fixed $p$, $C_{\rm p, s}$ is high when  satellite $s$ contributes to many of the 
 4-galaxy-normals laying  within 15$^{\circ}$  of the peak center. 
We note that over-densities that are located close to each other on the sphere generally share many of the satellites that contribute most to 4-galaxy-normals within 15$^\circ$ of the peak.
However the peak isolation criterion used in the previous step cures our peaks of such redundancies.}

\item  
For a given peak $p$, we order all  satellites by decreasing $C_{\rm p, s}$ to it.
 This is done for all the  isolated  peaks found in a given  4GND plot, 
 and for each of them we obtain an ordered list of contributing satellites.
 Examples of such lists are shown in Figure \ref{ex_cont} for two different peaks defined at a given timestep in PDEVA-5004. The x-axis shows satellite IDs in decreasing $C_{\rm p, s}$ order (see y-axis values). 

\end{enumerate}

\subsection{An extension to the method}
\label{PMExt}

\bt{We have extended  the 4GND plot method presented in \citealt{Pawlowski13} to thoroughly evaluate the properties and quality of the planar structure of satellites revealed by each over-density. }

\subsubsection{Peak strength analysis}
\label{PeakStre}
\bt{
In order to analyze the number of \bt{relevant density} peaks at each timestep and how this number evolves with time, to each peak we assign 
a \rb{'strength'},
 $C_p$, 
 defined as the normalized number (or \%) of 4-galaxy-normals within 15$^\circ$ of the respective peak  center; that is $C_{\rm p} \equiv \sum_s C_{\rm p, s}$,   where the contribution-number  $C_{\rm p, s}$ of the $s$ satellite to the $p$-th peak is defined in step (4) above.
For example, in Figure \ref{ex_cont}, $C_{1}$ ($C_{2}$)  would be obtained by summing  up  the $C_{1, \rm s}$ ($C_{2, \rm s}$) corresponding to all the satellites in the upper (lower) panel of the Figure.  
}

\bt{
By reckoning the number of peaks with $C_{\rm p}$ above given thresholds (i.e., the observational ones, for example), we can compare to observations. We can also calculate  how many peaks of a given strength there are at given timesteps in the simulations and how this number evolves with time. 
}

\subsubsection{Plane quality analysis}
\label{PlQuaAna}

 To analyze individually each over-density in terms of quality, as explained in Paper I,
  we start by fitting a plane to the 7 satellites that contribute the most \bt{to it}\footnote{This number $N_{\rm sat}$=7  is low enough to allow for an analysis of the ToI parameters behaviour as $N_{\rm sat}$ increases, and at the same time high enough that we begin with populated  planes. Note that taking instead $N_{\rm sat}=7 \pm 2$  to begin with does not alter our conclusions.}.
 Then, following the order of 
 $C_{\rm p, s}$ contribution, we iteratively  add one more satellite at a time   and   fit a plane to the new resulting satellite set, storing the \bt{ToI}  fitting outputs described at the beginning of this section. 
  This plane-fitting process is repeated until all the contributing satellites to the peak under consideration are used.
 In this way, for each  peak  found at a given timestep of the simulation, we obtain a collection  of
planes of satellites, each consisting of an increasing number of members. 

As said above, plane quality is measured through the $f_{\rm sat}$ and $c/a$ 
values. 
Being a two-parameter notion,
\bt{when comparing the quality of two planes, if in one of them $c/a$ is lower  and $f_{\rm sat}$ is higher than in the second, then the first plane has higher quality.}
\bt{Other cases when the qualities of two planes can be compared are when }
 either $f_{\rm sat}$ is constant (in which case the  plane with lowest $c/a$ has a higher quality),
   or   when  $c/a$ is constant (or at least it varies slowly with $f_{\rm sat}$),
in which case, the higher $f_{\rm sat}$, the better the quality.

As a practical implementation of these ideas, in this paper we show 
 how \bt{$b/a$ and }    $c/a$ 
 vary as the number  of satellites  
included in the plane-fitting increases; see e.g. Figures  \ref{pdeva_ba} and \ref{aqc_ca_fracA}, where the \bt{collection of planes
 associated to a given density  peak  is characterized by a line.
}

\begin{figure*}
\centering
\includegraphics[width=0.49\linewidth]{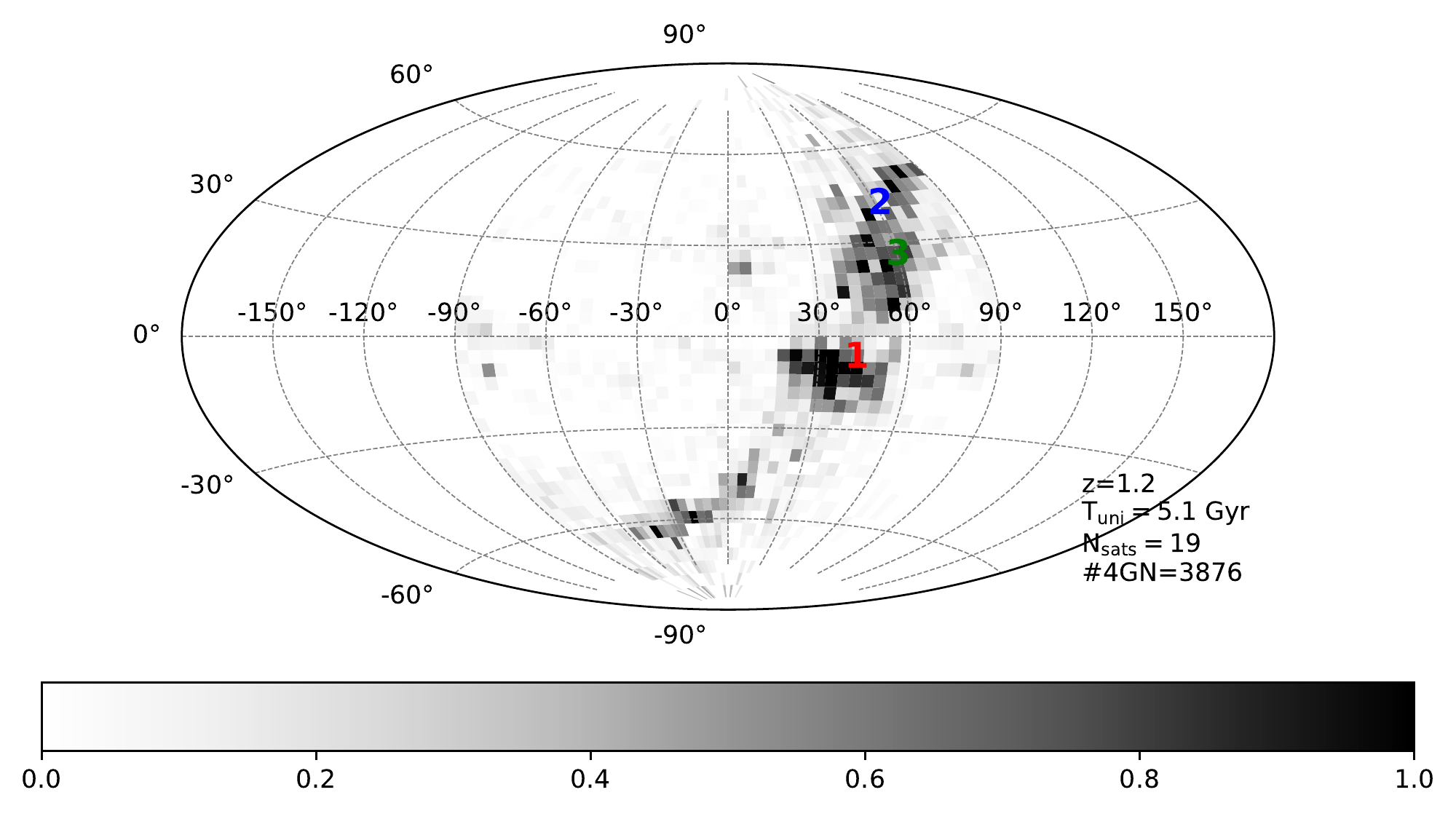}
\includegraphics[width=0.49\linewidth]{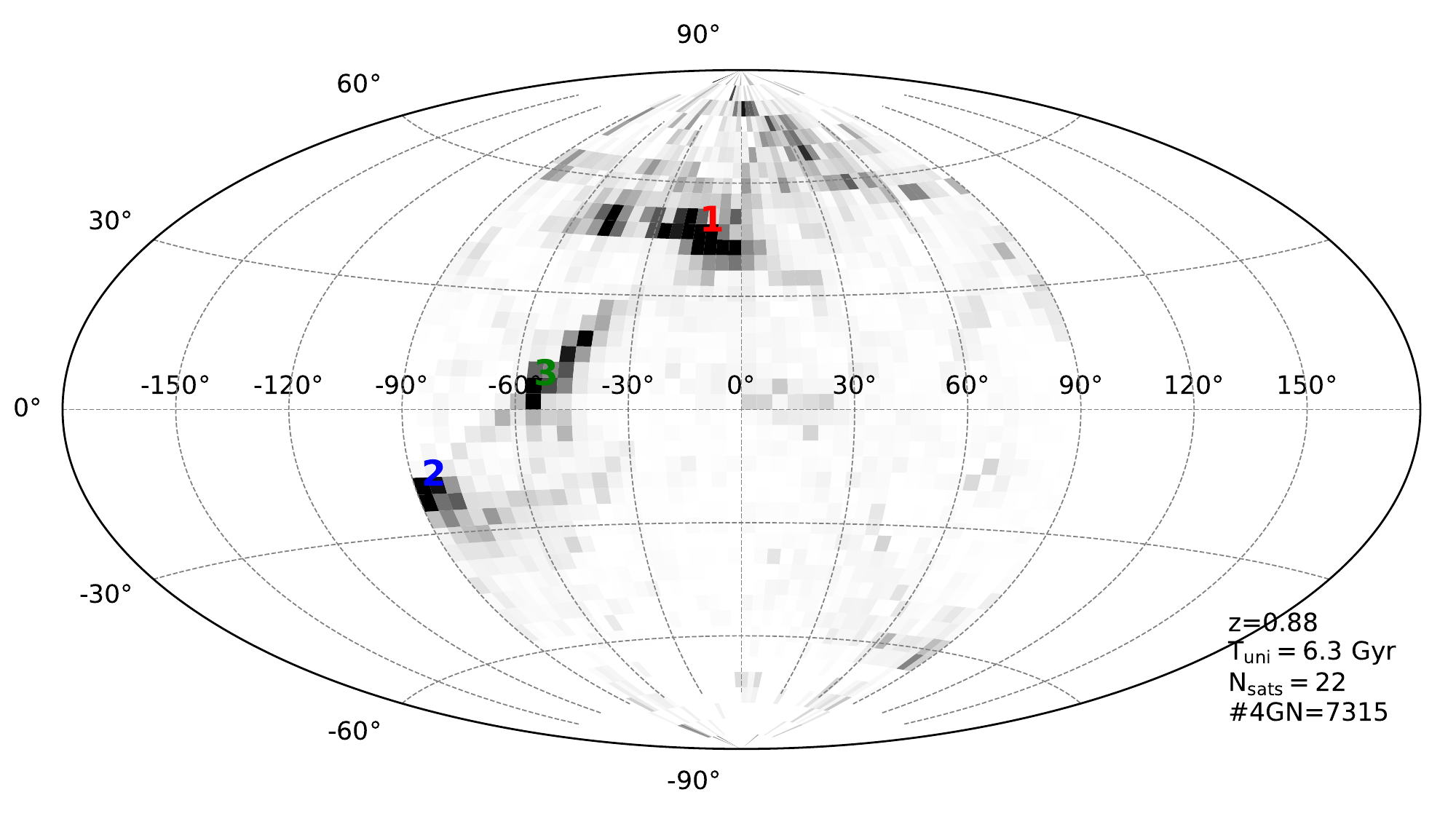}
\includegraphics[width=0.49\linewidth]{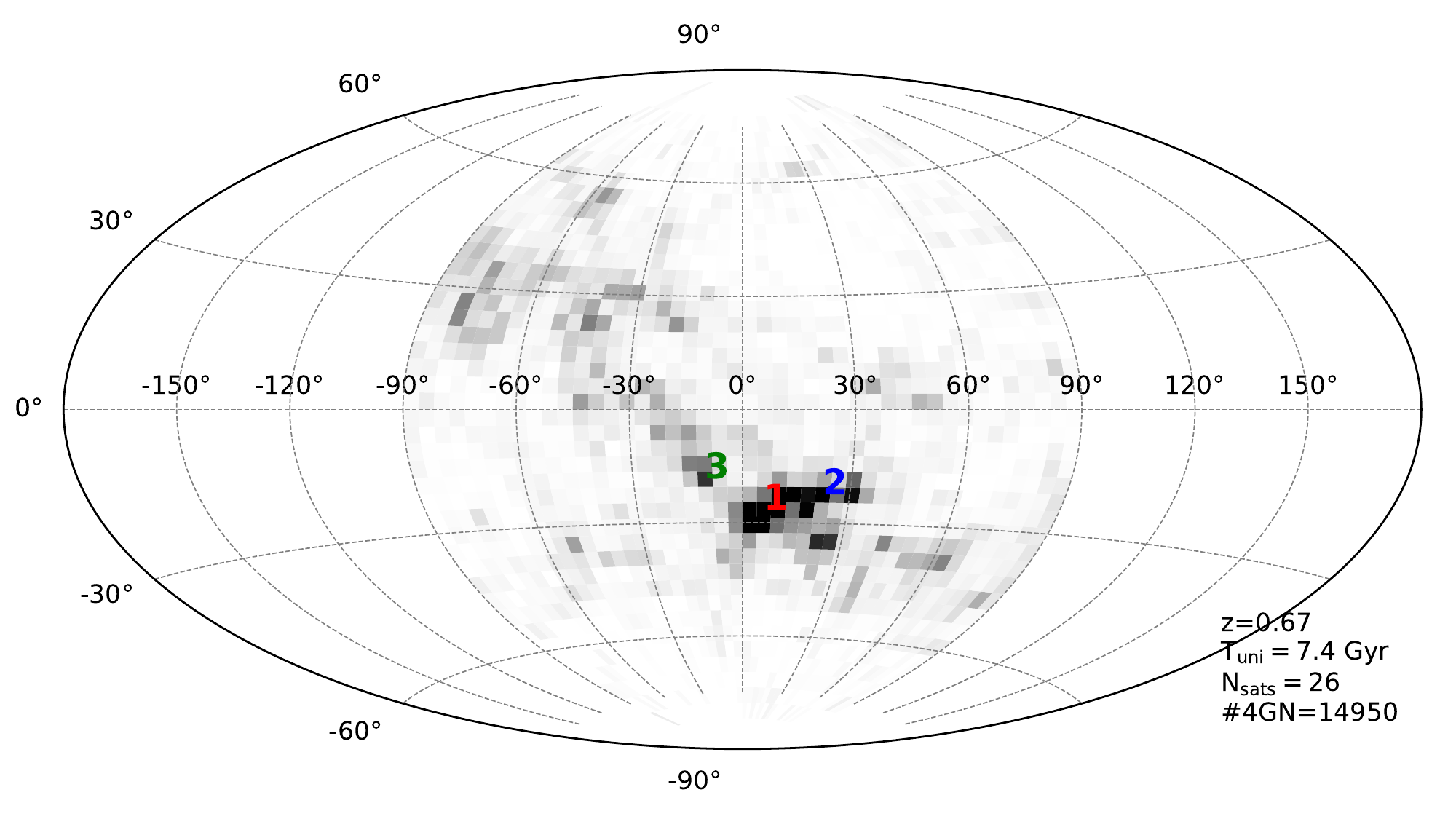}
\includegraphics[width=0.49\linewidth]{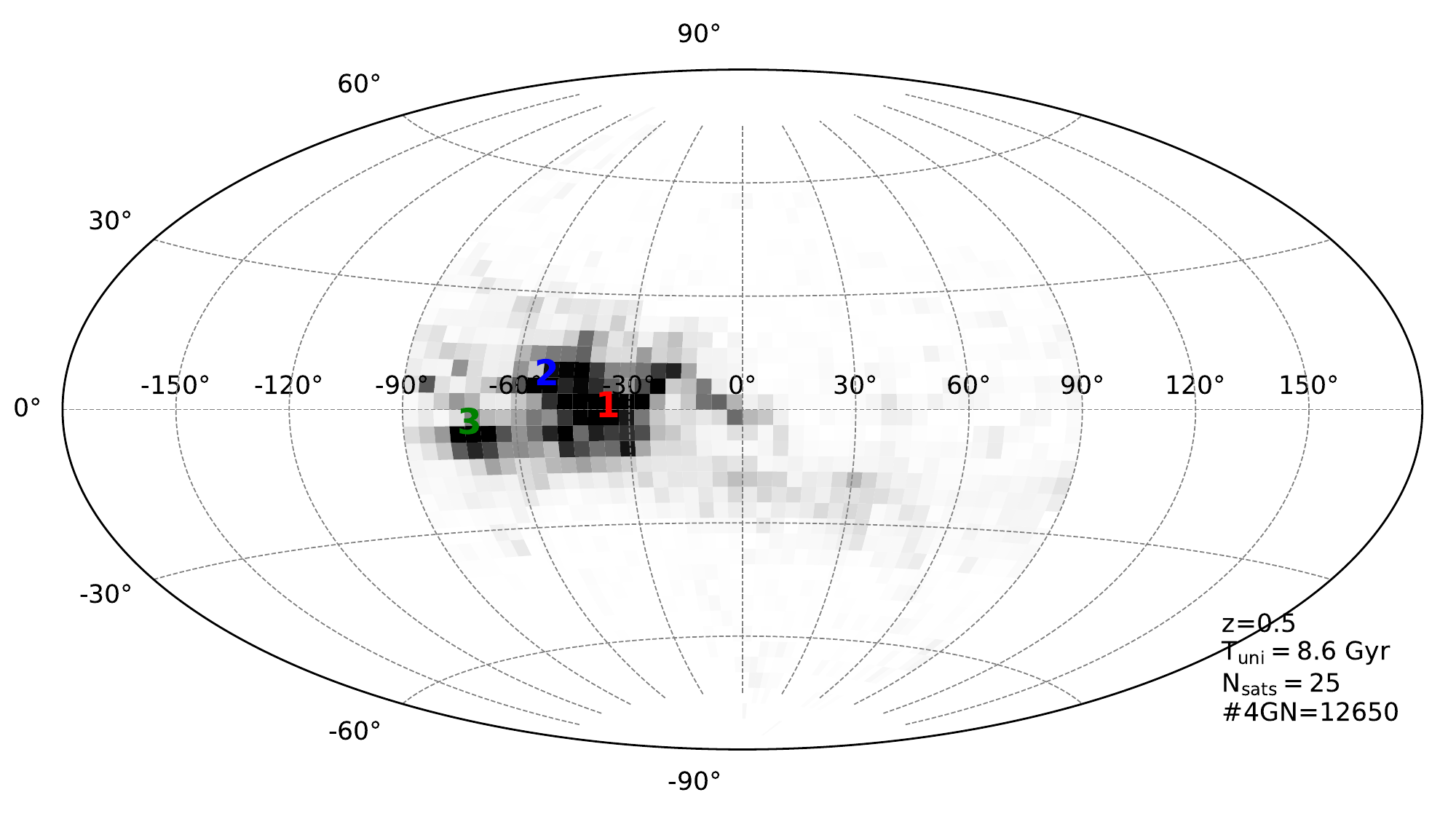}
\includegraphics[width=0.49\linewidth]{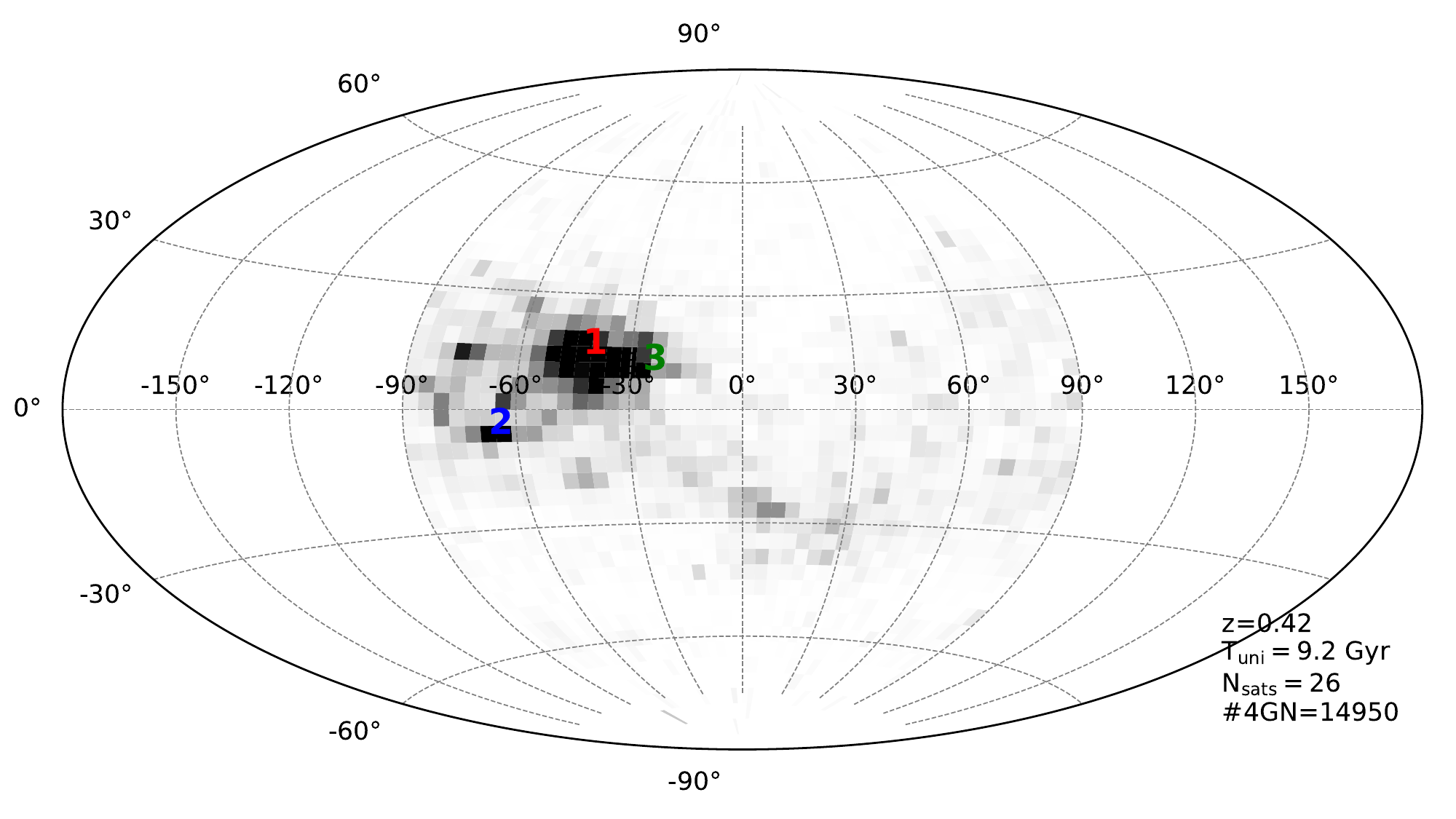}
\includegraphics[width=0.49\linewidth]{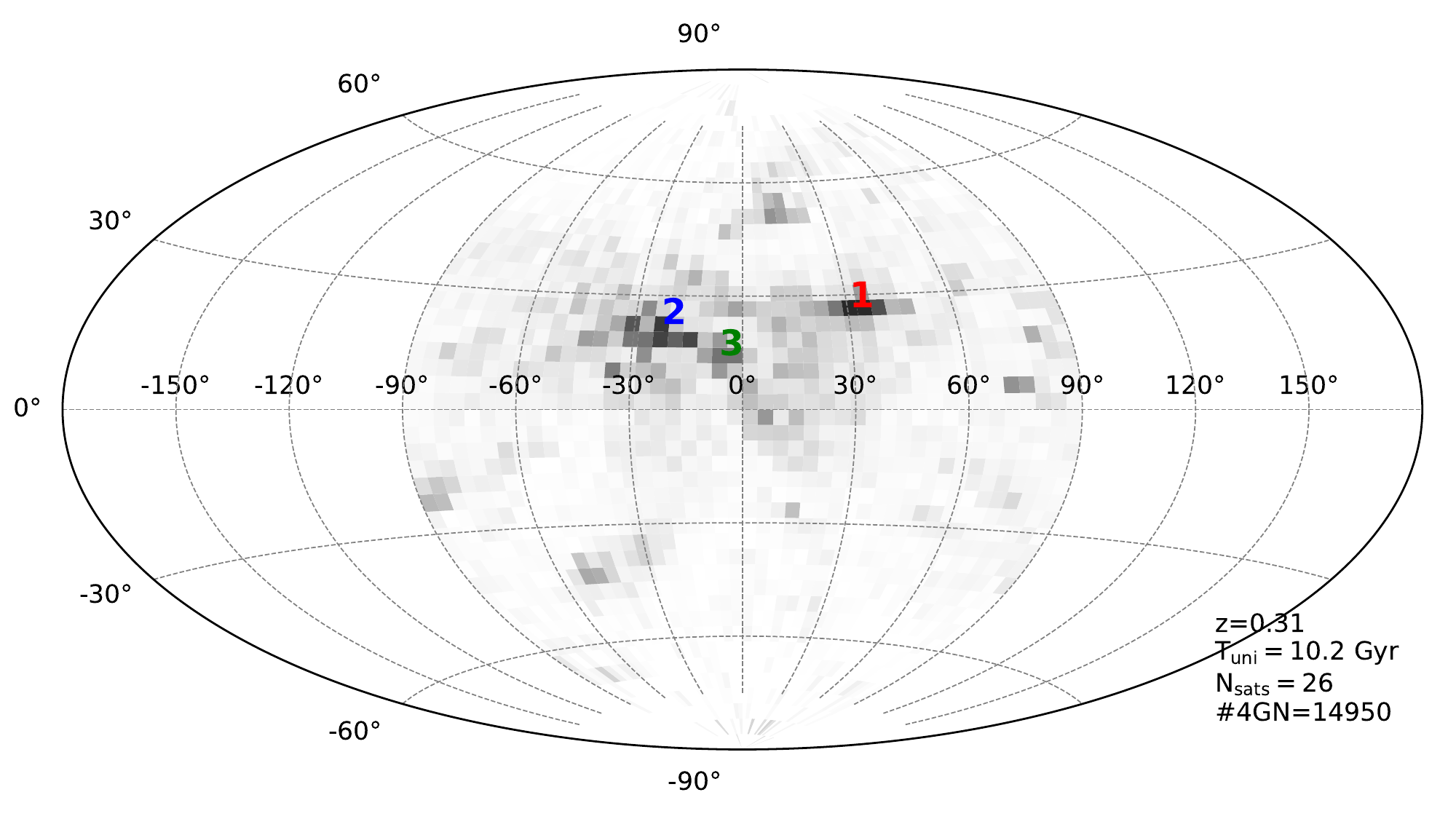}
\includegraphics[width=0.49\linewidth]{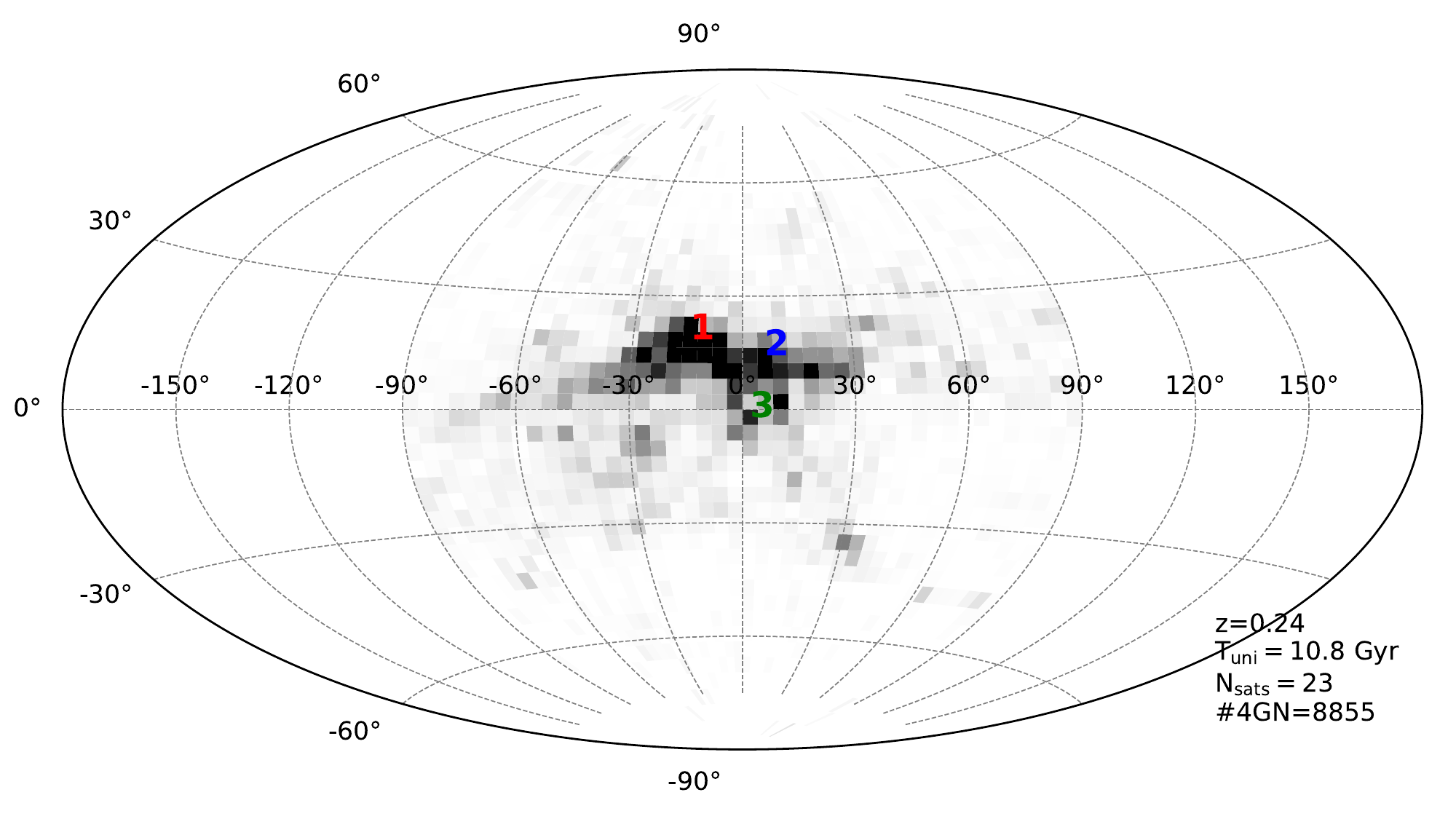}
\includegraphics[width=0.49\linewidth]{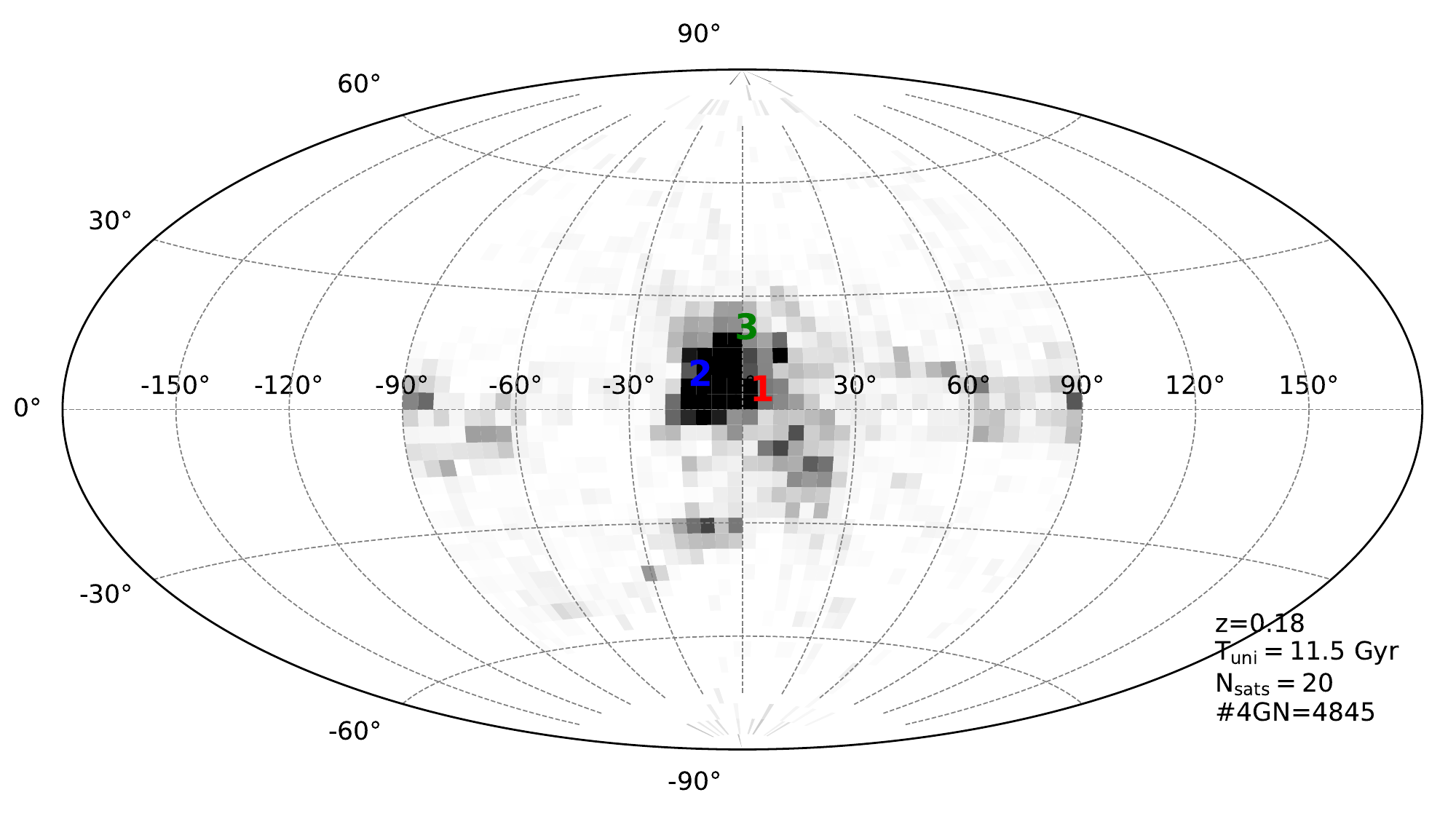}
\caption{
Examples of 4-galaxy-normal density plots (4GND plots) for galaxy  Aq-C$^\alpha$ (having applied the observational obscuration bias)  at different times. The legend shows the redshift $z$, the cosmic time it corresponds to in Gyr, 
 the total number of satellites considered and 
the total number of 4-galaxy-normals, at that timestep ($\#$4GN). 
 The main \bt{density} peaks, used for analyses in this work, are marked with numbers ordered according to 
the density of their central bin. 
  A color code is also used to identify their contributions in the next Figures. 
\bt{ The grayscale colorbar is common for all timesteps and its values are  proportional to the normalized  bin density.}
 }
\label{aqcdp}
\end{figure*}

\begin{figure*}
\centering
\includegraphics[width=0.49\linewidth]{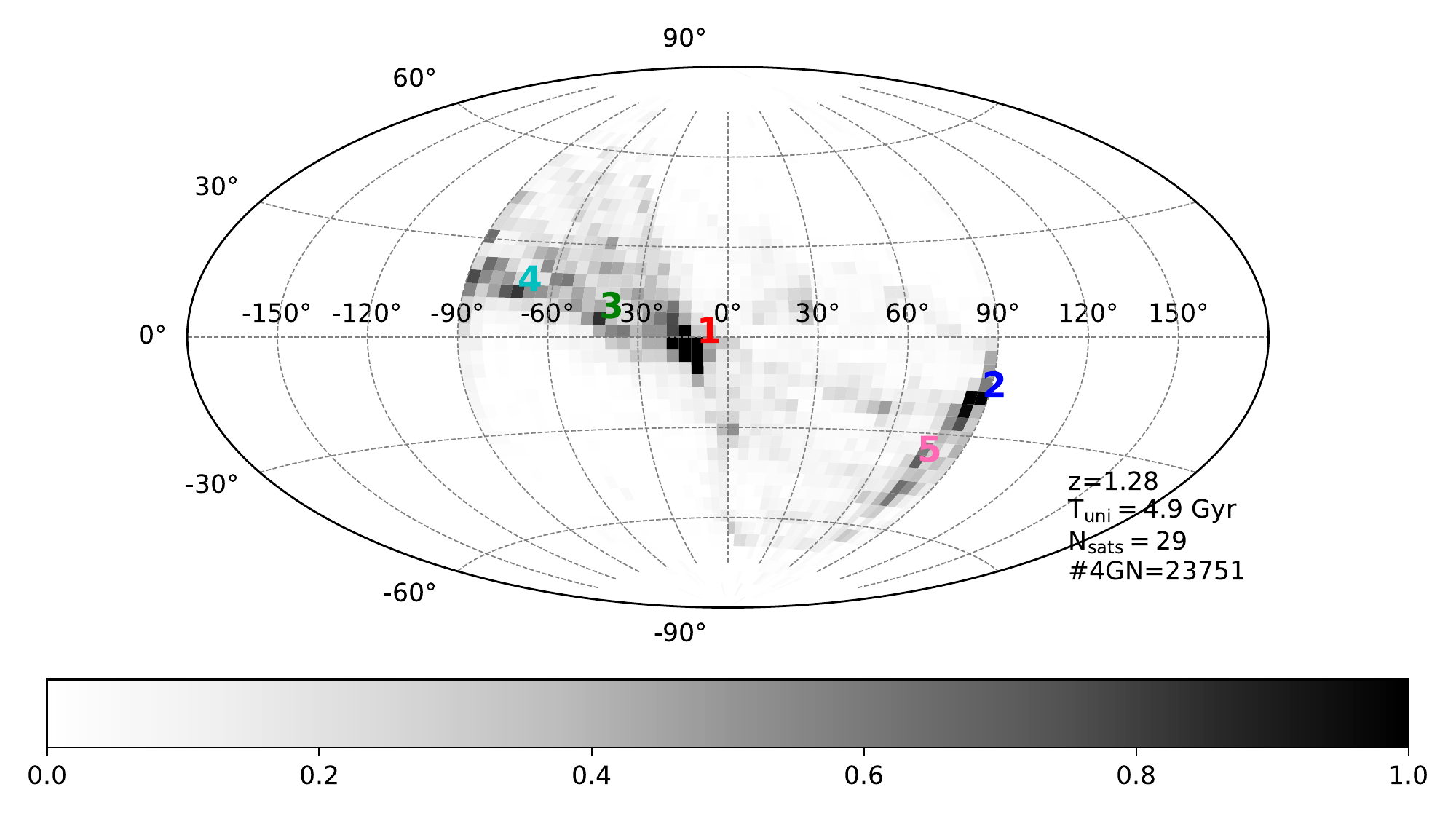}
\includegraphics[width=0.49\linewidth]{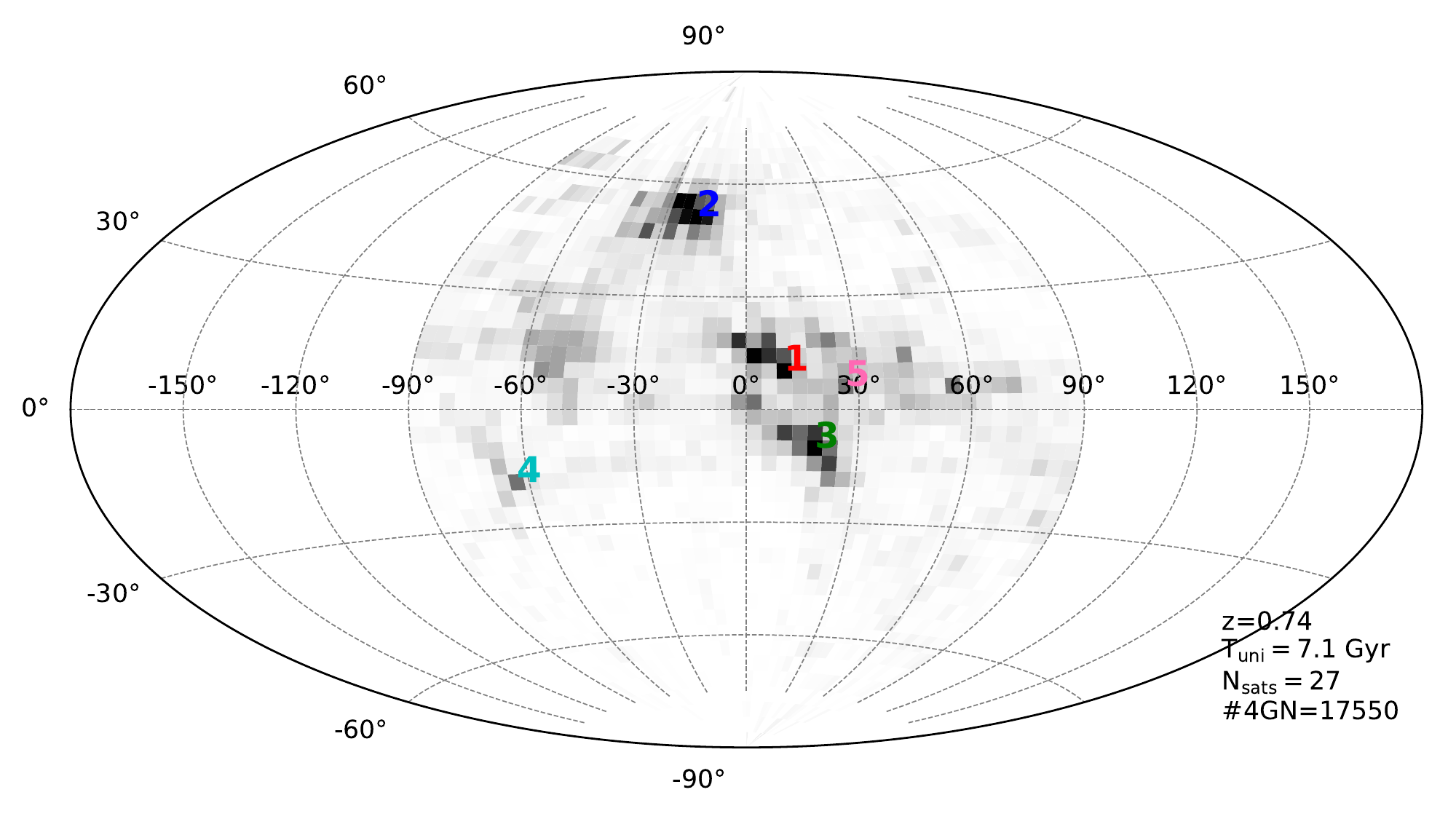}
\includegraphics[width=0.49\linewidth]{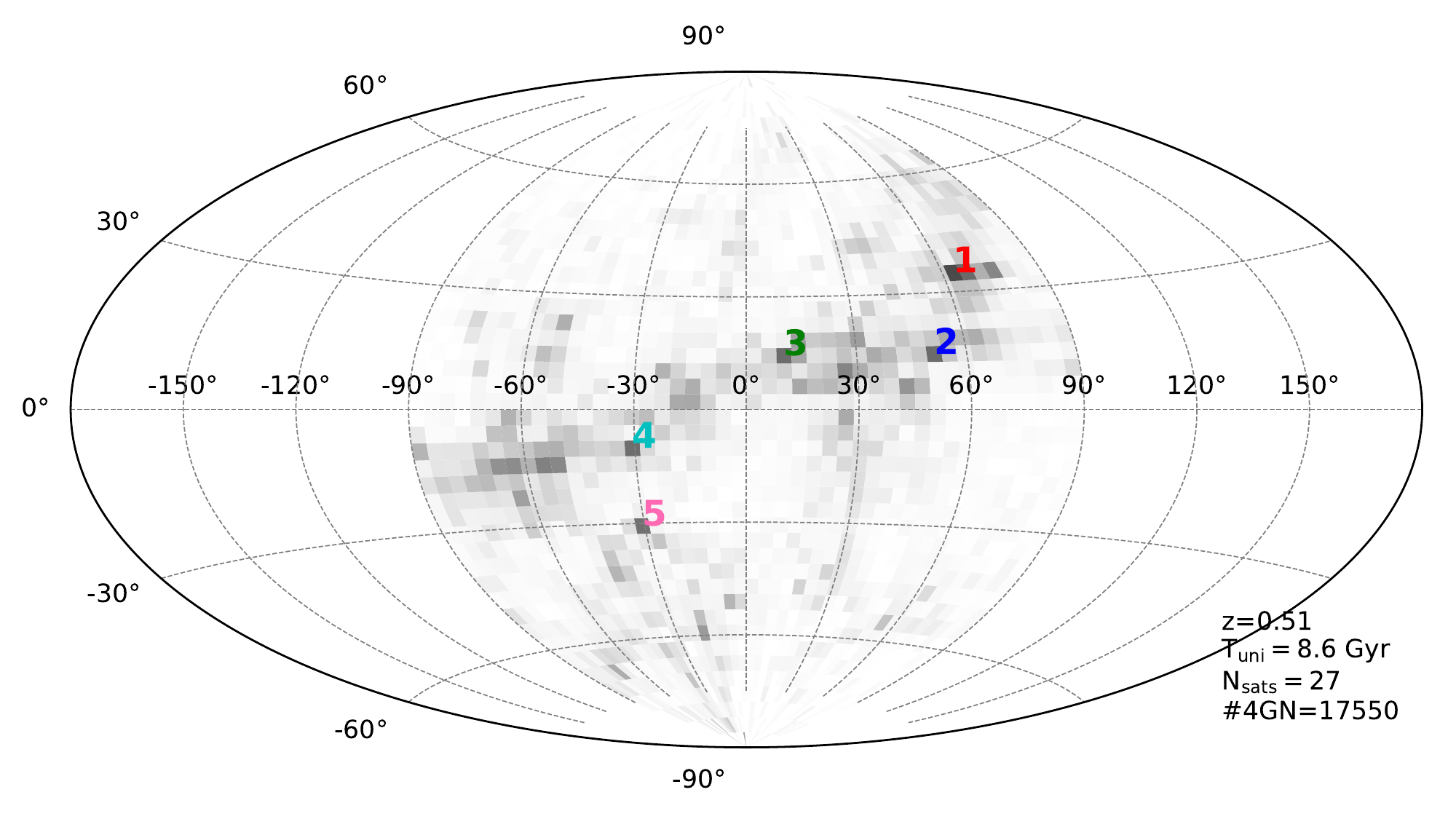}
\includegraphics[width=0.49\linewidth]{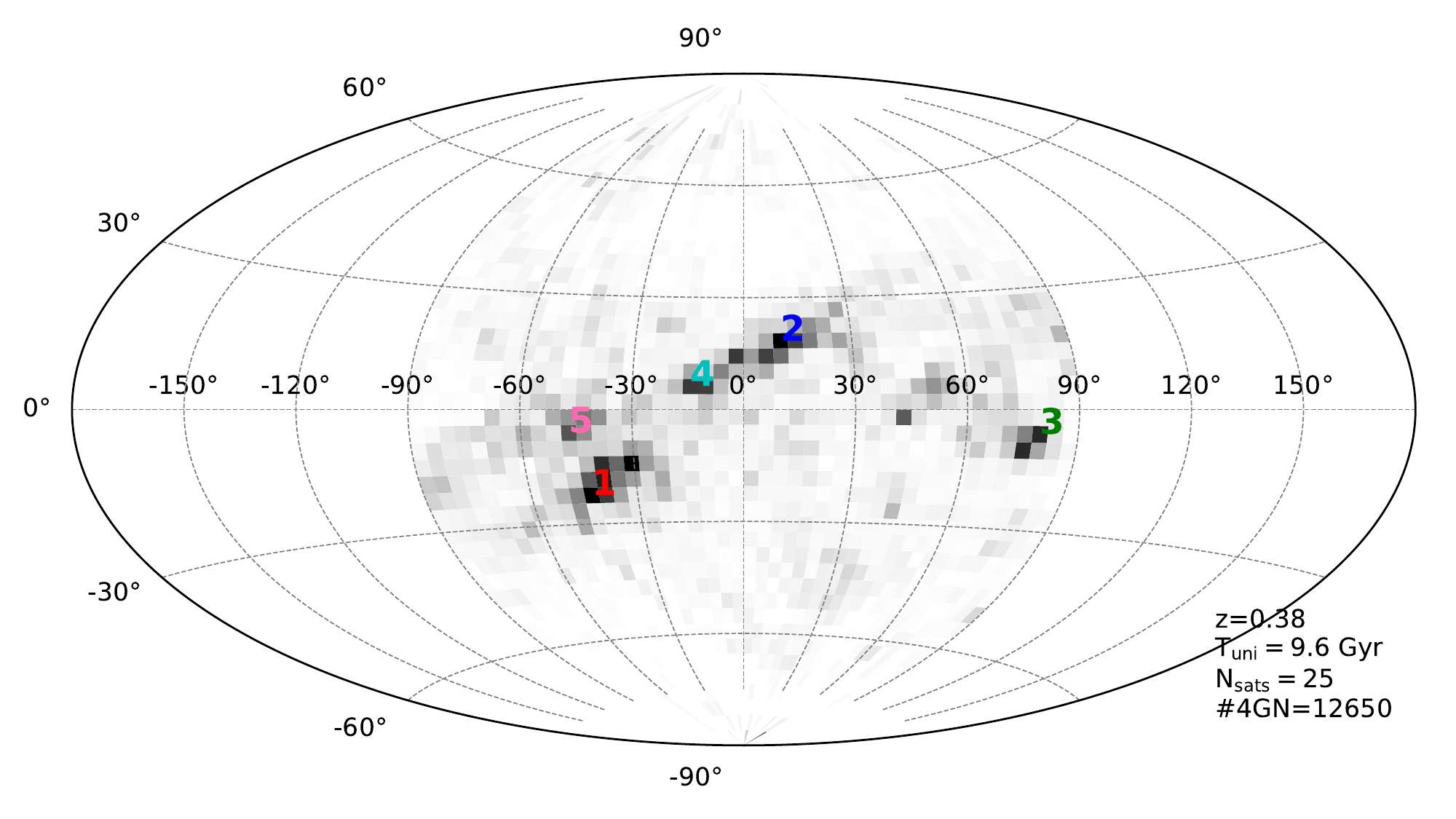}
\includegraphics[width=0.49\linewidth]{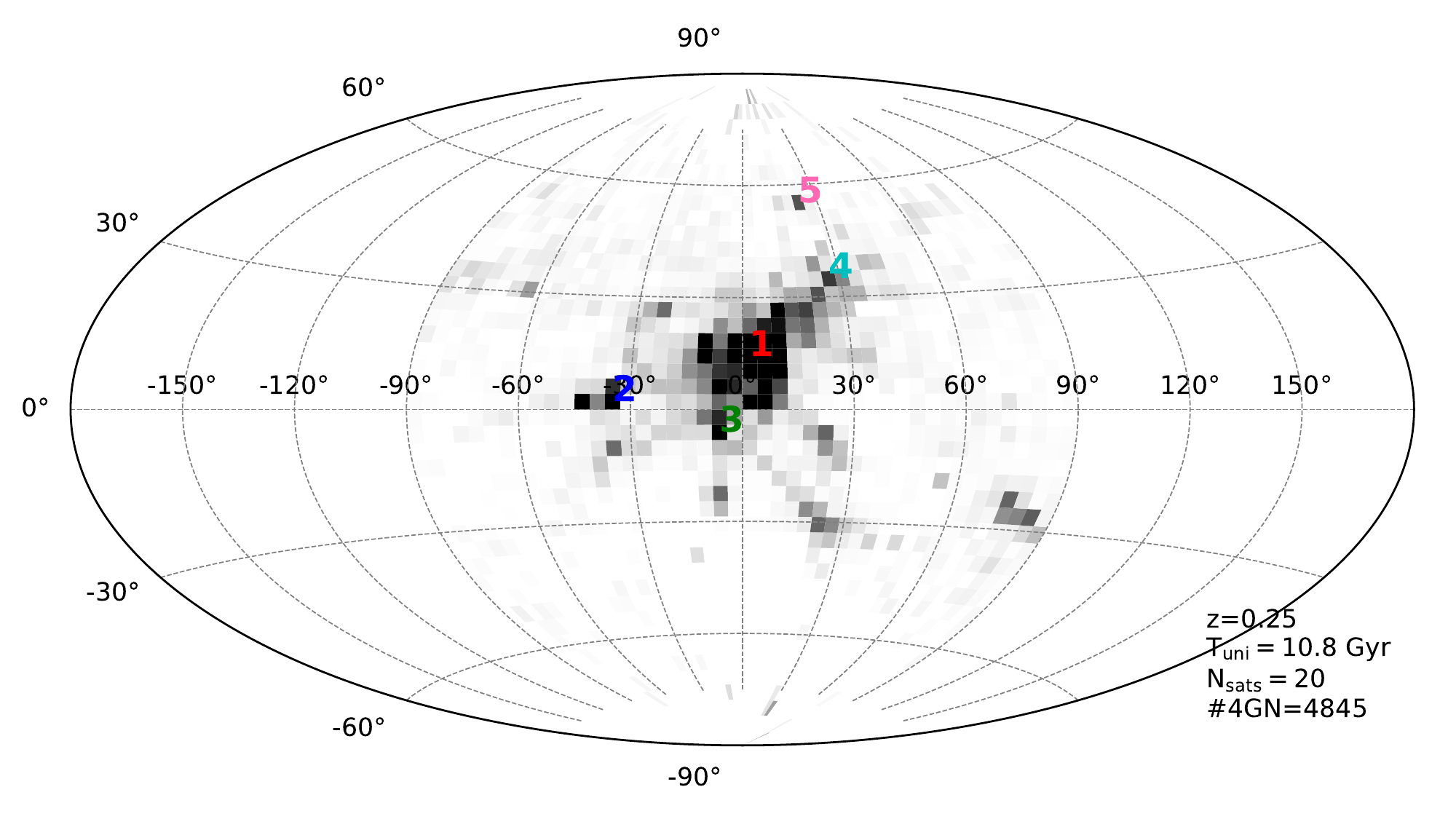}
\includegraphics[width=0.49\linewidth]{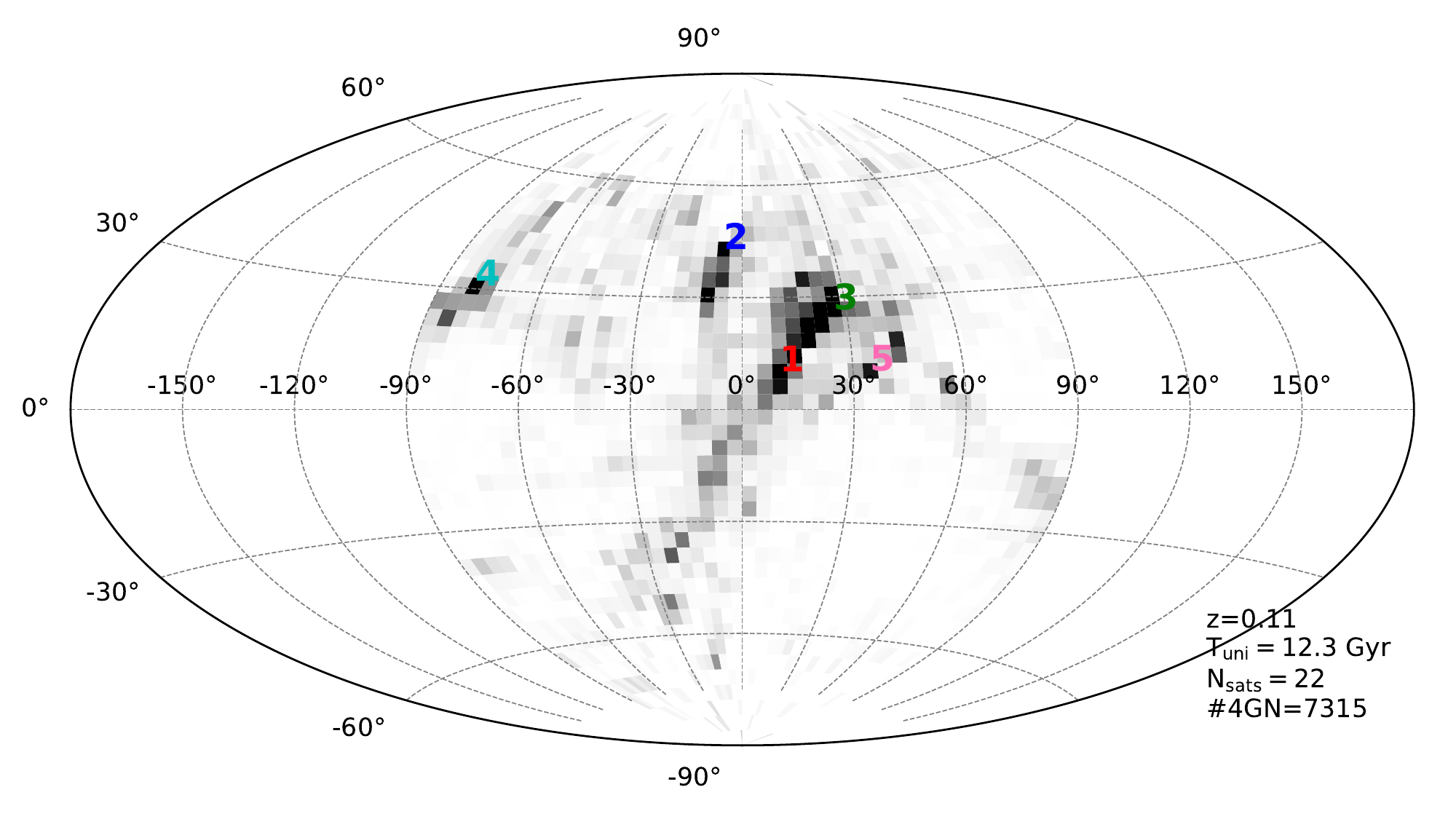}
\includegraphics[width=0.49\linewidth]{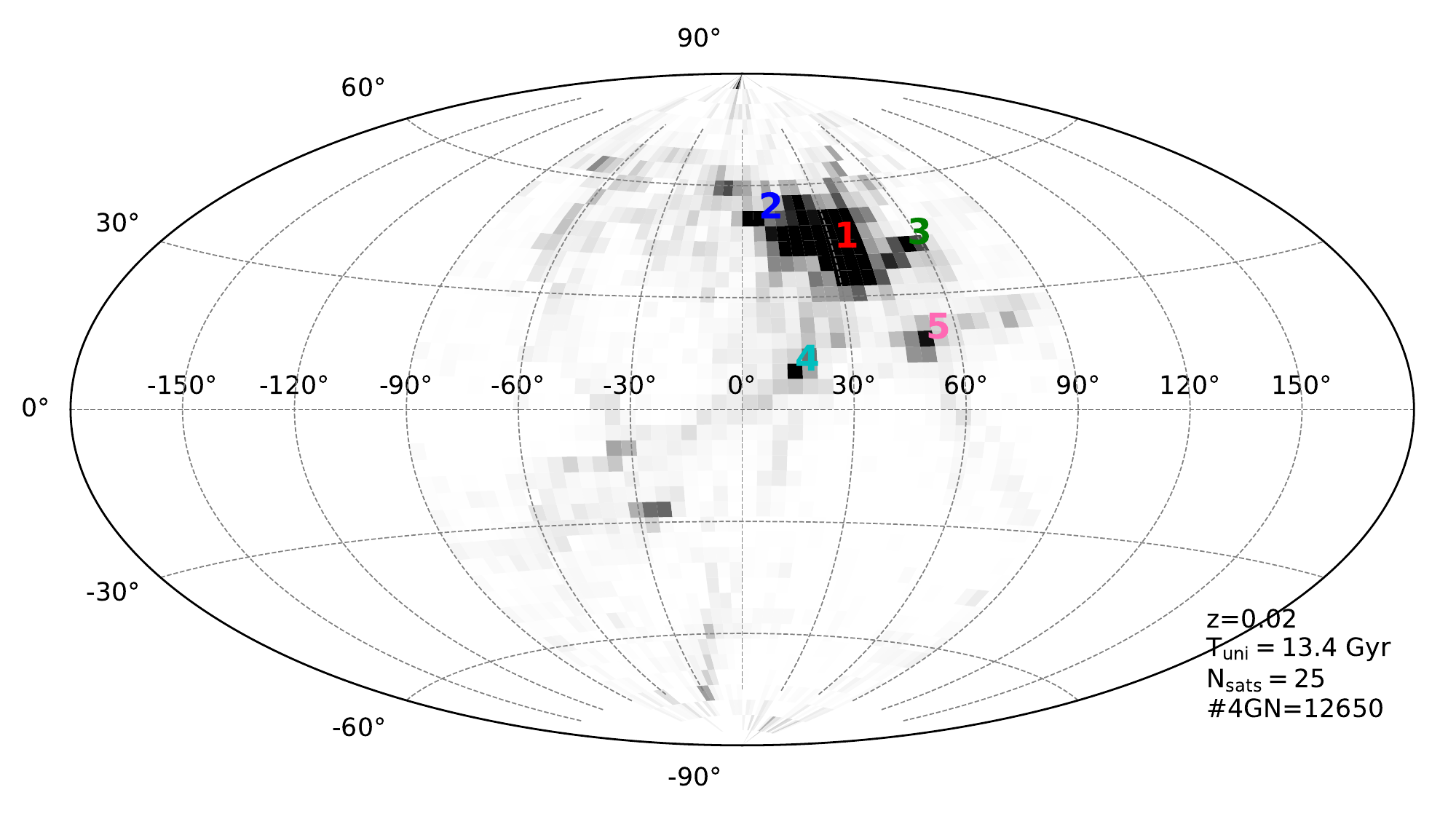}
\includegraphics[width=0.49\linewidth]{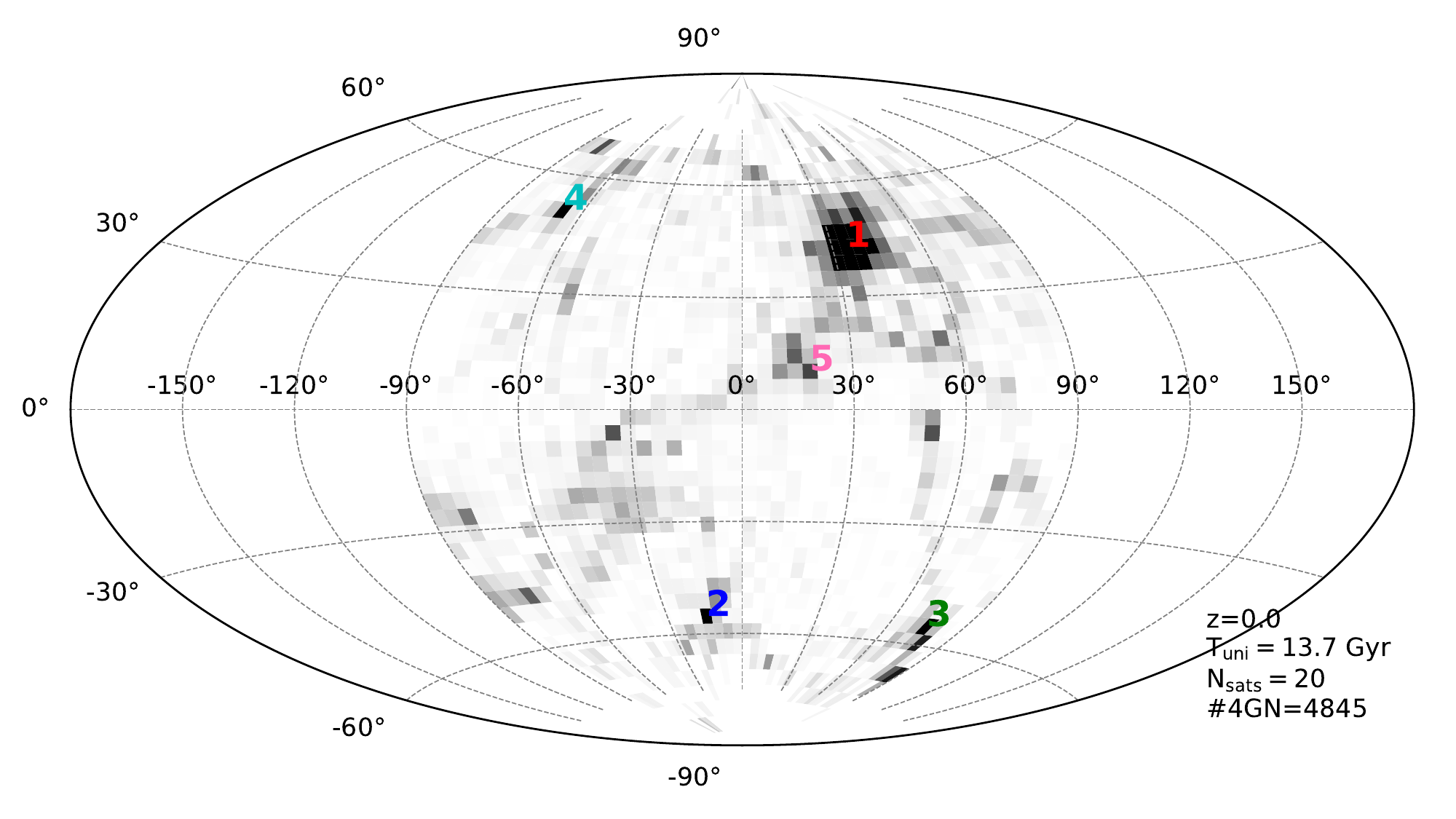}
\caption{
Same as \Fig{aqcdp} for  PDEVA-5004.}
\label{pdevadp}
\end{figure*}


\section{Results: Density Peak Analysis}\label{sec_simsn4p}

\subsection{4-galaxy-normal density (4GND) plots}

 In Paper I the extended 4GND plot  method has been applied  to the same  MW and M31 satellite samples  used in \citealt{Pawlowski13},
consisting of $N_{\rm tot}$=27 and 34 satellites for the MW and M31, respectively. 
 These are the confirmed satellites  within 300 kpc from their hosts, according to the
 \citealt{McConnachie12} ``Nearby dwarf galaxy database"\footnote{\url{http://www.astro.uvic.ca/~alan/Nearby_Dwarf_Database_files/NearbyGalaxies.dat}}
 \rb{as of June 2013\footnote{In Paper I, the most up-to-date sample of confirmed MW satellites is studied as well.}}. 
The MW shows \bt{one} important peak, while M31 shows two.
A detailed analysis of the corresponding planar configurations they point to is  presented in Paper I.

Figures \ref{aqcdp} and \ref{pdevadp}  show examples of  4GND plots  
for Aq-C$^\alpha$ and PDEVA-5004, respectively, where the  observational  Galactic obscuration bias has been applied.
The legend shows the redshift $z$, Universe age $T_{\rm uni}$,  total  number of satellites considered  $N_{\rm tot}$, and the total number of 4-galaxy-normals ($\#$4GN), at that timestep. 
 The main peaks, used for analyses in this work, are marked with numbers, ordered  according to 
the  their central bin density.
  Note that peaks are selected and ordered  independently at each timestep, and that a peak labelled $p=1$ is not necessarily related to another labelled the same way at a different timestep. A color code is also used to identify their contributions in the next Figures.
 The number, strength and location of over-densities changes with time
from showing   several intermediate/low over-densities at some moments, to a  clearly dominating one  at others (especially at the last timesteps analyzed).   
  This behaviour will be studied in the next sections.

Results  obtained when the observational obscuration bias is not applied, and therefore all satellites are taken into account, do not differ substantially from those shown in Figures \ref{aqcdp} and \ref{pdevadp}.
Just at some timesteps new features can be seen around the poles of the Aitoff diagram as compared to  its ``bias" counterpart figure, 
contributed by satellites orbiting in a plane close to that of the disc of the central galaxy.


\subsection{\bt{Satellite contribution-numbers to peaks} }
\label{PCN}

\begin{figure}
\centering
\includegraphics[width=1.02\linewidth]{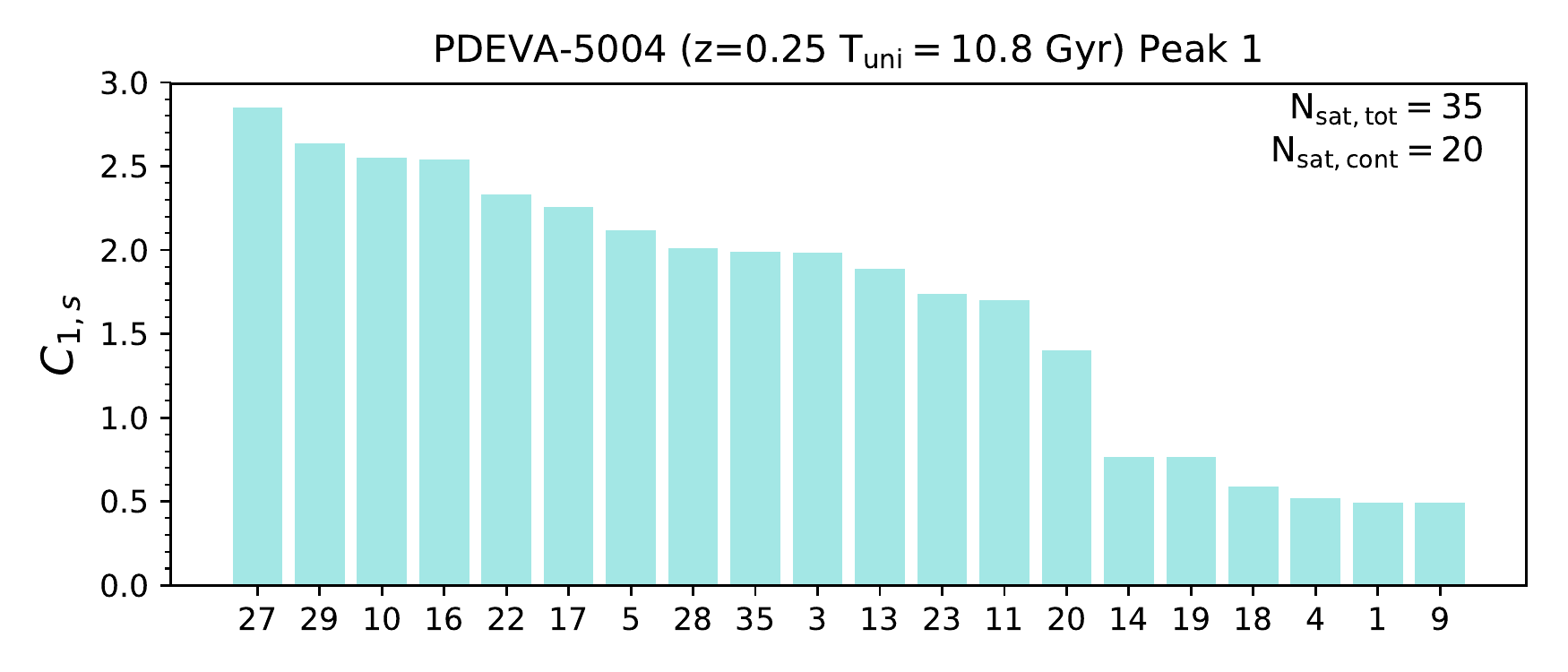}
\includegraphics[width=1.02\linewidth]{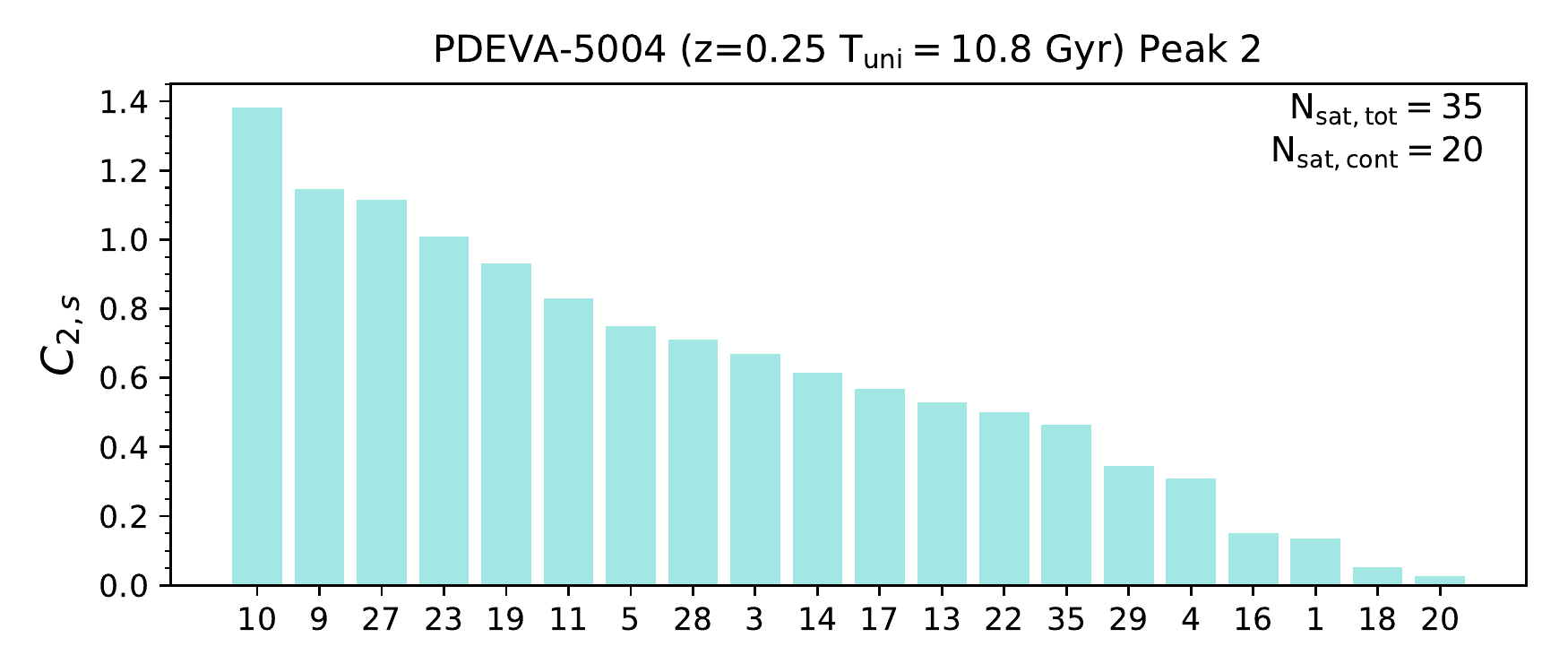}
\caption{
 Bar charts showing the 
contribution ($C_{\rm p, s}$) of satellites to 4-galaxy-normals in 15$^\circ$ around the first and second most important  over-densities in PDEVA-5004's 4GND plot  at $T_{\rm uni}=10.8$ Gyr (including the observational obscuration bias).   The x-axis shows contributing satellite IDs in decreasing $C_{\rm p, s}$ order. 
The total number of satellites considered at the given timestep, $N_{\rm sat,tot}$, and the total number of satellites contributing to 4-galaxy-normals to the given peak, $N_{\rm sat,cont}$, are stated in the right corner of the panels.   }
\label{ex_cont}
\end{figure}

 As explained at the end of Section \ref{N4p}, for each peak $p$ in a density plot  we obtain a list of satellites ordered according to their respective contribution-numbers $C_{\rm p, s}$.
This is the order in which satellites are added to the plane-fitting procedure explained in Sect. \ref{PlQuaAna}, to build the peaks' corresponding collection of planes. 
For illustration purposes, in Figure \ref{ex_cont}  we draw the $C_{\rm p, s}$ histograms corresponding to
 satellites $s$ contributing to 
 Peak 1 (top panel) and Peak 2 (bottom panel) 
of PDEVA-5004's  4GND plot  
at $\rm T_{uni}=10.8$ Gyr  (see Figure \ref{pdevadp}). 
 The x-axis shows the IDs of satellites; only the non-zero contributions have been plotted. 
Some satellites show a high contribution to one peak 
while others do not, meaning that they are involved in a low
number of 4-galaxy-normals close to the respective peak. 
In this particular case, we see that those satellites showing a high $C_{\rm p, s}$
relative to the main peak are not among those contributing the most to the second peak. 

\subsection{Peak Strength Analysis}
\label{PeStreAna}

\bt{
The  $C_{\rm p}$ peak strengths of Peak 1 and Peak 2 
(i.e., $C_1$ and $C_2$)
for the MW and M31 are given in Table \ref{c1mwm31}.  Errors are one sigma deviations over 100 random realizations of their radial distance uncertainties, as explained in Paper I. 
These are specially large in the case of M31 satellites; 
as  a result,   peaks in its 4GND plot are \bt{more} blurred \bt{as compared to the MW ones}, giving rise to lower $C_1$ and $C_2$ values in M31.
}

\begin{figure}
\centering
\includegraphics[width=\linewidth]{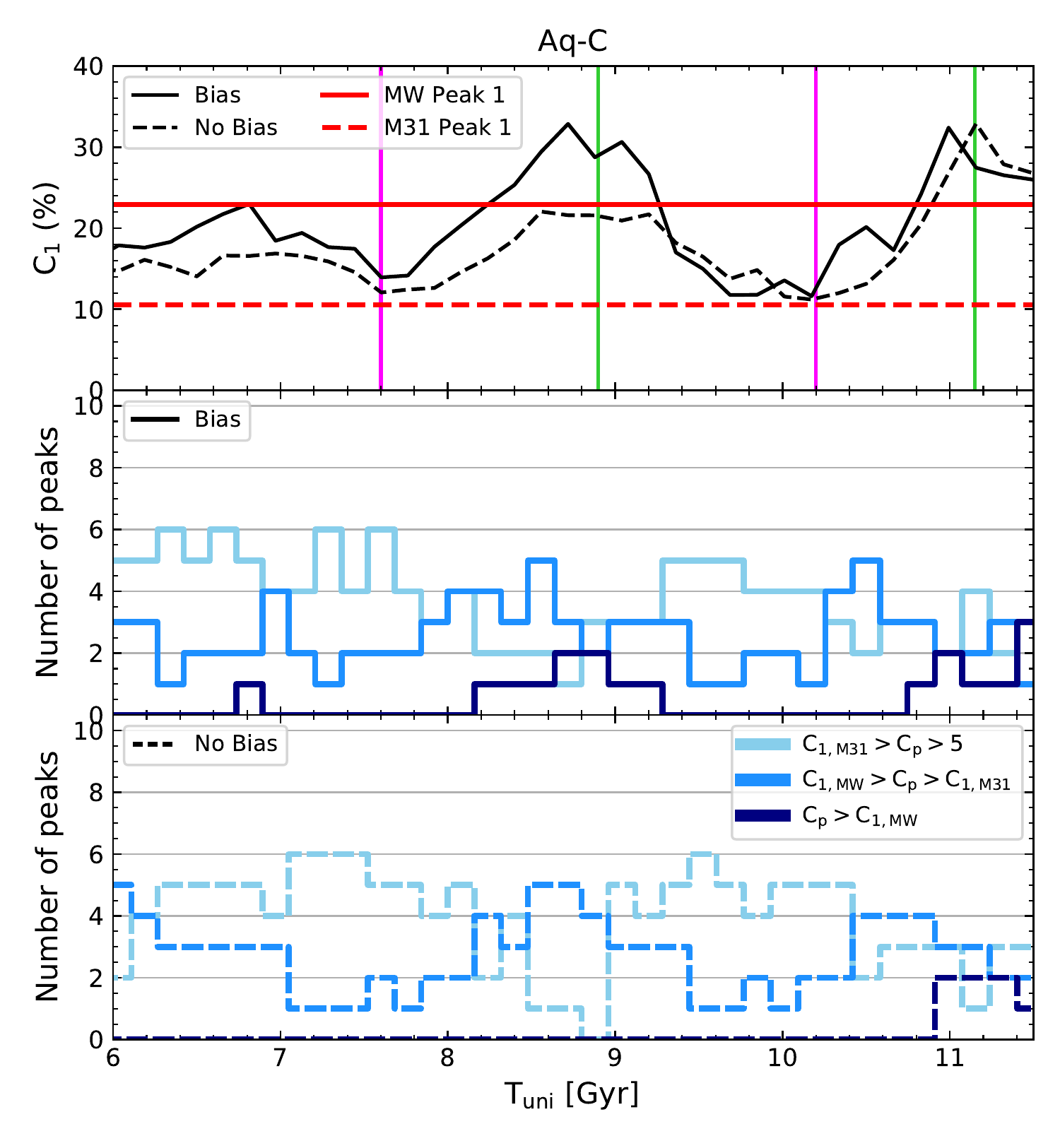}
\includegraphics[width=\linewidth]{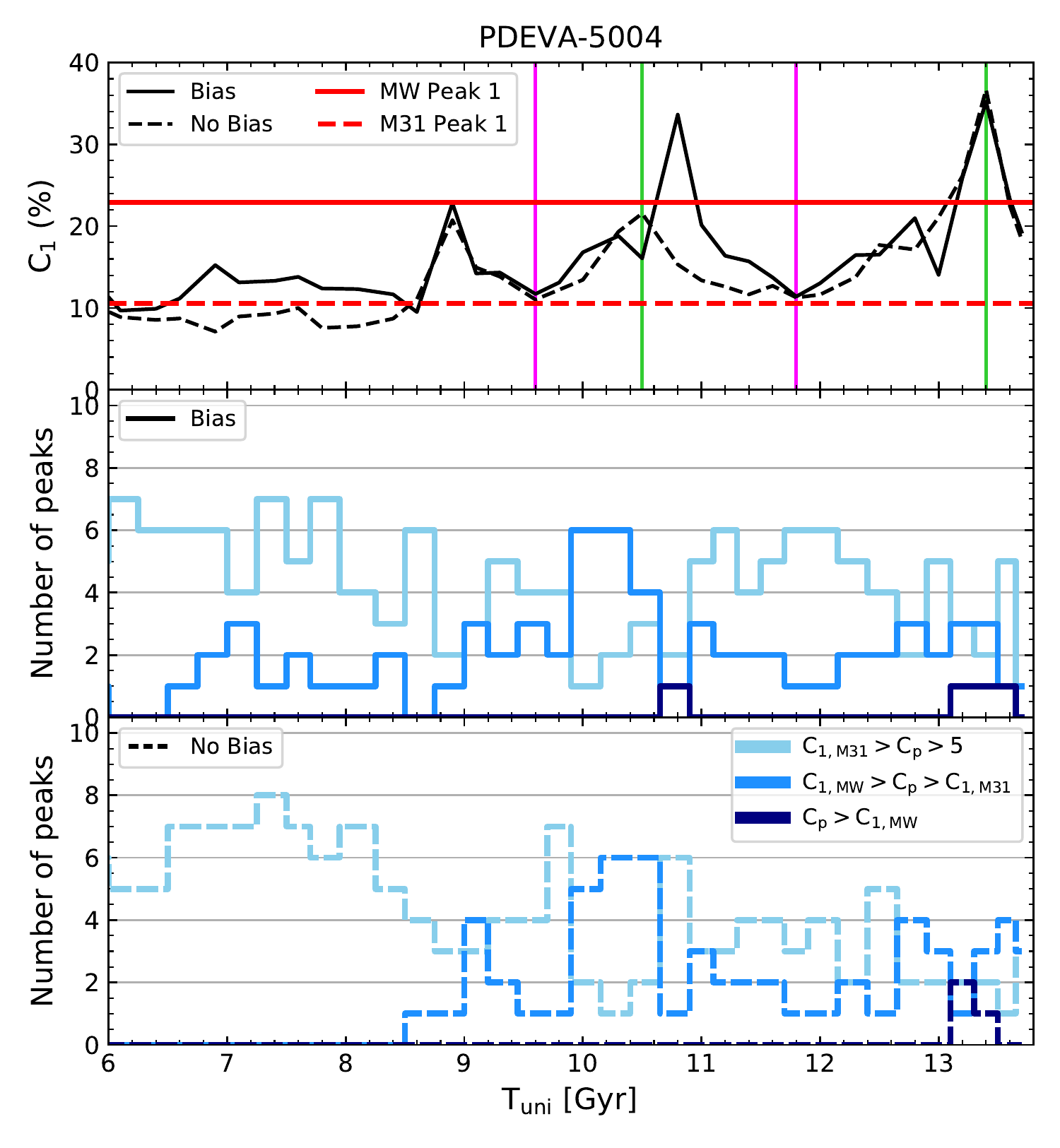}
\caption{
\bt{
\textit{Upper panels:} $C_1$ peak strength 
 as a function of the Universe age, for both biased and non-biased satellite samples.  Green (magenta) vertical lines mark the moments when non-biased $C_1$ reaches maximum (minimum) values.
\textit{Lower panels:} number of density peaks at each timestep with $C_1$
 strengths within given intervals.
\bt{These intervals are defined by the MW and M31 $C_1$ values, shown as horizontal lines in the upper panels (see also Table \ref{c1mwm31}).}
Top figure:  Aq-C$^\alpha$; bottom figure: PDEVA-5004.
}
}
\label{PeakStrength}
\end{figure}

\bt{
In the upper panels of Figure \ref{PeakStrength} we present the value of $C_1$  at each timestep 
for Aq-C$^\alpha$ and PDEVA-5004.
 $C_1$ fluctuates, reaching values that can be even higher than those of the MW or M31 at $z=0$.
 \bt{In general, the application of the observational obscuration bias enhances the strength value of the main peak. }
 }
 
 \bt{
The  Universe ages   T$_{\rm uni}$  where the respective $C_1$
\bt{(in the case where all satellites are considered, 'no bias')}
 have maxima (minima) are marked by green (magenta) vertical lines.
These  time intervals of local maxima and minima have an average  
duration of \bt{0.5 - 1} Gyr (consistent with the values \citet{Shao19} find in their analyses of the EAGLE simulation).
 This has been estimated from their FWHM, where we take the mean $C_1$ as floor value. These periods will be related with plane quality in the next sections.
}

\begin{table}
\caption{
\bt{Peak strengths $C_p$  for the main 2 peaks found in the MW and M31 4GND plots (see Paper I).
Peak strength is computed as $C_p \equiv \Sigma_s C_{s,p}$,
where $C_{s,p}$ is normalized to the total weighted number of 4-galaxy-normals (see Section \ref{PeStreAna}).
 Results shown are means and 1$\sigma$ standard deviations calculated from 100 random realizations using the radial distance uncertainties.
Peaks  \#1 are the strongest ones and  Peaks \#2 follow in strength.
 }
}
\begin{tabular}{l l l }
\hline
 & $C_1 \pm \sigma$ (\%)  & $C_2 \pm \sigma$ (\%) \\
 \hline
 \hline
 MW & 22.92$\pm$0.26 &  14.31$\pm$0.20 \\
 M31 & 10.53$\pm$ 0.62 & 10.52$\pm$1.62 \\
 \hline
\end{tabular}
\label{c1mwm31}
\end{table}

\bt{
Another interesting possibility that the peak strength $C_{\rm p}$ allows is to determine the number of peaks  with strengths  above given thresholds or within given  intervals, at different Universe ages.
\bt{This is shown in  the lower panels of  Figure \ref{PeakStrength},
 with respect to the strengths of the peaks labelled as \#1 in the MW and M31,
i.e., $C_{1,MW}$=22.9\% and $C_{1,M31}$=10.5\%.}
 At given times there are a few  peaks encompassing a high \% of 4-galaxy-normals \bt{(high $C_1$)} that then break into several different peaks with lower \bt{strengths}. These later on collimate into high $C_{\rm p}$ peaks again. 
That is, the number of peaks with $C_{\rm p}$ within given peak strength intervals fluctuates with time. 
 We recall that the background is also accounted for to normalize the peak strengths at fixed timesteps.
}

\bt{
Summing up, as measured with $C_1$,  the peak strengths in observations and simulations are 
 consistent within given time intervals.  
 Regarding the number of peaks, we see that when $C_1$ takes high values, the number of weak peaks decreases
and that  of stronger peaks increases. This happens specially at the last timesteps analyzed.
\bt{In particular}, the number of strong peaks
($C_1>C_{1,MW}$)
 is never very high (1-3 at most, occurring at late times), \bt{ in consistency with observations.} 
\bt{Again, we can see that the obscuration bias favors the emergence of strong peaks in both simulations.}
}


\section{Results: Plane quality analysis}
\label{PQA}

\subsection{Comparing to the MW and M31 satellite systems }

The quality of  planar configurations of satellites obtained from the two main density peaks in the MW and M31   have been analyzed in Paper I. 
\bt{The} extension to Pawlowski's 4GND plot method presented in Paper I 
\bt{has revealed a}
 richer and higher-quality plane structure in the MW and M31 than that   reported previously in the  literature. 
In particular, in both the MW and M31,
the quality of the 
planar structure  of satellites provided by Peak 1 is,
at any $N_{\rm sat}$, better than that corresponding to Peak 2 
(although in the case of M31 the differences between peaks are not that important when the error bands are taken into consideration).

In this  paper we focus  on the best quality planes that can be found in a satellite system at a given moment; 
therefore,
we will compare
our simulation results
 to the MW and M31 \textit{Peak 1} results, while 
those of  
  Peak 2 will not be used in this paper for comparison purposes.

\bt{As   said above,  it  was also shown in Paper I that there is no correlation between the stellar mass of an observed satellite and its $C_{\rm p, s}$ contribution to the main density peaks found. This allows for a fair comparison between simulations and observations despite the different satellite mass ranges involved.
}

\begin{figure*}
\centering
\includegraphics[width=0.8\linewidth]{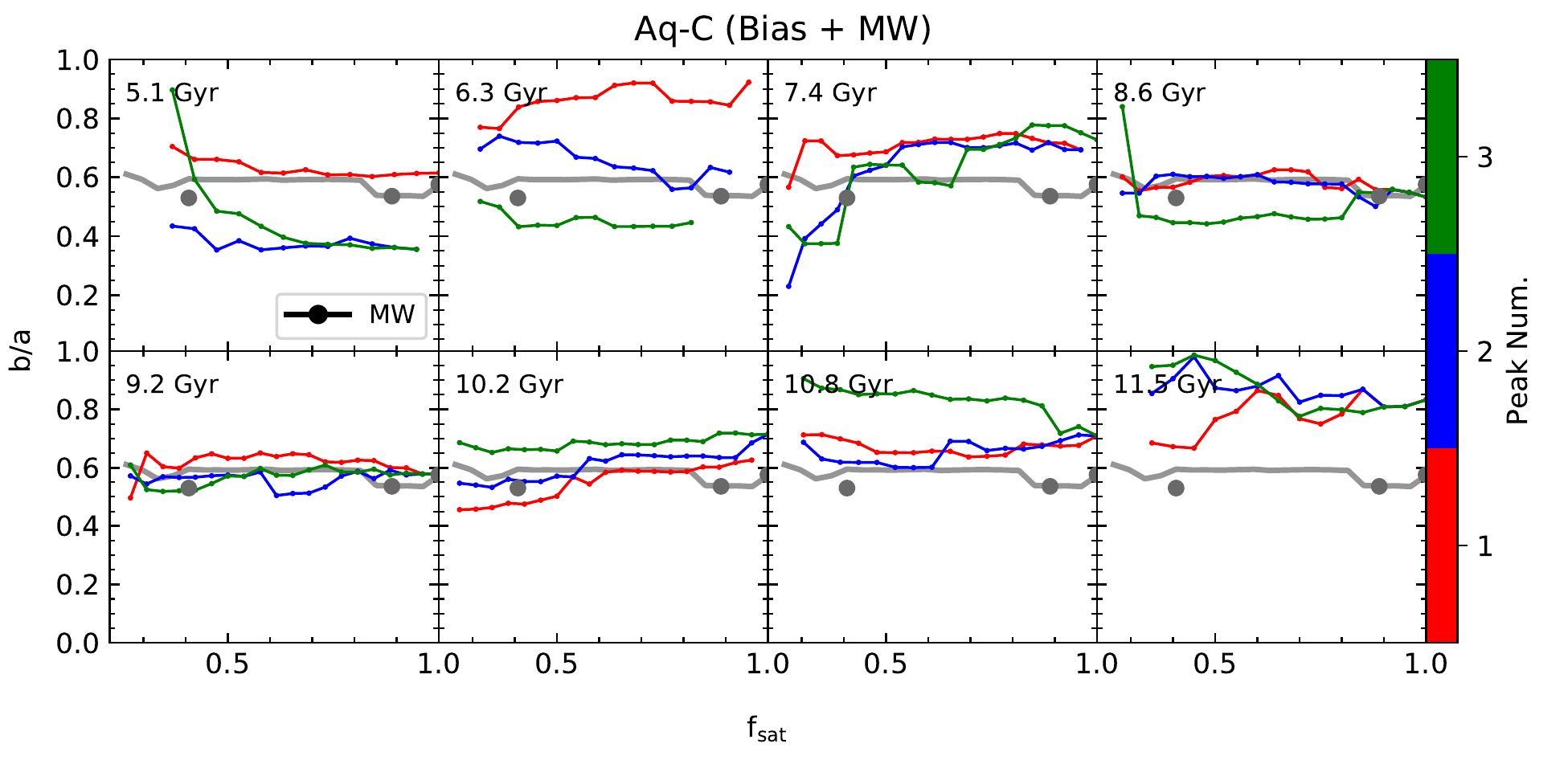}
\caption{
The intermediate-to-long axis ratio $b/a$ as a function of the fraction of satellites $f_{\rm sat}$ included in the plane at different Universe ages, T$_{\rm uni}$
for Aq-C$^\alpha$. We compare to the MW Peak 1 result   and hence have applied the observational Galactic obscuration bias to the simulation.
The results for the planar structures defined by the most-prominent density peaks  are shown as lines of different colors. 
 The color code and  numbering allow to find the corresponding peak in the 4GND plots (Figures \ref{aqcdp}).
 A gray solid line shows the result obtained from the MW's main density peak, and points
 show the specific values for  MW   observed planes of satellites  mentioned in the literature 
 (i.e., classical, VPOS-3, VPOSall, \citep{Pawlowski13}).
 }
\label{pdeva_ba}
\end{figure*}

\begin{figure*}
\centering
\includegraphics[width=0.8\linewidth]{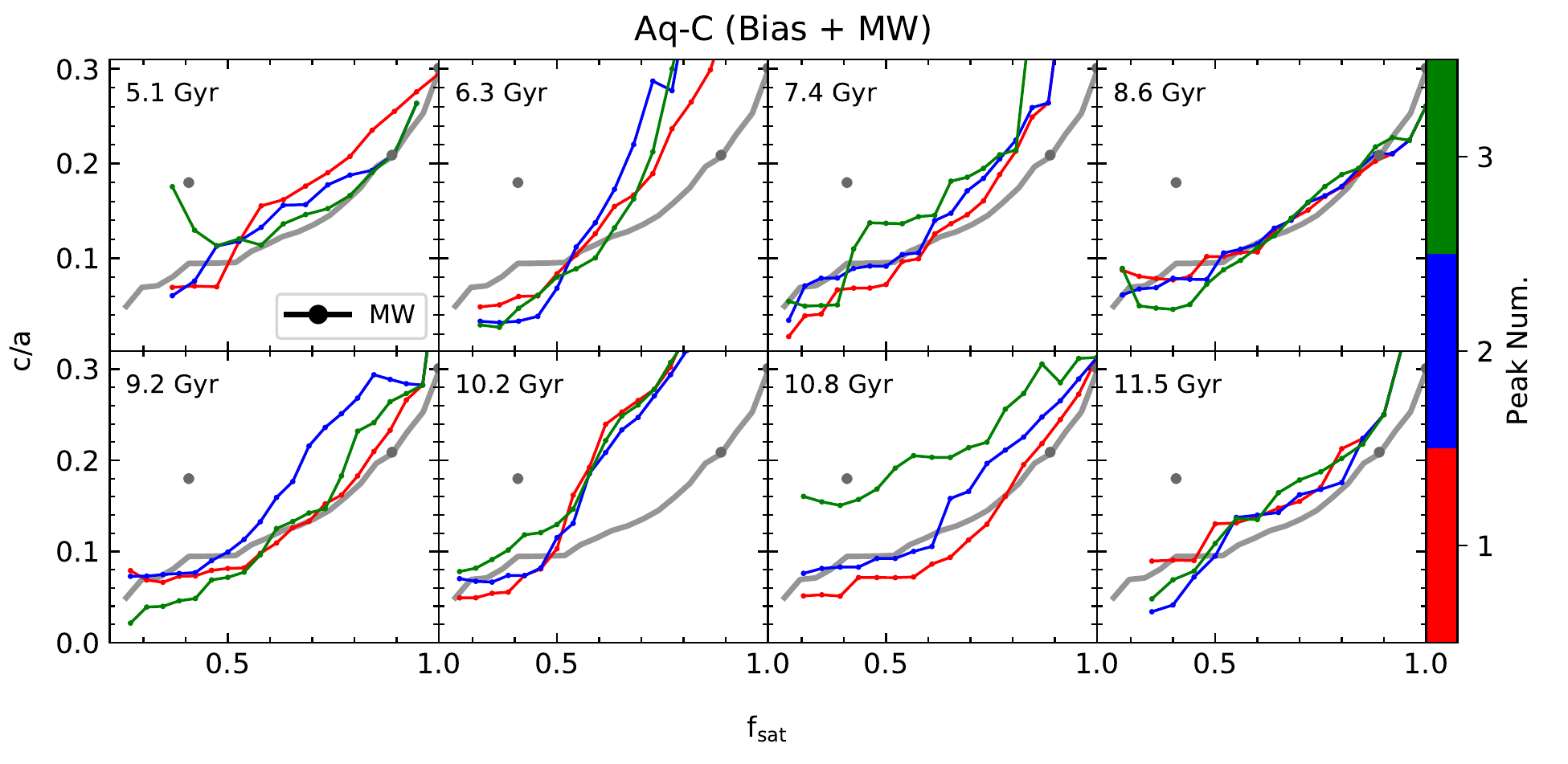}
\includegraphics[width=0.8\linewidth]{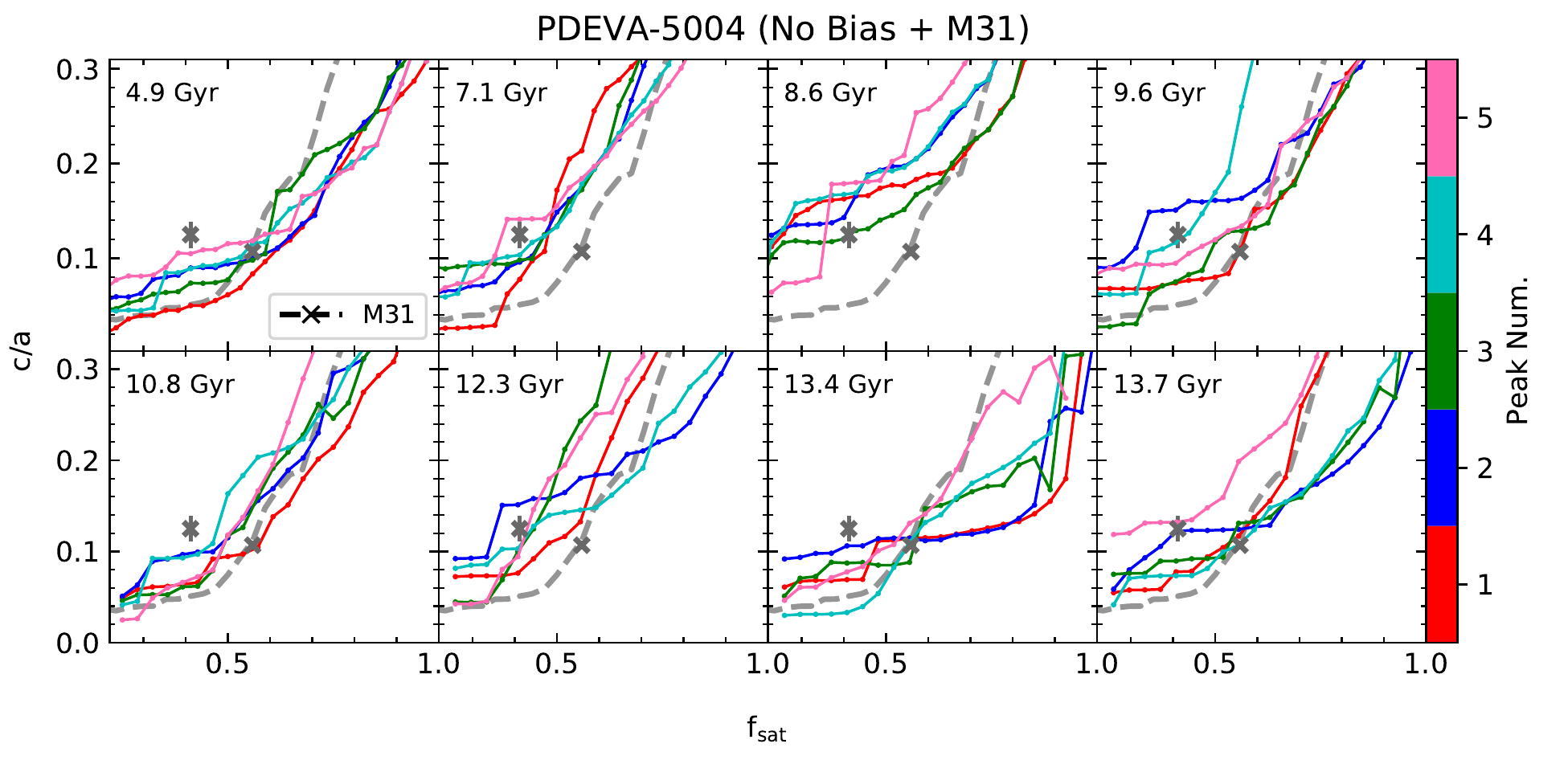}
\caption{
\bt{Examples of quality analysis of the planar structures found: 
The short-to-long axis ratio, $c/a$, as a function of
the fraction of satellites $f_{\rm sat}$ included in the plane  at different Universe ages.
\textit{Top}: 
Aq-C$^\alpha$ versus the MW. Simulated results include the observational obscuration bias. 
A gray solid line shows the result for the MW's main density peak, and points show the specific values for  MW   observed planes of satellites  mentioned in the literature (i.e., classical, VPOS-3, VPOSall). 
\textit{Bottom}: 
PDEVA-5004 versus M31. 
 A gray dashed line shows the result  for  M31's main density peak, and
 crosses  show the specific values for  M31   observed planes of satellites  mentioned in the literature (i.e., Ibata-Conn-14 and GPoA). 
 }
}
\label{aqc_ca_fracA}
\end{figure*}


\subsection{Quality of simulated planes in terms of the satellite fraction involved}

\subsubsection{$b/a$ vs. $f_{\rm sat}$}

Concerning the application of the method to the simulation data, we first   address the planar ($b/a \sim 1$) or filamentary ($b/a << 1$)  character of the best-fitting structures found with the ToI analysis,
 where $b/a$ is \bt{the intermediate-to-long axis ratio} in the ToI scheme. 
 As an  \bt{illustrative example}   of our results,  in Figure \ref{pdeva_ba} we plot 
  $b/a$   versus  $f_{\rm sat}$ for the main density peaks  found in Aq-C$^\alpha$ 
at different timesteps. 
In this figure and in the following ones that compare the changes  of a ToI output with $f_{\rm sat}$, each panel corresponds to a given timestep.  
Lines of  different colors stand for the
 collections  of  planes of satellites 
associated to the respective   peaks  \bt{numbered with the same color coding in Figure \ref{aqcdp}.  }
\bt{Based on Figure \ref{PeakStrength} and the 4GND plots shown in Figures \ref{aqcdp} and \ref{pdevadp}, the consideration of a number of  N$_{\rm peak}$=3  peaks for Aq-C$^\alpha$
and N$_{\rm peak}$=5  peaks for PDEVA-5004
seems a reasonable choice ensuring the exploration of  all possibly relevant planar configurations.}

 Observational data results are shown as gray lines and points. 
Points show the specific values for  MW/M31   observed planes of satellites at $z=0$ mentioned in the literature (i.e., MW: classical, VPOS-3, VPOSall;  M31: Ibata-Conn-14, GPoA; see values in Table 1 of Paper I).
\bt{When comparing to the MW, we show simulated results where the  obscuration bias is applied; when comparing to M31, we show results considering  all satellites\footnote{\bt{We acknowledge that for an even more accurate comparison, other observational biases could be applied to M31, such as that of the mask of the PAndAS survey which discovered most of its satellite galaxies \citep[see e.g.,][]{Gillet15}. Nonetheless, as PAndAS presents a very homogeneous panoramic coverage \citep[see Figure 2 in][]{Conn13},  we neglect any bias in this work.}}.
 }

\bt{We find no filamentary structures;
in fact, at all timesteps a planar structure exists whose $b/a$ is larger than the observational case, at all $\rm f_{sat}$.}
 The general behaviour of $b/a$, both for Aq-C$^\alpha$ and PDEVA-5004, 
 \bt{and in the 'bias' and 'no bias' cases,} 
 \bt{is that}  $b/a$ changes only slightly when new satellites are added to the fit,  giving rise to wide $f_{\rm sat}$ intervals where $b/a$ is almost constant.
  This means   that the planar character
\bt{of the spatial configuration of satellites } does not depend very much on the number of satellites involved. 
This behaviour is also found in the MW and M31 (see gray lines in Figure \ref{pdeva_ba}).


\subsubsection{$c/a$ vs. $f_{\rm sat}$}

Having confirmed that the structures found in our simulations are indeed planar,
we can proceed with the study of the quality of such planes
\bt{through $c/a$}. 
As explained previously,
quality is assessed by a two parameter notion ($c/a$, $f_{\rm sat}$) such that at a given $f_{\rm sat}$, the plane with the lowest $c/a$ presents the highest quality. 
%
In particular, 
\bt{we define that a
 \textit{strong}   consistency exists between  plane collections  from simulations  and observations
when there is one  colored  line  from simulations    with similar or lower $c/a$  values than that of the MW/M31  gray  line  at \textit{all} $f_{\rm sat}$.}
A \textit{weaker} condition refers to consistency  between an observed plane and one detected in simulations  
with the same  \textit{particular}  $f_{\rm sat}$.
In this case, the peak assuring consistency between data and simulations can vary from $f_{\rm sat}$ to $f_{\rm sat}$.

\bt{We have carried out the analysis of plane population and thickness ($c/a$ versus $f_{\rm sat}$)  for the main  peaks  found in Aq-C$^\alpha$ and PDEVA-5004, comparing to both the MW and M31. 
As examples of the results obtained, in  Figure \ref{aqc_ca_fracA} we present the results for Aq-C$^\alpha$ vs. MW (top panel; the observational obscuration bias is applied) and PDEVA-5004 vs. M31 (bottom figure; all satellites are considered).}

Independently of applying the observational obscuration bias or not, 
 both Aq-C$^\alpha$  and PDEVA-5004  present 
 thin and highly-populated planes at all timesteps.
  In general, $c/a$ is low ($\lesssim 0.3$) for all peaks at all timesteps when including up to $\sim80\%$ of satellites. This is already proving the  oblate spatial distribution of the entire satellite population in both simulations.  
In particular,
at \textit{all} timesteps
\bt{and in both simulations,}
 there are planar structures compatible
(in terms of $c/a$ and $f_{\rm sat}$)
 with  the M31 Ibata-Conn-14 plane 
and the MW classical plane.
 The GPoA value is also recovered in  almost all  timesteps.
The strong consistency condition is met  in many cases. For example,   we can find very similar or higher quality planar structures than that of the MW 
 in Aq-C$^\alpha$ (`bias' case) 
 at $T_{\rm uni}$ = 8.6, 9.2 and 10.8  Gyr;
and we find similar or higher quality planar structures than that of M31 in PDEVA-5004
at $T_{\rm uni}$ = 4.9, 9.6, 10.8, 13.4 and 13.7 Gyr.
 \\

\begin{figure*}
\centering
\includegraphics[width=0.70\linewidth]{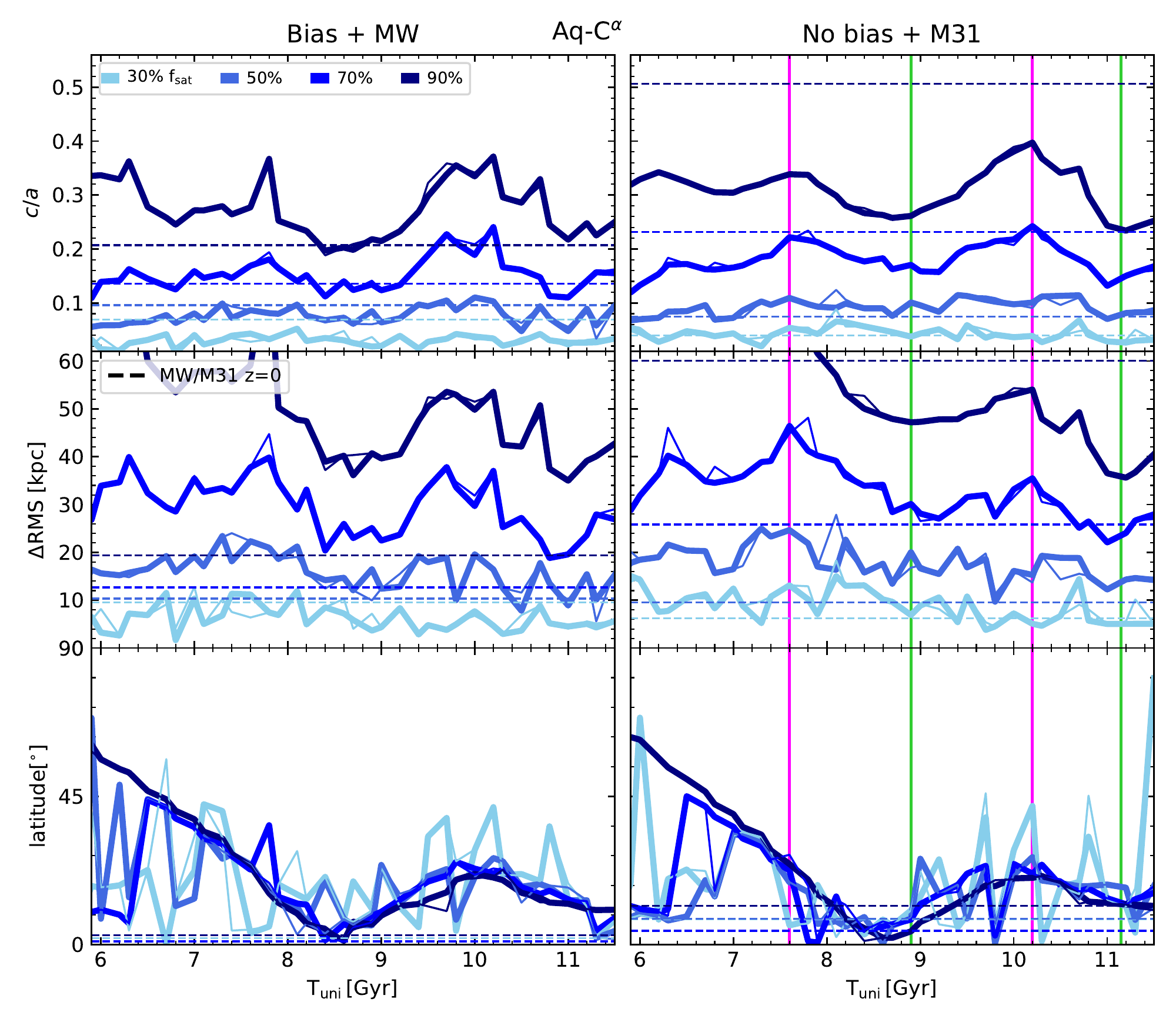}
\includegraphics[width=0.70\linewidth]{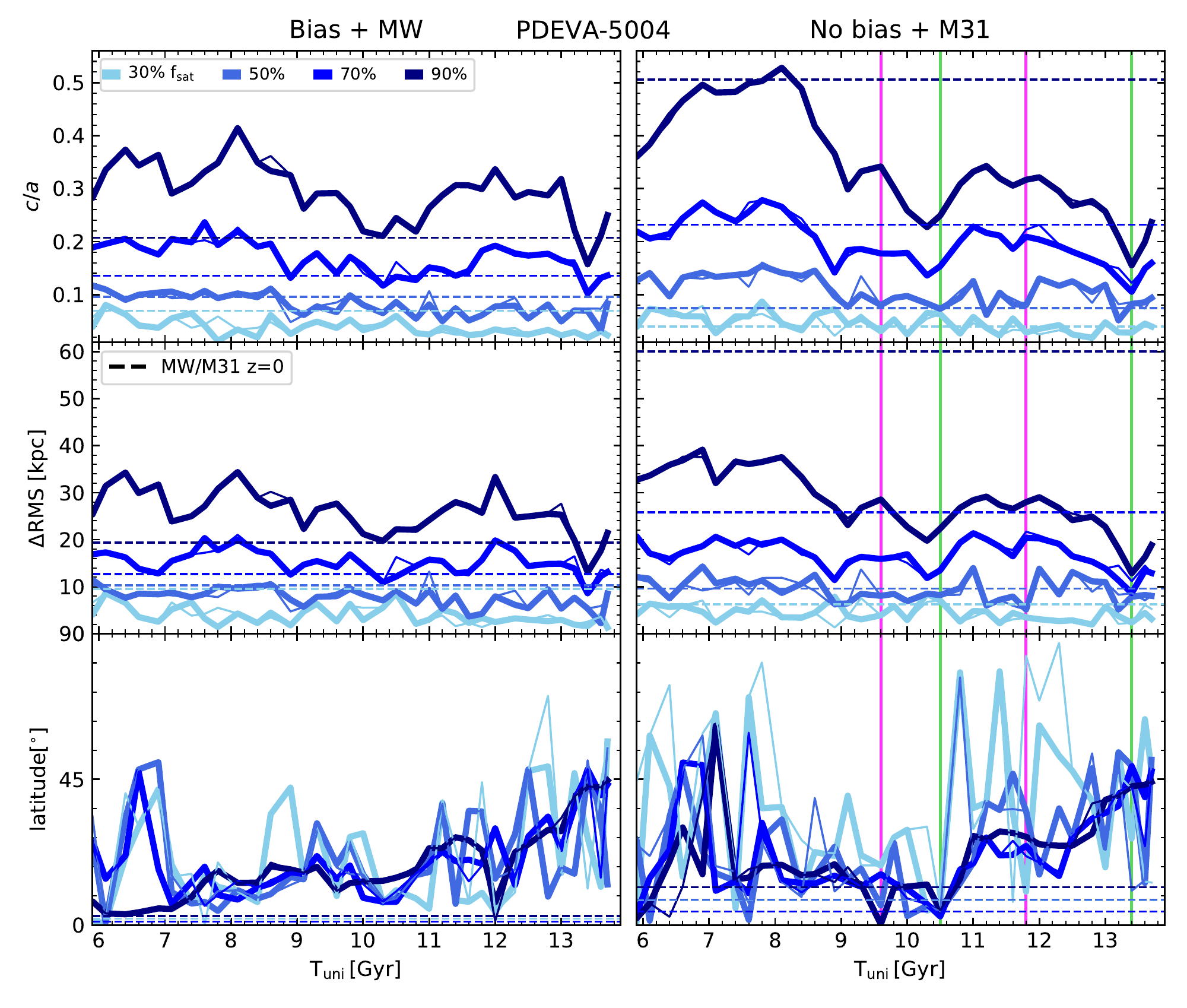}
\caption{\bt{
Short-to-long axis ratio $c/a$,  $\Delta$RMS height and plane of satellites inclination relative to the disc, for the best planes found at each timestep including
a fraction
 $\rm f_{sat}=$30\%, 50\%, 70\%, and 90\% of the total number of satellites.
 Top: Aq-C$^\alpha$. Bottom:  PDEVA-5004.
Left panels: Results having applied the observational obscuration bias compared to the MW $z=0$  values.
Right panels: Results considering all satellites compared to the M31  $z=0$ values.
 \bt{Observational values are shown as} horizontal dashed lines with the same color code.
 Very thin lines show the results obtained if we use a 2 times smaller bin size in the 4GND plot method. 
  \bt{Table \ref{table_n4p_fixfrac} provides the specific parameter values for observations and simulations as averaged in the last 1 Gyr analyzed.}
  }
 }
\label{fixfrac}
\end{figure*}



\subsubsection{The highest-quality plane of satellites at each timestep}

A more compact  and clearer  way of presenting the results on plane quality analysis showed in 
Figure  \ref{aqc_ca_fracA}
is looking at the ``best"  plane found at each timestep including a fixed fraction  of  the total number of satellites. 
 This best plane is selected as the one with the lowest $c/a$  at fixed $f_{\rm sat}$,
 \bt{ which can be easily read from  Figure   \ref{aqc_ca_fracA}}
 \bt{(note that this best quality plane does not necessarily  correspond to the same peak as $f_{\rm sat}$ changes)}.

\bt{In Figure \ref{fixfrac}  we show the properties of the "best" planes of satellites found at each timestep. In particular, we focus on  $c/a$,  $\Delta$RMS height and the inclination of the plane relative to the disk (latitude angle). }
 Different  shades of blue  stand for planes   with different   $f_{\rm sat}$=  30\%, 50\%, 70\%, 90\%. 
We show results  with and without applying the observational obscuration bias in the  left and right panels, respectively. 
\bt{In the right panels, green (red) lines  mark the values of the Universe ages where $C_1$ maxima (minima) appear in Figure \ref{PeakStrength}.}
For comparison, the results for the best planes in the MW and M31 at $z=0$ with the same $f_{\rm sat}$  are also shown as horizontal lines.
\bt{These values, together with   those  of simulated results averaged over the last 1 Gyr  analyzed, } are given in Table \ref{table_n4p_fixfrac}.  
 Note how at high  $f_{\rm sat}$ ($70\%$ and $90\%$ lines)  M31 presents very large $c/a$ and $\Delta$RMS, due to the system configuration in two almost perpendicular planes (see Paper I).

\begin{table*}
\centering
\caption{
\bt{Plane parameters
 of the  best (i.e., with lowest $c/a$) planes found in Aq-C$^\alpha$ and PDEVA-5004   including a fixed fraction  (X\%)  of satellites. 
We show $c/a$, $\Delta$RMS and $D_{\rm CG}$.
 Values are averaged over the last Gyr of the corresponding analysis period (see Figure \ref{fixfrac}).
 Fractions for the MW and M31 have been calculated relative to a sample size of $N_{\rm tot}$= 27 and 34 satellites, respectively. The results shown are the mean values and standard deviations resultant of 100 random realizations of radial distances within the observational uncertainties.}
 }
 \tiny
\begin{tabular}{ll | lll  | lll  | lll | lll }
\toprule
&    & \multicolumn{3}{c}{Aq-C$^\alpha$ (last 1 Gyr)} & \multicolumn{3}{c}{PDEVA-5004 (last 1 Gyr)}       & \multicolumn{3}{c}{MW ($z=0$)}         & \multicolumn{3}{c}{M31 ($z=0$)}        \\
&    & $c/a$ & $\Delta$RMS & $D_{\rm CG}$ &   $c/a$ & $\Delta$RMS & $D_{\rm CG}$ & $c/a$ & $\Delta$RMS & $D_{\rm CG}$ &    $c/a$ & $\Delta$RMS & $D_{\rm CG}$  \\
&    &   & kpc & kpc   & &kpc &  kpc  &   & kpc & kpc  &   & kpc & kpc  \\
\hline
\hline
\multirow{3}{*}{bias}    & 30\% &  0.03     &  5.30    &  23.70     & 0.03      & 2.46   & 12.43     &  0.07$\pm$0.001   &  9.54$\pm$0.15    &  13.23$\pm$0.15 &   & &    \\
                         & 50\% & 0.07      &  12.70    &   23.77     & 0.06      & 6.27   & 11.96      &   0.10$\pm$0.001   & 10.31$\pm$0.12    & 14.61$\pm$0.11    & &  & \\
                         & 70\% &  0.14     & 23.82     &  23.07    &  0.15     &  13.12  &  6.70     &  0.14$\pm$0.002   &   12.67$\pm$0.18   & 15.72$\pm$0.15      &  &  &  \\
                         & 90\% & 0.26      & 41.05     &  12.60     &  0.25     & 21.11   &  4.13    &    0.21$\pm$0.002  &   19.39$\pm$0.19   &  10.46$\pm$0.15    &  &  &   \\
 \midrule
\multirow{3}{*}{no bias} & 30\% & 0.04   &   6.72  &  27.92       & 0.03      &  3.45      &  7.76   &   &   &     & 6.24$\pm$0.003    &  1.21$\pm$0.35 &6.17$\pm$0.57  \\
                         & 50\% & 0.09   &  15.46  &  16.53       & 0.09      &    9.22    &  5.30   &   &   &     & 9.56$\pm$0.002  &  2.05$\pm$0.20 &6.29$\pm$0.58    \\
                         & 70\% & 0.17   &  26.02   &   17.73     & 0.15      &   13.14    & 5.83    &   &   &     & 25.83$\pm$0.008  & 3.30$\pm$0.83  &2.62$\pm$2.04   \\
                         & 90\% &  0.28  &  40.98   &   14.42     & 0.23      &   19.68      & 3.92  &   &   &     &  60.06$\pm$0.009  & 8.02$\pm$1.18  &10.17$\pm$6.27 \\        
   \bottomrule
\end{tabular}
\label{table_n4p_fixfrac}
\end{table*}

In terms of $c/a$ and $f_{\rm sat}$,  both Aq-C$^\alpha$  and PDEVA-5004  simulations  present high quality planes.
The best planes of satellites take $c/a$ values that change with time, reaching at some timesteps, and particularly near the respective last periods analyzed, values compatible with those  in the MW and M31 at $z=0$ involving the same fraction of satellites. 

\bt{We note that the fact that quality in our simulations increases towards low redshift is in contrast to \citet{Shao19} findings, who report thinner planes of satellites at early times in EAGLE.}
\bt{On the other hand, Figure \ref{fixfrac} shows biased $c/a$ results are systematically somewhat lower than non-biased ones in both simulations. This occurs because when applying the bias we are removing a fraction of the volume where satellites can be considered, which prevents plane-thickening. This result may indicate that the quality of the MW's planar structure of satellites can appear artificially enhanced because of Galactic obscuration.}

In terms of $\Delta$RMS, 
PDEVA-5004 reflects similar results and the same fluctuation patterns seen with $c/a$. 
Especially at low redshifts, very low $\Delta$RMS heights are found.
In Aq-C$^\alpha$,
\bt{despite the low $c/a$ values, 
 we find larger $\Delta$RMS values. This is because $\Delta$RMS is a dimensional quantity that therefore depends on the overall size of the system at issue. }
 This parameter very clearly decreases
  as the system evolves:
   Aq-C$^\alpha$  is still settling its size  until $\rm T_{uni} \sim$ 9 Gyr. 
At the last moment of our analysis ($\rm T_{uni}=11.5\,$ Gyr) the $\Delta$RMS heights of planes are
generally  compatible with their observed counterparts at $z=0$
 \bt{(except for the `biased' (versus MW) results involving $f_{\rm sat}$ 90\% and 70\%, 
 and the `non-biased' ones (versus M31) involving a 50\%).} 
 \bt{However we note that Aq-C$^\alpha$ has still 2 Gyrs to reach $z=0$,
 and the system could still evolve towards a lower $\Delta$RMS  value.}

\bt{
In the third rows of each panel of Figure \ref{fixfrac},  we plot the 
 angle formed at each timestep by the normal vector to the plane of satellites and 
 the  galaxy's  disc plane (we use Galactic latitude angle, with $|b|\in$[0, 90]$^\circ$). 
 We can see that while at low $f_{\rm sat}$ 
  curves show a fluctuating behaviour with T$_{\rm uni}$, the fluctuation level decreases as $f_{\rm sat}$ increases, and, finally, almost no fluctuations show up at $f_{\rm sat}$=90\%. 
  An important variation  in   this angle is an   indication  that the identities of  the satellite members of planes with given $f_{\rm sat}$ have changed. 
  Therefore these    results are indicating   that the  satellite members of the best quality planes change quite a lot at low $f_{\rm sat}$, while at $f_{\rm sat}$= 70\% or even 50\%, these identities are  kept to an important extent.
}

\bt{
 Moreover,  at times when $c/a$ reach  their minima (and  the main peak strength $C_1$ reach their maxima)
 the latitude angle  in both simulations  is small and sometimes close to 0$^\circ$
 (except for PDEVA-5004 at $z\approx 0$);  that is, satellite planes are near   to perpendicular to the galaxy's disc. 
 This result suggests that the best quality of satellite planes  is 
in many cases 
 reached at time intervals when no (or rather low) galaxy disc torques act upon the satellites belonging to the plane that best fits the satellite set \citep[see e.g.,][]{Danovich15}.}\footnote{
\bt{
 We notice the 
 robustness of our results against the bin size in the 4GND plots. 
The very thin blue-series lines  in Figure \ref{fixfrac} show results calculated with half the bin size used to calculate the  thicker  lines therein. Differences are small and unimportant.
}
}

Finally, we focus on the  distances $D_{\rm CG}$, or offsets, from the center of the galaxy to the previously presented best  planes found in Aq-C$^\alpha$ and PDEVA-5004.
Contrary to other plane-identification methods used in simulation studies \citep[see e.g.][]{Gillet15,Buck15,Ahmed17}, 
in the 4GND plot method planes are not constrained to pass through the center of the main galaxy.
Table \ref{table_n4p_fixfrac} shows  the averaged over the last 1 Gyr   offsets to the best planes with  $f_{\rm sat}=$30, 50, 70, 90\%.
\bt{For comparison,} the  MW VPOS-3 (f$_{\rm sat}=$24/27=88\%) and M31 GPoA (f$_{\rm sat}=$19/34=56\%) planes present an offset of  10.4 and 1.3 kpc from the center of  the MW and M31, respectively \citep[see Table 3 in][]{Pawlowski13}.
Also, the collection of planes defined by the second peak in the \bt{4GND} plot of M31  presents distances $D_{\rm CG}$ between $ \sim $  15  and 35 kpc (see Paper I). 
 Compared to  \bt{these,} the plane offsets measured in both simulations have reasonable values, passing   close to the center of the main galaxy.
\\


This section  reveals that  there are indeed   preferential planar configurations of satellites at given moments in both Aq-C$^\alpha$ and PDEVA-5004 simulations. 
  These planes  are thin and highly populated, compatible on average with all  characteristics of the observed planar structures found in the MW and in M31 at $z=0$,  and even defining   higher quality  planes at particular given times.
A rough comparison to the
planes of satellites 
reported in previous studies with  zoom-in  hydro-simulations \citep{Gillet15,Ahmed17,Maji17a}  
 focusing on the $c/a$, $\Delta$RMS and $N_{\rm sat}$ parameters, strongly suggests that the planes found in Aq-C$^\alpha$ and  PDEVA-5004  have,
 within some time intervals, a higher quality and reveal a higher degree of  spatial ordering  in the satellite distribution.
However this comparison 
is not completely unbiased due to the different
\bt{types of simulations  and}
 methods for plane-identification used
  in our study and in
  others.
  \bt{ First, only comparisons between zoom-in simulations that meet the conditions listed in section 2 of this paper make sense:   the dynamical effect of a  massive, MW-like disk potential on satellite planes could be an important piece of the puzzle. Second, in the 4GND plot method no priors are assumed:  we do not choose the 11 most massive (or most luminous) satellites among the simulated satellite sample,  planes are not required to pass through the center of the host galaxy, or to be thinner than a given RMS thickness, etc. }




\section{Co-orbitation?}
\label{Coorbit}

\begin{figure}
\centering
\includegraphics[width=0.5\linewidth]{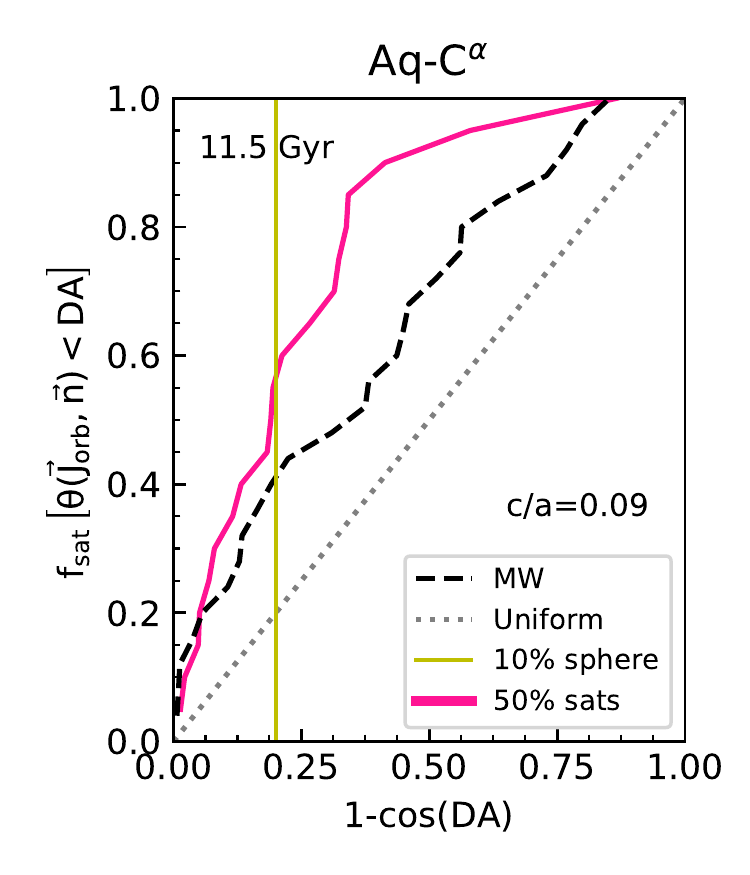}
\includegraphics[width=0.445\linewidth]{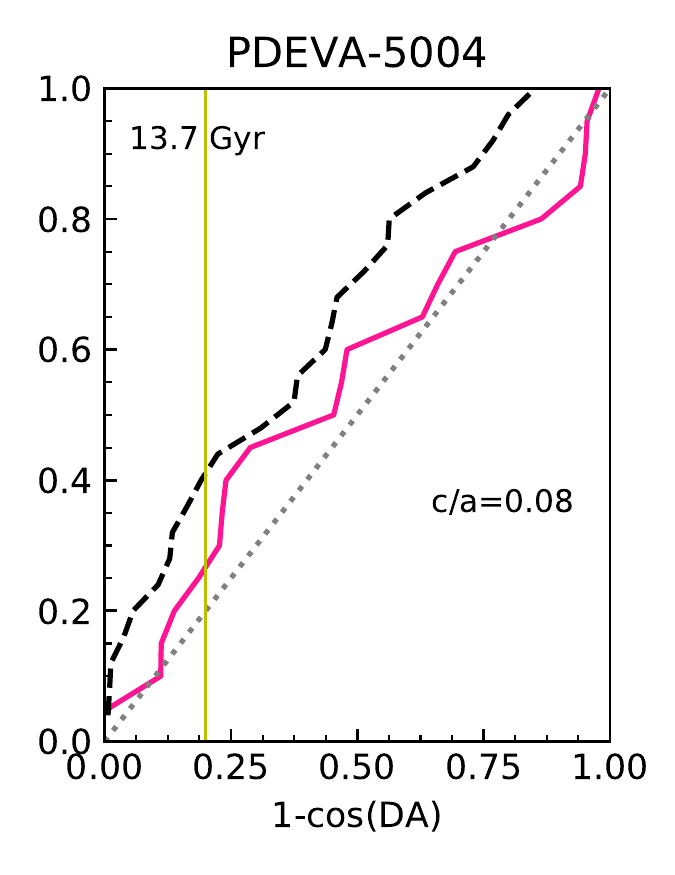}
\caption{\bt{
Fraction of satellites with orbital poles enclosed within an angular distance 'DA' 
 measured from the normal to the best (i.e., lowest $c/a$) plane  including a fraction $f_{\rm sat}=$50\% of the total number of satellites (see Figure \ref{fixfrac}).
 Left: Aq-C$^\alpha$; right: PDEVA-5004; at their last analyzed timesteps. Results include the observational obscuration bias.
  The dotted line shows the result for a uniform distribution of orbital poles on the sphere, and the dashed line the result for the confirmed satellites in the MW, using data from   \citealt{Pawlowski13b,Fritz18}.
A yellow vertical line marks an angle of DA=36.78$^\circ$.
MW satellites with orbital poles enclosed by this angle as measured from the VPOS are considered to co-orbit in \citealt{Fritz18}.
 }
 }
\label{pdeva_uno}
\end{figure}

\begin{figure*} 
\centering
\includegraphics[width=0.49\linewidth]{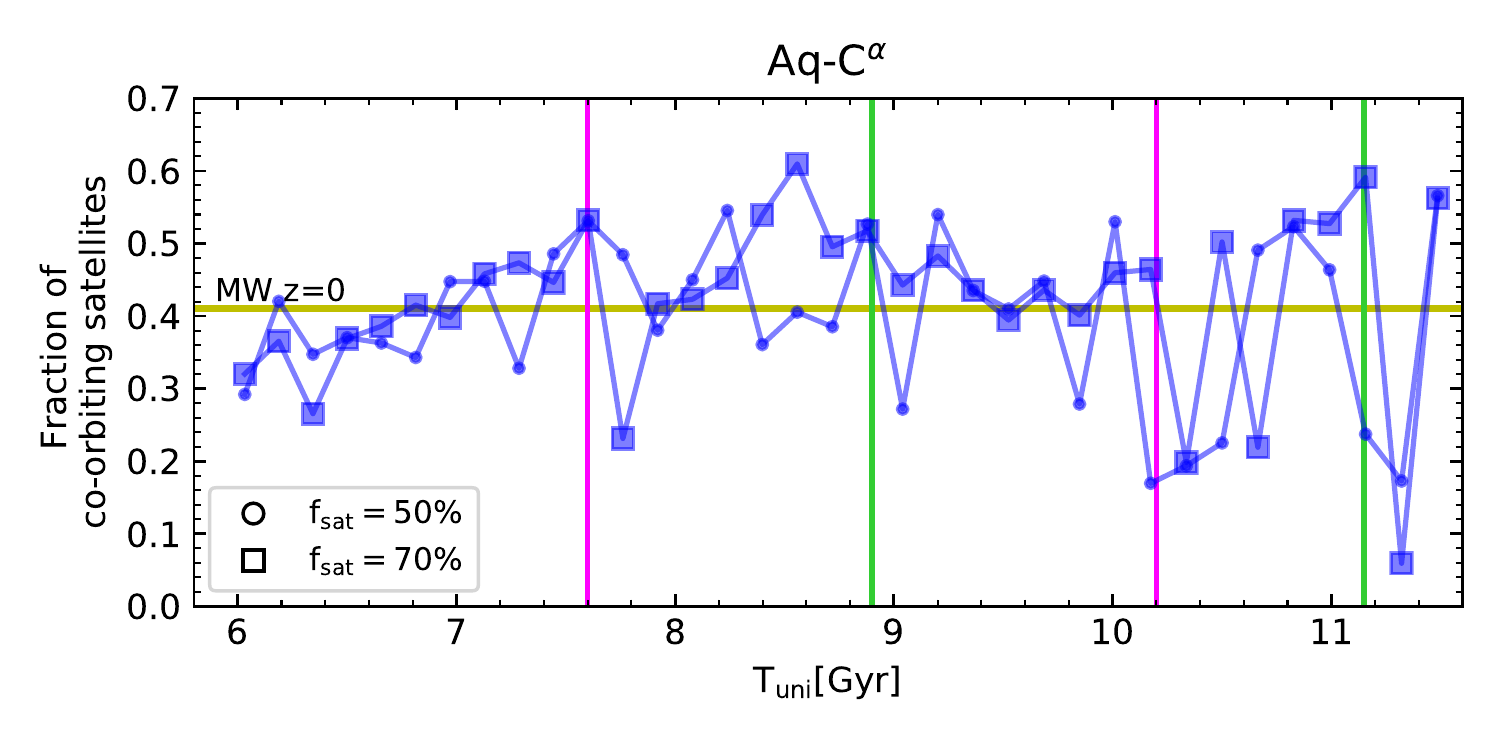}
\includegraphics[width=0.49\linewidth]{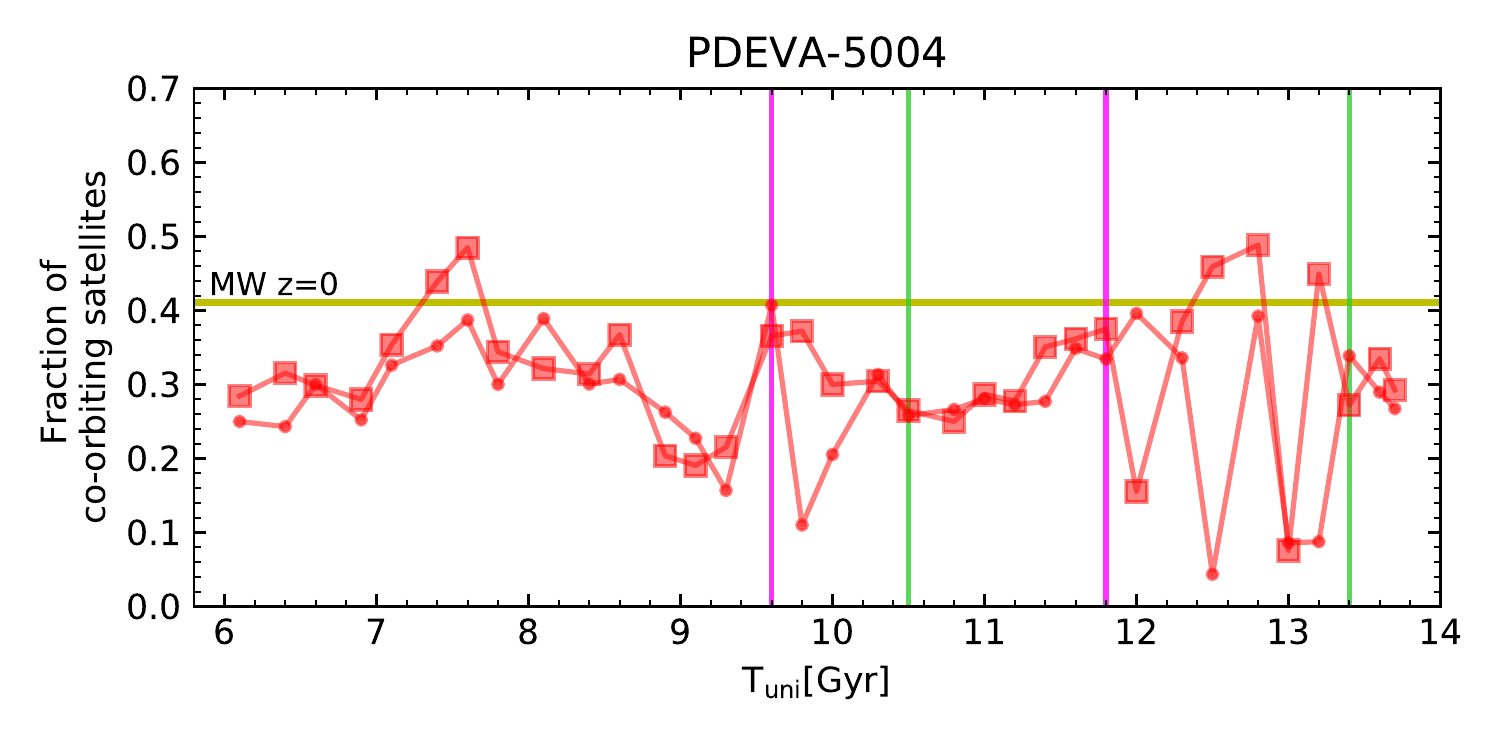}
\caption{
\bt{The fraction of co-orbiting satellites in the best quality planes involving $f_{\rm sat}$ = 50 and 70\% of satellites at each timestep, versus the Universe age. 
Left: Aq-C$^\alpha$; right: PDEVA-5004. The observational obscuration bias has been applied to obtain these results.
Co-orbiting satellites are those  with orbital angular momentum vectors  within 36.78$^\circ$  around  the normal to the plane. Green (magenta) vertical lines mark time intervals where the main peak strength  $C_1$ reaches maximum (minimum) values (see Figure \ref{PeakStrength}).
An horizontal line marks the corresponding fraction of co-orbiting satellites in the MW at $z=0$ according to Figure \ref{pdeva_uno}.
} }
\label{CoorbvsTuni}
\end{figure*}

\bt{One relevant feature of the main plane } of satellites observed in the MW is that  it  presents a relatively high degree of coherent rotation within the plane (see Section \ref{intro}). 
This means that the orbital angular momentum vectors (i.e., orbital poles, $\vec{J}_{\rm orb}$) of the  constituent satellites are   aligned with the normal to the plane.
Orbital angular momentum is defined as
 $\vec{J}_{\rm orb} = \vec{r} \times m\vec{v}$, 
 where $\vec{r}$ and $\vec{v}$ are the position and velocity of the center-of-mass of the satellite relative to the center-of-mass of the host disc galaxy.

To  study if the satellites included in  the high-quality planes detected in the simulations with the extended 4GND plot method 
are co-orbiting within the plane,   we first  compute the  $\vec{J}_{\rm orb}$  vectors  of the satellites at each timestep and project them on the sphere. 
 Then,   we quantify the clustering of  $\vec{J}_{\rm orb}$ vectors around a given $\vec{n}$ direction (where   $\vec{n}$ is the normal vector  to a given plane),   and evaluate the fraction of satellites that are kinematically-coherent (i.e. co-orbit)  within the  plane.
To this end, we take this direction  $\vec{n}$  as a reference axis,   and measure the angular distance $DA$
to each individual satellite orbital pole.
 In order to do this systematically at each timestep, we take  as reference axis   the normal  $\vec{n}$   to  the best plane (i.e., with lowest $c/a$)   including 50\%  of the total number of satellites at the respective timestep (see Figure \ref{fixfrac}).

\bt{This is exemplified in Figure   \ref{pdeva_uno}, for the last timestep analyzed in each simulation.}
 The x-axis shows $1-\cos(DA)$, 
 ranging from 0 to 1 
 \rb{ as we do not differentiate  between co-rotation or counter-rotation with the  disc of the central galaxy, and $DA$ can be a maximum of 90$^\circ$. }
 The y-axis shows the fraction of the total number of satellites with  $\vec{J}_{\rm orb}$ enclosed by a certain angular distance $DA$  from the reference axis $\vec{n}$.
  Since it is only possible to compare to   MW data (no proper motion data is available for M31 satellites), the results shown include the observational obscuration bias.
The dashed line shows the MW case, where we use the latest available data for the confirmed MW satellites 
 (Table \rb{4} in \citet{Fritz18} calculated from \textit{Gaia}-DR2 data, or alternatively, 
for the satellites missing there, 
  Table 4 in \citet{Pawlowski13b})\footnote{We consider in this plot a final sample of 25 out of 27 MW satellites as there is no published  proper motion   for Canis Major or Bootes III.}.
For  comparison, a dotted line illustrates the  expectation from a  uniform distribution of orbital poles,
 and  a yellow vertical line is depicted at $DA=36.78^\circ$ which encloses 10\% of the sphere surface
 \rb{(or 20\% when it is not distinguished if satellites co- or counter-rotate with respect to the disc)}:
  this is the angle  around the VPOS within which it is considered in \citet{Fritz18} that MW satellites  co-orbit
   ($\gtrsim$ 40\% of MW satellites co-orbit, \rb{see Paper I}).  
\bt{We see at these example timesteps that while for Aq-C$^\alpha$  the fraction of co-orbiting satellites is $\sim$45\%, for PDEVA-5004 it is much lower, and below the MW fraction.}

\bt{We use the previous analysis to show in}  Figure \ref{CoorbvsTuni} 
  the fraction of co-orbiting satellites in the best quality planes involving $f_{\rm sat}$ = 50 and 70 \% of satellites at each timestep.  
Figure  \ref{CoorbvsTuni}  indicates that  while in some cases there is consistency with the MW or even a higher degree of co-orbitation (particularly so in the Aq-C$^\alpha$ case), in others we can see that  the fraction of co-orbiting satellites around the  $\vec{n}$  direction is very low,   despite these directions defining the highest-quality   spatial planar arrangements found at that moment. 
\bt{Moreover, the abrupt changes in the fraction of  co-orbitating satellites  from one timestep to another is a consequence of the different identities of satellites constituting the best quality planes at close times. 
Important differences in the fraction of co-orbiting satellites are also found between the 50\% versus de 70\% case at a same timestep:  an indication that while one of them shows a kinematical coherence, the other does not. 
}

\bt{
In this respect, it  is interesting to compare the time locations of the $C_1$ maxima and minima (see Figure \ref{PeakStrength}) 
with a measure of the co-orbitation of the involved satellites.
Figure \ref{CoorbvsTuni} indicates that,  in both simulations, vertical green  (magenta)  bands not always correspond to a high (low) fraction of co-orbiting satellites.
}

These results imply that, in general, the best planes in positions found with the  4GND plot method may just be  fortituous spatial alignments of satellites, and therefore \textit{transient} structures
\citep[see also][]{Gillet15,Buck16,Fernando17,Shao19}.

 In order to efficiently detect  kinematically-coherent planes of satellites in these simulations, a deeper and more precise  analysis of the persistence or not of good quality planes of satellites across time is needed,
 which  demands using the full six-dimensional phase-space information of satellites  for plane searching. 
In a forthcoming paper  (Santos-Santos et al., in prep.;  Paper III)  a new method is developed to address the plane of satellites persistence issue 
\rb{(see also Paper I for its application to MW data)}.

%
%

\section{Discussion}
\label{Discu}

\subsection{Quality analysis in terms of $N_{\rm sat}$}
\label{NsatA}

\bt{
The use of normalized quantities such as $C_{\rm p}$ in the peak analysis makes results independent of the total sample sizes $N_{\rm tot}$.
In this line, for the quality analysis of planes we have used
 $f_{\rm sat}$, a $N_{\rm tot}$ independent quantity allowing a clean comparison of samples of different size.
}

\bt{
The analysis has been repeated through $c/a$ versus $N_{\rm sat}$ (i.e., the absolute number of satellites instead of its fraction $f_{\rm sat}$).
To this end,
following \citet{Riley2019},
 the total number of satellites $N_{\rm tot}$ in simulations and observational samples has been matched at each timestep.
 In particular, for each simulated satellite, we compare its distance to its host with the  galactocentric distance of all  MW (or M31) satellites,  and select 
from these 
 the optimal  match (without replacement).
 In this way, a sample of observed galaxies is built with $N_{\rm tot,obs}=N_{\rm tot,sim}$.
This matching is needed for a proper comparison because in simulations $N_{\rm tot}$ depends on time (see Figure \ref{ntotvsTuni}), while in the $z=0$ satellite system of the MW and M31   $N_{\rm tot}$ is a fixed number.
Results
 are qualitatively the same as those obtained in terms of $f_{\rm sat}$.
 
We note that without $N_{\rm tot}$-matching, results on consistency with observations  can be easily obtained from Figure \ref{aqc_ca_fracA} by translating the observational curve rightwards an amount $\frac{N_{\rm tot, obs}}{N_{\rm tot, sim}}$.
Inconsistency or not would be due to the value of this number, with possible inconsistencies resulting  from the different size of simulated and observed  satellite samples; 
which justifies why this exercise is needed.
}

\subsection{Can the peak strength be used as a measure of plane quality?}
\label{PeakasQ}

\bt{
Green (magenta) vertical lines in Figure \ref{fixfrac}  mark the T$_{\rm uni}$ values where the main peak strength $C_1$ have maxima (minima)   in Figure  \ref{PeakStrength}.
A very relevant result is that maxima occur at T$_{\rm uni}$ values when the $c/a$ value  of the $f_{\rm sat}$=90\% curve is minimum, that is, when the quality shows a maximum. And conversely, magenta lines are close to  maxima of $c/a$,
 that is, bad qualities.
To find out whether this behaviour keeps  at other $f_{\rm sat}$ values,  and whether we can use peak strength $C_p$
to measure quality at given $f_{\rm sat}$,  we have calculated the main peak strength as a function of $f_{\rm sat}$, i.e., $C_1(f_{\rm sat})$\footnote{$C_1(f_{\rm sat})$ is calculated as explained in subsection \ref{PeakStre}, except that we do not sum over all the satellites. We stop summing up when a given satellite fraction, $f_{\rm sat}$, is reached. For example $C_1$(50\%) in the upper panel of Figure \ref{ex_cont}, would just involve the 10 first satellites (which are ordered by decreasing $C_{1, \rm s}$).},
 and compared it to the $c/a$ of the respective best plane found with same $f_{\rm sat}$ at that timestep. 
}

\begin{figure}
\centering
\includegraphics[width=\linewidth]{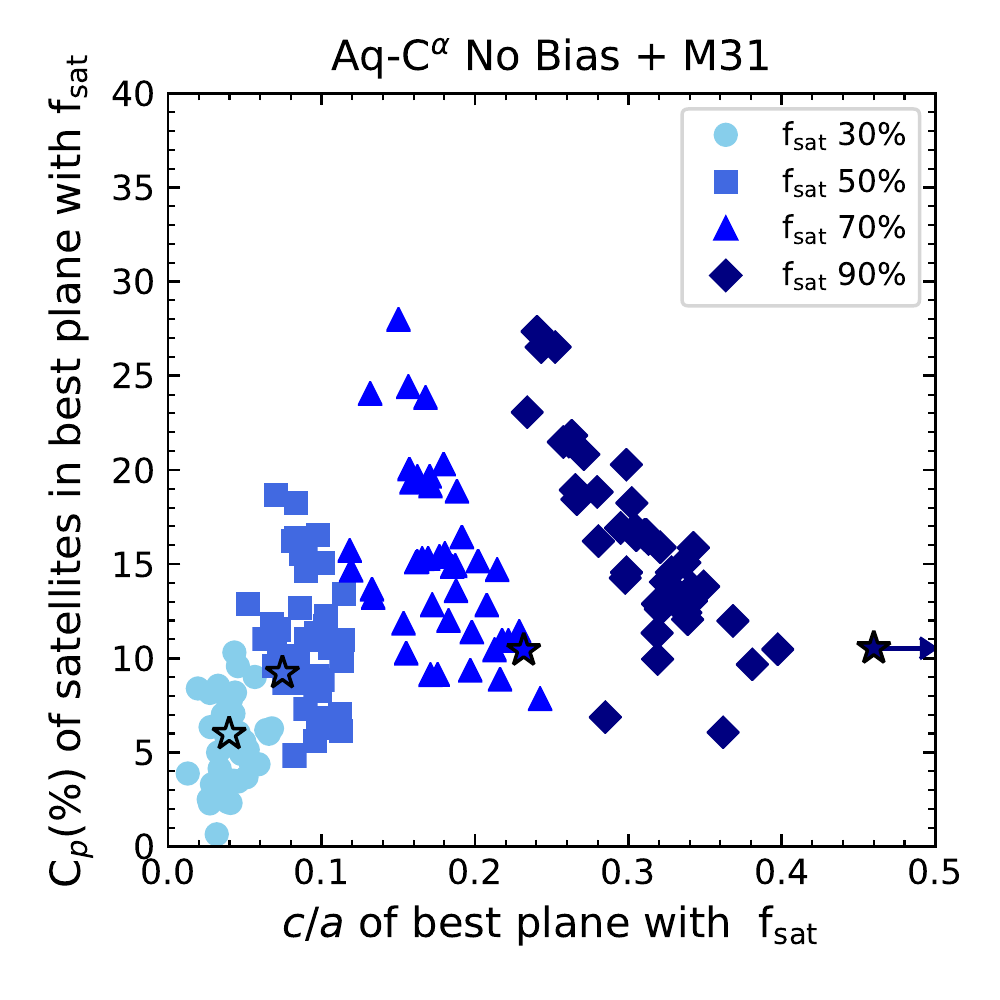}
\caption{
\bt{
Main peak strength considering different  $f_{\rm sat}$,  $C_1(f_{\rm sat})$, versus the short-to-long axis ratio $c/a$ of the best plane with  $f_{\rm sat}$ satellites. Different blue shades stand for different  $f_{\rm sat}$ values, with the same colorcoding as in Figure \ref{fixfrac}.  Stars show the M31 observational values at $z=0$. 
}
 }
\label{CpvsCa}
\end{figure}

\bt{
As an example,
in Figure \ref{CpvsCa} we see for Aq-C$^\alpha$  ('no bias' case, where all satellites are considered)
that a correlation exists at all $f_{\rm sat}$. There is, however, an important dispersion particularly at low $f_{\rm sat}$. 
The same qualitative behaviour is found for PDEVA-5004, and in the 'bias' case for both simulations.
Therefore we conclude that although indicative, the peak strength $C_p$ is not an accurate enough measure of the quality of 
its collection of planes.
}

\bt{
It is interesting to note that, when absolute numbers  of satellites  ($N_{\rm sat}$, without $N_{\rm tot}$-matching) are used instead of fractions  ($f_{\rm sat}$) to analyze quality,
the correlations shown in Figure \ref{CpvsCa} disappear  or are  very  weak. This is a clear indication that the mutual relationships  between   quality and peak strength are best manifested when the quality analysis is made in terms of  $f_{\rm sat}$, rather than $N_{\rm sat}$.
}

\subsection{Does radial compactness of the satellites affect quality?}
\label{RadialCompact}

\begin{figure*}
\centering
\includegraphics[width=\linewidth]{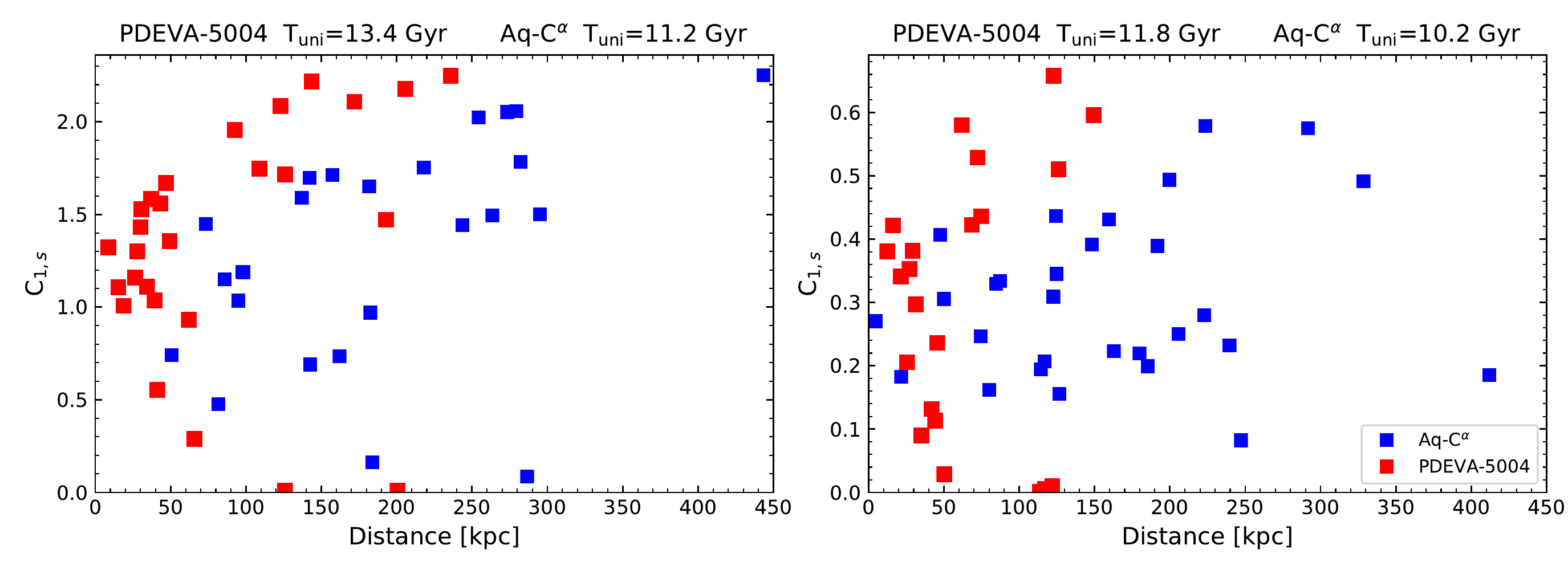}
\caption{ 
\bt{The contribution of each satellite $s$ to the main peak ($C_{1,s}$), versus the radial distance of the satellite to the center of its host disc galaxy. 
The left panel corresponds to moments in the evolution of the respective simulations where $C_1$ is very high ($C_1\sim33$\% for both simulations, see Figure \ref{PeakStrength}). The right panel corresponds to moments when $C_1$ is relatively  lower ($C_1\sim10$\% for both simulations).  }}
\label{radcomp}
\end{figure*}

\bt{
In this subsection we analyze the possible correspondence between plane quality and the radial compactness of a satellite system.}

\bt{
Figure \ref{radcomp} shows $C_{1,s}$, i.e., the contribution of each satellite $s$ to the main peak, versus the radial distance of the satellite to the center of its host disc galaxy. 
The left panel corresponds to moments in the evolution of the respective simulations where $C_1$ is very high ($C_1\sim33$\% for both simulations, see Figure \ref{PeakStrength}), while the right panel corresponds to moments when $C_1$ is relatively lower ($C_1\sim10$\% for both simulations).  
}

\bt{
There is a clear  correlation for further away satellites to contribute more to a given peak (higher $C_{\rm p,s}$).
At fixed $C_1$, this effect is slightly intensified in systems where there is an agglomeration of satellites at shorter distances (more compact systems). This is expressed with systematically higher $C_{\rm p,s}$ values, but a similar slope in the correlation. See for example the left panel, where in PDEVA-5004 --which is a more radially compact system than Aq-C$^\alpha$-- satellites show higher $C_{1,s}$ values.
Moreover, if we compare PDEVA-5004 satellites in the left and right panels (i.e., timesteps where the  $C_1$ is different  but there is a similar radial compactness of satellites) 
we see that  the $C_{1,s}$   is dramatically lower for the system with lower $C_1$.
}

\bt{
While a spatial configuration with several central satellites and a few further away ones (i.e., a compact system) increases the probability of 4-galaxy-normals pointing in a similar direction, from these results it is clear that this geometric effect is not the driving reason for having a high clustering of 4-galaxy-normals or high $C_1$. 
We conclude that radial compactness does not have a determinant role in setting plane quality.
}


\section{Summary and conclusions}\label{conclu}

 \bt{
To address the so-called ``Planes of satellites problem" (see discussion and references in Section 1)
 we have applied for the first time the 4-galaxy-normal density (4GND) plot method \citep{Pawlowski13} 
to hydrodynamical simulations. An  extension of the method, sketched in 
\citep[][Paper I]{SantosSantos19}, is presented in detail and discussed in this paper.
}

\bt{
Our choice has been zoom-in simulations because the higher resolution (as compared to large-volume simulations) one can reach by using this method allows 
MW-type galaxies with  massive, extended discs to form, so that the dynamical effects of the disc potential on satellite planes (torques, destruction of satellites on  radial orbits) can be accounted for.
Another  \bt{advantage of zoom-in simulations }
is that  they offer the possibility that  these MW-like disc galaxies are 
surrounded by a number of resolved satellites high enough 
\bt{to be comparable}  to current samples of confirmed MW or M31 satellites \citep[$\sim $ 30, see][]{Pawlowski13}.
Well-behaved discs and a large enough satellite number are conditions not recovered, by the moment, by  larger volume cosmological simulations.
}

\bt{
The extension to the 4GND method
is designed to identify, systematically catalog and   study in detail
 the quality of the predominant spatial planar configurations  of satellites
 revealed by  over-densities in the  4GND plots.
It allows to extract the best quality planes out of the number of combinations we can form with $N_{\rm sat}$ \bt{satellites}  from a sample of size $N_{\rm tot}$, with a low computational cost. }
Quality is  evaluated through the outputs of a  Tensor of Inertia  analysis \citep[ToI,][]{Metz07} using a normal-regression fitting technique.
Quantitatively, planes (i.e., best-fitting solutions with high concentration ellipsoid medium-to-long axes ratio, $b/a \sim 1$),
 have a good quality if they are populated relative to the sample size $N_{\rm tot}$
(high f$_{\rm sat} \equiv N_{\rm sat}/N_{\rm tot}$) and thin (low 
short-to-long axis ratio $c/a$).
Being a two-parameter notion, the quality of two or more planes can be compared 
 if one has lower $c/a$ and higher $f_{\rm sat}$ than another, or
if either  $f_{\rm sat}$ or $c/a$ are constant.

\bt{
Density  peaks are determined by local, isolated maxima in the  4GND plot.
We have defined the  peak strength, $C_p$, as the \% of 
weighted 4-galaxy-normals within 15$^\circ$ of the peak center.
Peaks can be compared with each other across times through their respective strengths, and with those of the MW and M31 satellite systems as well. 
Different satellites contribute differently to $C_p$. 
The satellite $s$ contribution to peak $p$ (i.e., $C_{\rm p,s}$) is defined as the normalized, weighted count of $s$  contributions  to  4-galaxy-normals placed within 15$^{\circ}$ of the peak center.
}
\rb{Satellites are ordered by decreasing $C_{\rm p,s}$ to a peak, and a plane is fitted to groups of increasing $N_{\rm sat}$ satellites following this order. This yields, for each density peak, a \textit{collection} of planes.}

In Paper I  we report on the application of the extended 4NDP method to the confirmed
MW and M31 satellites.
The method extension reveals a richer planar structure, 
allowing to find planes of satellites around the MW and M31 with higher qualities than those previously reported with a given  $N_{\rm sat}$.
\bt{We find }
 a second populated, high quality  plane around M31.
Another important result,
 in view of the narrow range of satellite mass distributions 
 \bt{that can be currently afforded} in hydrodynamical simulations, 
 is that  satellite mass plays no role
 in determining a satellite's  membership or not to the respective best-quality planes. 
This enables us to perform, through the  extended 4GND plot  method,  
 comparisons between    results  from simulations  and   MW/M31
 data 
  (where the satellite mass range is wider).

 In this paper we present results of a detailed search for  positional planar structures  of satellites  in  two different \bt{(initial conditions, subgrid modelling and numerical approaches)} zoom-in cosmological hydrodynamical simulations of  isolated     MW-like disc galaxies: Aq-C$^\alpha$ and PDEVA-5004. 
They meet the conditions \bt{of having} a central host galaxy with an extended disc, an overall quiet  merger  history,   a numerous ($\sim 30$) satellite population, and more than 50 baryonic particles per satellite. No other particular selection criteria have been applied.
 In particular,  
we focus our analysis  on the best quality planes  with a given satellite fraction $f_{\rm sat}$. 
 We  compare to the best quality planes found in the MW and M31 systems at \bt{$z=0 $ with} given $f_{\rm sat}$, i.e., those found from their so-called respective \#1 Peaks as defined in Paper I.

\bt{
The analysis goes from the halo virialization time to low redshifts, along the slow phase of host mass assembly. 
The halo mass growth histories of \bt{Aq-C$^\alpha$ and PDEVA-5004}   
  \bt{present as usual two phases, and  no other relevant particularities.}
Satellite samples have been  identified at $z$=0.5.
The number of satellites $N_{\rm tot}$ considered in the analysis  at each timestep varies. In particular,   satellites are not  considered    when they are  orbiting beyond 450 kpc from the center of the host galaxy,  after they have been accreted by the host galaxy, or if they are within the avoidance volume when we apply an observational obscuration bias to compare with the MW. 
The distributions of \bt{satellite } radial distances 
to the \bt{center-of-mass} of the main galaxy vary with time. Expressed in terms of $f_{\rm sat}$, Aq-C{$^\alpha$}'s (PDEVA-5004's) radial \bt{distance} distribution is very close to \bt{that of } M31 (the MW), at particular times. 
}

The analysis of density peaks 
  reveals that  $C_1$ \bt{varies with}
 T$_{\rm uni}$, with time intervals where it is comparable or even higher than 
\bt{the MW value,} 
 $C_{1, \rm MW} = 22.91 \pm 0.26 $ \%.
 The number of peaks that strong changes with time too, reaching values of 1 - 2 at most and only in those 
time intervals when $C_1$ is high. In this regard, 
\rb{density peaks found in simulations}
are similar to those observed in strength and number, 
\rb{during} given time intervals.

 We find   \textit{planar} (i.e., high $b/a$ and low $c/a$) configurations of satellites at all studied timesteps  in both simulations.
 Indeed, no  filamentary  (i.e., $b/a<<1$) 
 configurations have been detected in the period analyzed.
The extended 4GND plot method allows to identify, at many timesteps in both simulations, i) planes of satellites with qualities that are compatible with the observed ones at $z=0$ including a specific $f_{\rm sat}$, and also in some cases, ii) planar structures that are compatible with the observed ones for \textit{all} $f_{\rm sat}$.

 We study the best quality planes
  (i.e.,  with lowest $c/a$)
   including  a fixed $f_{\rm sat}$ 
 found at each timestep. 
 In both simulations,
their $c/a$ values   change with time,  independently of the  $f_{\rm sat}$ considered.
 Planes compatible  with the observed ones in the MW and M31 at $z=0$ are found at different timesteps or time intervals. 
\bt{Interestingly, these timesteps turn out to coincide with the time intervals where  $C_1$ shows maxima. 
More specifically, a correlation has been found between $C_1$ and $c/a$ at fixed $f_{\rm sat}$, but with important dispersion.
\bt{Therefore } $C_1$ can be used as an estimation of plane quality, but not to \bt{accurately} measure it.
Another interesting result is that the highest quality planes are often close to perpendicular to the host disc plane. Ideally, perpendicular planes of satellites would not suffer torques from
\rb{the disc.}
}

\bt{
No new information on quality is provided when using   $\Delta$RMS as quality indicator   once the satellite system size has settled. 
 And no new information on quality (in terms of comparison to observational satellite planes) is obtained either when using absolute satellite numbers $N_{\rm sat}$ instead of fractions $f_{\rm sat}$ = $\frac{N_{\rm sat}}{N_{\rm tot}}$, with $N_{\rm tot, obs}$ matched to $N_{\rm tot, sim}$. 
}

\bt{
Interestingly, when the observational obscuration bias is applied, slightly higher peak strengths are measured, as well as somewhat lower $c/a$ and $\Delta$RMS values than in the non-biased case.  The  is  due to  solid angle restrictions in the biased case, where satellites orbiting at low latitudes are neglected. 
}

\bt{
No clear, conclusive  signal on the correspondence between plane quality and the radial compactness of a satellite system has been detected in this work. While satellites at larger distances from the host galaxy 
provide somewhat higher $C_{\rm p,s}$ than nearby ones,
the spatial satellite configurations that show the highest $C_1$ (and therefore highest overall plane qualities) are not those most compact.
}

We further have investigated if satellites composing the high-quality planes of satellites found with
 the extended 4GND plot positional method 
  present a common  orbitation within the plane they describe.
 We find that in some cases the fraction of co-orbiting satellites is very low, which we interpret as a sign of these positional-planes consisting partly of interlopers: satellites that fall within the plane accidentally. 
 Therefore planes found with  this   method based on positional data   in general do not constitute a kinematic unit   and, in some cases,  could be  non-persistent in time.
\bt{
Because  of this, 
the search for the physical reasons favouring or destroying positionally detected  high-quality planes cannot be meaningfully addressed.
}
  In Santos-Santos et al. in prep. (Paper III), a  new methodology  is introduced where the full six-dimensional
phase-space information of satellites is used, leading to 
 the determination of persistent, kinematically-coherent planes of satellites in both simulations.


 Summing up, the application of  the 4GND plot method \citep{Pawlowski13} with its extension presented here,
\citep[see also][Paper I]{SantosSantos19}
   to two zoom-in hydrodynamical simulations of MW-type disc galaxies, leads us to the following conclusions:

\begin{itemize}[noitemsep,leftmargin=*]
\item
\bt{Satellites  are organized in planar configurations  (not  filamentary, i.e.,  $b/a\sim 1$)   in both the Aq-C$^\alpha$ and PDEVA-5004   simulations,  at all timesteps analyzed.  
 }
\bt{The plane short-to-long axis ratio $c/a$, and the plane population ($f_{\rm sat}$) measure plane quality. }

\item
\bt{The strengths of the strongest peaks in the 4GND plots ($C_{1}$)
vary  with time.}
\rb{Their values are consistent, along given periods of cosmic evolution, with that measured for the MW, and always higher than  for M31.}

\item 
\bt{The $c/a$ ratios of the best quality planes found
including a  fixed  $f_{\rm sat}$  
vary with cosmic time, and during some periods reach a high quality.
 Along these  good quality time intervals they
are compatible with   the observed planar structures found in the MW and in M31 at $z=0$.
These periods coincide with those when $C_1$ reaches maximum values.
The time-scale for these plane quality changes is  $\sim 0.5-1$ Gyr.
}

\item 
\bt{ $c/a$ and $C_1$ 
show correlations with increasing dispersion as $f_{\rm sat}$  decreases.
 $C_1$ can be used as a quality indicator, but not to accurately measure it.  }

\item 
\bt{The application of the observational obscuration bias enhances plane quality, either measured by peak strength or plane thickness.
}

\item
\bt{The orientations of planes of satellites with respect to the disc of their host galaxy change with time. 
  In most cases, planes are close to perpendicular to the disc during periods of good quality.  
}

\item
\bt{The compactness of the distribution of satellite-host radial distances 
does not have a driving role at setting the quality of   planes of satellites. 
}

\item
\bt{In agreement with previous findings, the fraction of co-orbiting satellites found in high quality positionally-detected planes is rather low, suggesting that these planes 
do not represent a kinematic unit and are not persistent in time. 
}

\item
\bt{The plane persistence issue in observations and simulations 
cannot be  properly addressed unless
 the  full six-dimensional phase-space information is considered.
 Such a methodology will   be developed in a forthcoming paper.
}

\end{itemize}

\bt{The general conclusion of this paper is that even if two galaxies do not make a  statistical sample, the fact that these two so different MW-like galaxies
(whose selection method is quite general) do have, at given time intervals, high quality positional satellite planes,
would suggest that these planes can be expected not to be infrequent in $\Lambda$CDM L* disc galaxies in periods when they are free of major merger events in their assembly history.}

\acknowledgements
\bt{We thank the anonymous referee for useful comments and suggestions that have helped improve this work.}
This work was supported through  MINECO/FEDER (Spain)  AYA2012-31101,  AYA2015-63810-P and MICIIN/FEDER (Spain) PGC2018-094975-C21 grants.
\rb{ISS acknowledges support by the Arthur B. McDonald Canadian Astroparticle Physics Research Institute. 
This work used the Ragnar cluster funded by Fondecyt 1150334 and Universidad Andrés Bello and Geryon cluster (Pontificia Universidad de Chile). We used a version of Aq-C-5 that  is part of the CIELO Project run in Marenostrum (Barcelona Supercomputer Centre).
This project has received funding from the European Union’s Horizon 2020 Research and Innovation Programme under the Marie Skłodowska-Curie grant agreement No 734374- LACEGAL. ISS acknowledges funding from the same Horizon 2020 grant}
for a secondment at the Astrophysics group of Univ. Andr\'es Bello (Santiago, Chile), and  from the Univ. Aut\'onoma de Madrid for a stay at the Leibniz Institut fur Astrophysik Potsdam (Germany). ISS thanks Dr. Patricia Tissera and Dr. Noam Libeskind for kindly hosting her.


\bibliography{archive_planes}{}
\bibliographystyle{aasjournal}

\end{document}